\documentclass[amsmath,amssymb,aps,prd,nofootinbib,twocolumn,superscriptaddress]{revtex4-2}
\usepackage[utf8]{inputenc}
\usepackage{mathrsfs}
\usepackage{bm}
\usepackage{url}
\usepackage[normalem]{ulem}
\usepackage{mathtools}
\usepackage{array}
\newcolumntype{P}[1]{>{\centering\arraybackslash}p{#1}}
\newcolumntype{M}[1]{>{\centering\arraybackslash}m{#1}}
\usepackage[caption=false]{subfig}
\usepackage{dcolumn}
\usepackage{graphicx, epsfig}
\usepackage[dvipsnames]{xcolor}
\usepackage{mathrsfs}
\usepackage{bm}
\usepackage{yhmath}
\usepackage[caption=false]{subfig}
\usepackage[normalem]{ulem}
\usepackage{mathtools}
\usepackage{bigints}
\usepackage{float}
\usepackage[colorlinks = true, linkcolor = purple, urlcolor  = blue, citecolor = blue, anchorcolor = blue]{hyperref}
\usepackage{float}
\usepackage{multirow}

\def\t13{\mathrel{{\theta_{13}}}}
\def\y12{\mathrel{{\tan^2 \theta_{12}}}}
\def\c2{\mathrel{{\chi^2 }}}

\newcommand{\be}{\begin{equation}}
\newcommand{\ee}{\end{equation}}
\newcommand{\ba}{\begin{eqnarray}}
\newcommand{\ea}{\end{eqnarray}}

\begin{document}

\title{Gravitational wave triggered searches for high-energy neutrinos from binary neutron star mergers: prospects for next generation detectors
}

\author{Mainak Mukhopadhyay}
\email{mkm7190@psu.edu}
\affiliation{Department of Physics; Department of Astronomy \& Astrophysics; Center for Multimessenger Astrophysics, Institute for Gravitation and the Cosmos, The Pennsylvania State University, University Park, PA 16802, USA
}
\author{Shigeo S. Kimura}%
\affiliation{Frontier Research Institute for Interdisciplinary Sciences; Astronomical Institute, Graduate School of Science, Tohoku University, Sendai 980-8578, Japan
}

\author{Kohta Murase}
\affiliation{Department of Physics; Department of Astronomy \& Astrophysics; Center for Multimessenger Astrophysics, Institute for Gravitation and the Cosmos, The Pennsylvania State University, University Park, PA 16802, USA
}
\affiliation{Center for Gravitational Physics and Quantum Information, Yukawa Institute for Theoretical Physics, Kyoto University, Kyoto 606-8502, Japan
}
\affiliation{
School of Natural Sciences, Institute for Advanced Study, Princeton, NJ 08540, USA
}
\date{\today}

\begin{abstract}
The next generation gravitational wave (GW) detectors -- Einstein Telescope (ET) and Cosmic Explorer (CE) will have distance horizons up to $\mathcal{O}(10)$ Gpc for detecting binary neutron star (BNS) mergers. This will make them ideal for triggering high-energy neutrino searches from BNS mergers at the next generation neutrino detectors, such as IceCube-Gen2. We calculate the distance limits as a function of the time window of neutrino analysis, up to which meaningful triggers from the GW detectors can be used to minimize backgrounds and collect a good sample of high-energy neutrino events at the neutrino detectors, using the sky localization capabilities of the GW detectors. We then discuss the prospects of the next generation detectors to work in synergy to facilitate coincident neutrino detections or to constrain the parameter space in the case of non-detection of neutrinos.
We show that good localization of GW events, which can be achieved by multiple third generation GW detectors, is necessary to detect a GW-associated neutrino event or put a meaningful constraint ($\sim 3\sigma$ confidence level) on neutrino emission models.
Such an analysis can also help constrain physical models and hence provide insights into neutrino production mechanisms in binary neutron star mergers.
\end{abstract}

\maketitle

\section{Introduction}
The era of multimessenger astronomy is rapidly progressing, with planned upgrades to IceCube - IceCube-Gen2~\cite{IceCube-Gen2:2020qha}, KM3NeT~\cite{KM3Net:2016zxf}, the Pacific Ocean Neutrino Experiment (P-ONE)~\cite{P-ONE:2020ljt}, TRIDENT~\cite{Ye:2022vbk} and Baikal-GVD~\cite{BAIKAL:2013jko} bolstering the neutrino sector and the next generation gravitational wave (GW) detectors like Einstein Telescope (ET)~\cite{Maggiore:2019uih} and Cosmic Explorer (CE)~\cite{Reitze:2019iox} complementing in the GW sector. 
The former gives us an opportunity to look for high-energy neutrino events, while the latter is expected to have a distance horizon of at least $\sim$ a few $10$ Gpc for binary neutron star mergers. However, beyond the individual capabilities of the detectors in regards to their respective messengers, it is also important to consider their joint capabilities in observing messengers from various astrophysical sources in order to maximize our understanding through observational insights.

With the detection of the binary neutron star (BNS) merger event GW170817 by the LIGO and Virgo collaborations~\cite{LIGOScientific:2017vwq} in the GW channel and subsequent observations of a faint gamma-ray burst~\cite{Goldstein:2017mmi,LIGOScientific:2017zic} and other electromagnetic signatures~\cite{LIGOScientific:2017ync,DES:2017kbs,Coulter:2017wya,J-GEM:2017tyx,Valenti:2017ngx,Lipunov:2017dwd}, the power of multimessenger observations using GW detectors and electromagnetic telescopes was well-established. However no high-energy neutrinos (or high-energy gamma rays) were observed from this event~\cite{ANTARES:2017bia,HESS:2017kmv,Super-Kamiokande:2018dbf}, even though BNS mergers are considered to be sources of high-energy neutrinos~\cite{Kimura:2017kan,Biehl:2017qen,Ahlers:2019fwz,Kimura:2018vvz,Carpio:2020app,Fang:2017tla,Carpio:2020wzg,Decoene:2019eux}. 
This naturally leads to the question of what will the landscape of joint GW and neutrino observations look like in the next era with the upcoming powerful detectors in both sectors. In particular, ideally even if a coincident detection of high energy neutrino from a single nearby BNS merger event cannot be made, the large distance horizons of future detectors might enable us to stack signal events to eventually collect coincident neutrino events from comparatively distant sources or in their absence help us constrain the physical models.

This forms the basis of our present work. We discuss the possibilities of synergic observations between the neutrino and GW detectors. Dedicated searches for high-energy neutrinos coincident with BNS mergers have been performed by the IceCube collaboration~\cite{IceCube:2020xks}.
In~\cite{Bartos:2018jco} search techniques for common sources of GW and high-energy neutrinos were discussed using a Bayesian approach. An archival search for the $80$ confident events reported in GWTC-2.1 and GWTC-3 catalogs was conducted by the IceCube collaboration using an unbinned maximum likelihood analysis and a Bayesian analysis using astrophysical priors, resulted in no significant neutrino associations~\cite{IceCube:2022mma}.

A dedicated triggered search pipeline called Low-Latency Algorithm for Multi-messenger Astrophysics (LLAMA)~\cite{Countryman:2019pqq} using LIGO/Virgo candidates was developed to look for high-energy neutrinos from the candidates.
The main framework for the pipeline involves receiving significant events generated by detection pipelines from LIGO/Virgo as input along with the reconstructed skymaps. The pipeline (LLAMA) receives this information to collect the neutrino and GW localizations from IceCube and LIGO/Virgo respectively, to run a joint analysis. The results of the joint analysis is then distributed for other multi-messenger followups. The joint analysis technique used in the pipeline considers a high-energy neutrino coincident with a GW signal if - (a) the high energy neutrino event is detected within a time window of $t_{\rm GW} \pm 500$ s, where $t_{\rm GW}$ is the time of the GW event and (b) the likelihood density between the neutrino signal and the $90$\% confidence region associated with the GW signal is greater than $10^{-4} \rm deg^{-2}$.

In principle, depending on the astrophysical system of interest, an analysis can be performed where one of the neutrino or GW observations can be used as a trigger for the other observation. For sources that are known to emit heavily in neutrinos, e.g., core-collapse supernovae, a neutrino triggered search might be more efficient~\cite{Mukhopadhyay:2021gox,Mukhopadhyay:2022qmo}. For the case of compact binary mergers, the maximum energy is radiated in GWs, and the GW signals might contribute in triggering neutrino searches. Binary neutron star mergers form an interesting class of astrophysical sources that are known to emit heavily in GWs and may also have significant high energy neutrino emission depending on the presence of jet and ejecta~(e.g., \cite{Kimura:2018vvz}). We particularly focus on BNS mergers for this work.  

The next generation GW detectors with large distance horizons would observe several hundreds of BNS merger events per day, providing triggers for high-energy neutrino searches in the next generation neutrino detectors. However, with the increase in the number of triggers, the backgrounds for the neutrino detectors can also be overwhelming. In this work, instead of using all the GW events as triggers for neutrino searches, we calculate distance limits from whcih to collect \emph{meaningful} triggers, where meaningful in this context implies triggers for which stacking the neutrino observations would lead to a comparatively low background. We propose to do this by quantifying the fraction of sky area covered using the sky localization capabilities of the GW detectors. We present our results for the prospect of coincident detection of high-energy neutrinos from BNS mergers along with the operation time of the detectors. In case of a non-detection this would imply strong constraints on the physical models associated with BNS mergers. We also give estimates of backgrounds associated with such limiting distance triggered stacking searches, which provides an optimal operation time and a threshold for the fraction of sky area covered.

A related work was also performed for the low energy thermal neutrinos from BNS mergers, where the potential of the next generation GW detectors was examined in the context of searching for $\mathcal{O}(10)$ MeV neutrinos from BNS mergers at the next generation sub-megaton scale water Cherenkov detectors, such as Hyper-Kamiokande~\cite{Kyutoku:2017wnb,Lin:2019piz}. Besides being on the opposite end of the energy spectrum, these have short emission timescales of $\sim 1 - 10$ s. However, the timescales associated with high-energy neutrino emission from BNS mergers can be diverse and as long as $\sim 10^7$ s~\cite{Fang:2017tla}, which would then require a completely different analysis strategy to collect low background coincident neutrino events from BNS mergers as described in this work. It is also important to note that the current work is complementary to the triggered-stacking searches that the current neutrino detectors perform~\cite{Countryman:2019pqq,IceCube:2020xks,IceCube:2022mma}, which use likelihood search techniques along with various astrophysical priors.

The paper is organized as follows. In Sec.~\ref{sec:strategy}, we discuss the main method and the ingredients required to perform the analysis. Using the ingredients, the main results including the detection prospects for coincident high energy neutrino events using GW triggered searches and the effects of backgrounds are discussed in Sec.~\ref{sec:res}. We summarize in Sec.~\ref{sec:conclsn} and discuss the implications of our work in Sec.~\ref{sec:disc}.
\section{Strategy}
\label{sec:strategy}
Let us assume a given astrophysical source emits total energy ${\mathcal E}^{\rm tot}$. Consider the energy emitted in the GW sector is ${\mathcal E}_{\rm GW}$, in high energy neutrinos is $\mathcal{E}_\nu^{\rm HE}$, and in all other channels $E^{\rm misc}$. Note that these other channels ($\mathcal{E}^{\rm misc}$) can also include electromagnetic and exotic emissions. Thus, we have ${\mathcal E}^{\rm tot} = {\mathcal E}_{\rm GW} + \mathcal{E}_\nu^{\rm HE} + \mathcal{E}^{\rm misc}$. Note that independent of the specific models, $\mathcal{E}^{\rm misc} > \mathcal{E}_\nu^{\rm HE}$ because the amount of generated gamma rays is comparable to that of high-energy neutrinos whether one considers hadronuclear or photohadronic interactions~\cite{Murase:2013rfa}.
Since the focus of this work is to discuss the prospects of GW triggered high energy neutrino observations, we consider that the source emits strongly in gravitational waves and use that as the trigger. In particular, BNS mergers, besides strongly emitting in GWs, might also have considerable neutrino emissions~\cite{Kimura:2017kan,Biehl:2017qen,Ahlers:2019fwz,Kimura:2018vvz,Carpio:2020app,Fang:2017tla,Carpio:2020wzg,Decoene:2019eux}. We focus on BNS mergers and discuss the strategy in the context of such mergers. 

For BNS mergers, typically a small fraction ($\alpha$) of the total energy is emitted in GWs, that is, ${\mathcal E}_{\rm GW} \sim \alpha {\mathcal E}^{\rm tot}$, where $\alpha \sim 1$\%~\cite{Shibata:2002jb,LIGOScientific:2017vwq,Radice:2016rys,Shibata:2017xht,PhysRevLett.120.111101,Shibata:2019wef,Radice:2020ddv}. The total energy $\mathcal{E}^{\rm tot}$ in this case refers to the orbital or the binding energy of the BNS system\footnote{It is also important to note that the majority of the energy budget does not go into radiation and is mostly utilized as the spin energy of the compact object formed.}. The energy in the GWs (${\mathcal E}_{\rm GW}$) is the observed quantity and has all the uncertainties based on the detector properties and the source. The quantity ${\mathcal E}_{\rm GW}$ will depend on the signal-to-noise ratio (SNR) of the observation, which in turn, depends on the noise curves or sensitivity of the GW detector and the distance of the source. 

The total energy emitted in high energy neutrinos can be assumed to be a fraction of the gravitational-wave energy (${\mathcal E}_{\rm GW}$). Thus, we can define
\be
\label{eq:enuhe}
\mathcal{E}_\nu^{\rm HE,true} = f_\nu {\mathcal E}_{\rm GW}\,,
\ee
where $\mathcal{E}_\nu^{\rm HE,true}$ is the energy emitted in high-energy neutrinos at the source and $f_\nu$ is an effective parameter and is defined as the energy radiated in high energy neutrinos as a fraction of the GW energy emitted by the same source. 
This will be an important parameter in our analysis. A higher value of $f_\nu$ would imply a larger total energy emitted in high-energy neutrinos leading to more optimistic prospects. 

The high-energy neutrinos can be produced in the relativistic jets launched after the merger, and we need to take into account the relativistic beaming effect when discussing the neutrino emission from the jets. The isotropic-equivalent energy emitted in high-energy neutrinos at the source is given by 
\be
\mathcal{E}_\nu^{\rm HE,iso} = \frac{\mathcal{E}_\nu^{\rm HE,true}}{f_{\rm bm}} = \bigg( \frac{f_\nu}{f_{\rm bm}} \bigg) \mathcal{E}_{\rm GW}\,,
\ee
Often $f_{\rm bm} \sim 1$\% is used, which is roughly consistent with the afterglow observations~\cite{Fong:2015oha}. 
It is expected that $\mathcal{E}_\nu^{\rm HE,true}$ is only a tiny fraction of ${\mathcal E}_{\rm GW}$. However the neutrino emission and the gravitational wave emissions can have completely different beaming factors. Finally, it is worth mentioning that the definition of $f_\nu$ to be a ratio against the GW energy emitted in neutrinos is just a definition. In general, the two emissions have different physical sources and are not related, so $f_\nu>1$ is possible in principle but highly unlikely for realistic astrophysical environment.

One can also choose $\mathcal{E}_\nu^{\rm HE,true}$ from physical models. For example, according to neutrino emissions from BNS mergers considered in~\cite{Kimura:2018vvz,Fang:2017tla} one can find that $\mathcal{E}_\nu^{\rm HE,true} \sim 10^{48}$ erg. Assuming ${\mathcal E}^{\rm tot} = 5 \times 10^{54}$ erg, then gives $f_\nu \sim 10^{-4}$.
\begin{figure*}
\centering
\subfloat[(a)] {\includegraphics[width=0.49\textwidth]{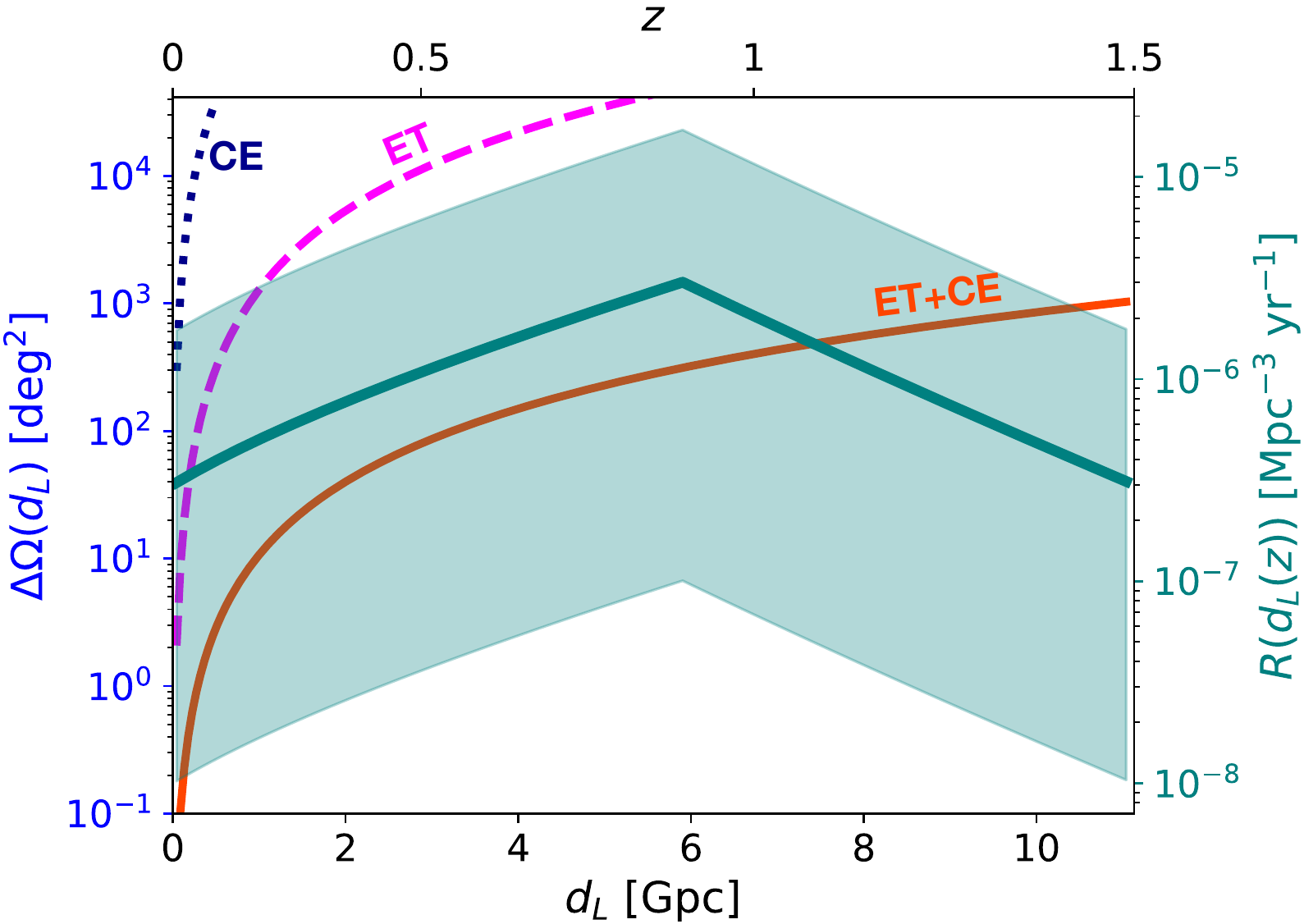}\label{fig:angerr_rates}}\hfill
\subfloat[(b)] {\includegraphics[width=0.47\textwidth]{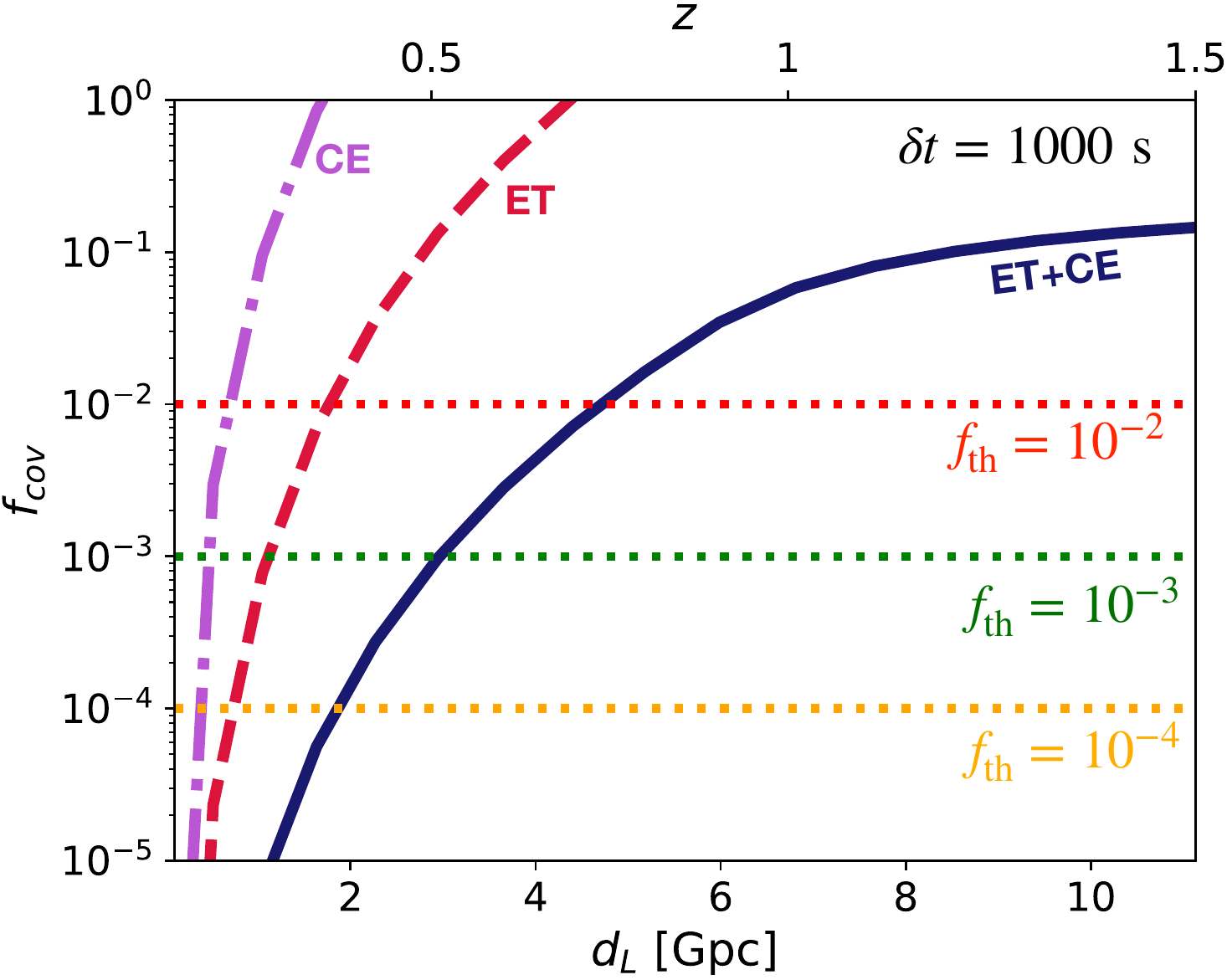}\label{fig:delt_allskyerr}} \hfill
\caption{\label{fig:2} (a) The left axis shows the the size of the localization area in the sky for a single BNS merger event ($\Delta \Omega$) as a function of the luminosity distance $d_L$ for CE (dashed dark blue line), ET (dashed pink line), and ET+CE (solid orange line). The right axis shows the BNS merger rate $R(z)$. The fiducial rate is shown as a solid teal line and the area between the upper and the lower limits of $R(z=0)$ is shaded. (b) The fraction of sky area covered $f_{\rm cov}$ as a function of $d_L$ for CE (dot-dashed purple line), ET (dashed orange line), and ET+CE (solid dark blue line) (see Eq.~\ref{eq:fth} for details). The horizontal dotted lines show the threshold for the fraction of sky area covered $f_{\rm th}$ equal to $10^{-2}$ (red), $10^{-3}$ (green), and $10^{-4}$ (orange). The time duration assumed for the plot is $\delta t = 1000$ s.
}
\end{figure*}

Below we give brief details about the next generation GW and neutrino detectors that we use for this work.
\subsection{GW detectors}
The GW detectors we consider in this work are the upcoming third generation detectors: Einstein Telescope (ET) and Cosmic Explorer (CE). Following Ref.~\cite{Chan:2018csa}, we use the results of the geometrical configuration known as ET-D for Einstein Telescope. This consists of 3 interferometers, each having an opening angle of $60^\circ$ and are rotated by $120^\circ$ relative to the other, resulting in a configuration of an equilateral triangle. The interferometer arm length is $10$ km. Since the location of both ET and CE is undecided yet, the results used assume that ET is located at (longitude,latitude) $= (10.4^\circ, 43.7^\circ)$ and CE at (longitude,latitude) $= (-119.41^\circ, 46.45^\circ)$. The corresponding antenna patterns assumed for each detector are given in Ref.~\cite{Regimbau:2012ir}. Both ET and CE are sensitive between 
$\mathcal{O} (10)\ {\rm Hz}$ to a few kHz~\cite{Maggiore:2019uih,Evans:2021gyd} and have an amplitude spectrum density of $\sim 10^{-25} \rm Hz^{-1/2}$. The scenario of ET and CE working together (referred to as ET+CE hereafter) is also considered to illustrate the prospects of triggered searches.

Given the properties of the GW detector, the size of the localization area $\Delta \Omega$ depends on the distance to the source~\cite{Chan:2018csa}. In Fig.~\ref{fig:angerr_rates} on the left axis, we show $\Delta \Omega$ as a function of the luminosity distance $d_L$ for ET (dashed pink), CE (dotted dark-blue) and ET+CE (solid orange). As the distance to the source increases, the localization area becomes larger as expected. It is interesting to note that although CE has a longer arm and hence is sensitive to higher distances for BNS mergers, it has a comparatively poor sky localization capability and is limited to $< 1$~Gpc for covering all sky (sky localization area is $4 \pi$) as compared to the ET which can use triangulation techniques to have a much better $\sim \mathcal{O}(10^3)\ \rm deg^2$ at $1$~Gpc sky localization capability. The combination of ET+CE is the most effective and has a localization capability $\sim \mathcal{O}(10^3)\ \rm deg^2$ even at $d_L\sim10$ Gpc, which is a couple of orders of magnitude better than that of ET. Finally, the sky-localization areas can be less optimistic~\cite{Baral:2023xst} than what is considered in our current work.
\subsection{Neutrino detectors}
For neutrino detectors, we consider the planned upgrade to IceCube - IceCube-Gen2. The current IceCube detector has 86 strings deployed at a distance spacing of 120 metres resulting in roughly a cubic-kilometer detector volume. The planned IceCube-Gen2 will add $120$ strings to the existing configuration at distances of $240$ metres resulting in a detector volume of $\sim 8\ \rm km^3$. This combined with better PMTs will likely boost its sensitivity by a factor of $10^{2/3}$ (roughly $5$).
For this work, we use the publicly available declination-dependent effective area used in the IceCube ten-year point source (PS) search~\cite{IceCube:2021xar} and scale it by the factor ($10^{2/3}$) to obtain our results for the future IceCube-Gen2.
\subsection{Binary neutron star merger rates}
In this section, we discuss the rate of BNS mergers. The redshift dependence of the rate is adapted from that of short gamma-ray bursts (GRBs)~\cite{Wanderman:2014eza}. This is a reasonable assumption given that short GRBs might primarily originate from NS-NS mergers\footnote{NS-BH (neutron star-black hole) mergers are also suggested to produce short GRBs, but the NS-BH merger rate is likely lower than BNS merger rates based on GW observations~\cite{LIGOScientific:2021djp}, although the rate still has a large uncertainty.}. 
The redshift dependent rate of BNS mergers is given as~\cite{Wanderman:2014eza}
\be
\label{eq:bnsrate}
R(z) \equiv \mathcal{N} R(z=0) 
\begin{cases}
{\rm exp}\big((z-0.9)/0.39 \big)\,, z \leq 0.9\,,\\
{\rm exp}\big(-(z-0.9)/0.26 \big)\,, z >0.9\,,
\end{cases}
\ee
where the normalization $\mathcal{N}$ is chosen such that the rate at $z=0$ is given by the fiducial rate, $R(z=0) = 300\ \rm Mpc^{-3} \rm yr^{-1}$. We choose a conservative value for the fiducial case. However, this rate has associated uncertainties and can be over a wide range between $10\ \rm Mpc^{-3} \rm yr^{-1}$ (lower limit) and $1700\ \rm Mpc^{-3} \rm yr^{-1}$ (upper limit)~\cite{KAGRA:2021duu}. In Fig.~\ref{fig:angerr_rates}, on the right axis, we show $R(z)$ as a function of luminosity distance, $d_L$, where the fiducial rate ($R(z = 0) = 300$ Mpc$^{-3}$ yr$^{-1}$) is shown as a thick solid line and the area between the upper and lower limits is shaded with teal. We see the break at $z=0.9$ as the peak according to Eq.~\eqref{eq:bnsrate}. The cumulative rate is given by, $R_{\rm BNS} (z) = \int_0^z dz^\prime \ R(z^\prime) $. Given that the rate decreases post $z=0.9$, the cumulative rate is approximately constant for $z \gtrsim 0.9$. 
\subsection{Sky localization for GW detectors}
\label{subsec:frac_sky}
With the next generation GW detectors reaching distance horizons of $\mathcal{O} (10\ \rm Gpc)$, the number of events detected would be $\mathcal{O}(100)$ a day~\cite{galaxies10040090}. Hence, the entire paradigm of performing triggered neutrino searches based on GW detections depends on the localization of the source based on the GW observations. We focus on quantifying the possibilities and limitations based on the GW detectors and the sources in this section.

The total fraction of the sky area covered by the error regions given by the localization of BNS merger events by the GW detectors in a given time period is
\be
\label{eq:fth}
\int_0^{d_{\rm GW}^{\rm lim}} d\big( d_{\rm com} \big) \frac{\Delta \Omega \big(d_L \big)}{4 \pi} R\big( z \big) 4 \pi d^2_{\rm com} \delta t = f_{\rm cov} (d_{\rm GW}^{\rm lim})\,,
\ee
where $d_{\rm com}$ is the comoving distance, $f_{\rm cov}$ is the fraction of the sky area covered\footnote{Note that $f_{\rm cov} = 1$ would imply all sky ($4 \pi$) area being covered by the error regions as a result of the localization of BNS merger events by the GW detectors in a given time period.}, $d_{\rm L}$ is the luminosity distance that is a function of the redshift $z$ and comoving distance $d_{\rm com}$, $d_L = (1+z) d_{\rm com}$, $\Delta \Omega$ is the size of the localization area in the sky for a single BNS merger event for a given GW detector, and $\delta t$ is the duration of neutrino emission in the observer frame.

The integral is performed over $d_{\rm com}$ from $0$ to the given upper limit $d_{\rm GW}^{\rm lim}$. For a chosen value of $f_{\rm cov}$, which we call the threshold value $f_{\rm th}$, we evaluate the above equation to deduce the upper limit, hence obtaining the limiting distance $d_{\rm lim}^{\rm GW}$. Thus, $f_{\rm th}$ in this case is a parameter that then helps us in obtaining distance limits such that a collection of signal events is not overwhelmed by backgrounds. The identification of individual sources requires nearby merger events. However, for this work, we do not discuss identifying individual sources. Instead, we focus on prospects for high-energy neutrino detection by stacking BNS merger events using GW triggers.

The time duration of neutrino emission, $\delta t$, is an important quantity and can have a wide range of variation. For example, high energy neutrino emissions from a jet interacting with the ejecta post a merger was considered in Ref.~\cite{Kimura:2018vvz}. In this case, the neutrino emission peaks at around $\delta t \sim 2$ s post the merger, and thus, this scenario would be very optimistic. In contrast, a BNS merger may lead to the formation of a long-lived, millisecond magnetar surrounded by a low-mass ejecta. A portion of the magnetar's rotational energy is deposited in the nebula behind the ejecta, eventually leading to efficient pion production using the thermal and/or nonthermal photons in the nebular region which can then produce high energy neutrinos~\cite{Fang:2017tla}. In this case the timescales associated with the peak in the neutrino fluence is of the order $\delta t \sim 10^{6} -10^{7}$s, which would be a conservative scenario.

As a fiducial value, we assume $\delta t = 1000$ s and show the fraction of sky area covered by CE (dot-dashed purple), ET (dashed red), and ET+CE (solid dark blue) with the luminosity distance in Fig.~\ref{fig:delt_allskyerr}. The choice for the fiducial value of $\delta t = 1000$ s can be motivated from theoretical models~\cite{Kimura:2017kan,Matsui:2023ohr}, which suggest $\delta t \sim 100 - 1000$ s. Furthermore, current GW and neutrino searches also use a typical time window of $\delta t = 1000$ s for their analysis~\cite{Baret:2011tk,IceCube:2022mma,ANTARES:2023wcj}.
Depending on the chosen value of $f_{\rm th}$, one can estimate the distance limit for a given GW detector from which triggers can be used to have reduced backgrounds. Similar to the localization capabilities, we have the ET+CE combination to be the most effective followed by ET and CE. For CE, the fraction of sky area covered needs to be relatively larger than ET and ET+CE. For $f_{\rm th} \sim 1$\% of the total sky area, CE is limited to $\sim 0.7$ Gpc. The case can be improved for ET where choosing a threshold of $f_{\rm th} \sim 0.1$\% of the total sky area puts the distance limit at $\sim 1.1$ Gpc and the combination of ET+CE can have a limit at $\sim 1.8$ Gpc where the $f_{\rm th}$ has a very small value of $f_{\rm th}\sim 0.01$\% of the total sky area. 

It is important to note here that the results shown are in the absence of backgrounds. Including backgrounds and demanding a particular signal to background ratio puts constraints on the chosen value of $f_{\rm th}$ corresponding to the operation timescales of the GW detectors. This will be discussed in detail in Sec.~\ref{subsec:bkg}.
\section{Results}
\label{sec:res}
In this section, we outline the technique to highlight the prospects for the next-generation GW detectors to act as triggers for the next-generation neutrino detectors. Such observations also require some assumptions on the neutrino emission from the source, which is also discussed. We also present our main results and give estimates for background events in the following subsections.

\subsection{Distance limit for triggers}
\label{subsec:dist_cuts}
\begin{figure*}
\centering
\includegraphics[width=0.99\textwidth]{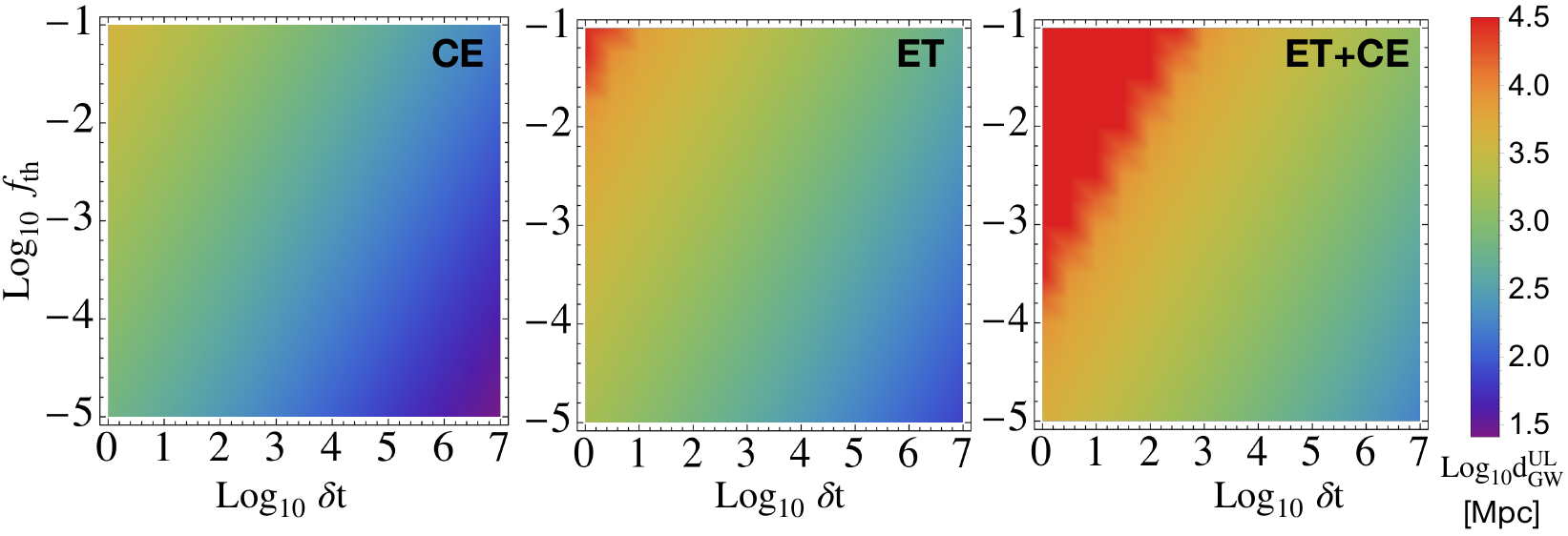}
\caption{\label{fig:et_combined_fth_deltat} Density plot showing the distance upper limit for GW detectors ($d_{\rm GW}^{\rm UL}$) (see Eq.~\ref{eq:dul_gw}) for CE (\emph{left}), ET (\emph{middle}), and ET+CE (\emph{right}) on the $\delta t$ - $f_{\rm th}$ plane. 
}
\end{figure*}
From the previous section and Eqs.~\eqref{eq:bnsrate} and~\eqref{eq:fth}, it is evident that the choice of the threshold for the fraction of sky area covered, $f_{\rm th}$, and the duration of neutrino emission, $\delta t$, are two relevant parameters in implementing the distance cuts for the triggers with the GW detectors. The limiting distance $d^{\rm lim}_{\rm GW}$ is defined as the distance such that from Eq.~\eqref{eq:fth}, $f_{\rm cov}(d^{\rm lim}_{\rm GW}) = f_{\rm th}$, that is, it is the upper limit of the integral in Eq.~\eqref{eq:fth} that results in the desired value of $f_{\rm th}$. There also exists a maximum distance horizon for the GW detectors beyond which all triggers can be used. We define this as ($d_{\rm GW}^{\rm hor}(z^{\rm hor}_{\rm GW})$) which is set to\footnote{This choice for the horizon distance is motivated by the distance horizon of ET for BNS mergers~\cite{galaxies10040090}. Although CE can be sensitive to BNS mergers from higher redshifts, the sky localization is poor. Furthermore, the rate of BNS mergers decreases post $z \sim 0.9$ as can be seen from Fig.~\ref{fig:angerr_rates}, and thus, the choice of $z_{\rm GW}^{\rm hor}$ has little influence on our conclusion.} $z^{\rm hor}_{\rm GW} \sim 3.5$.

Thus, the final limiting distance ($d_{\rm GW}^{\rm UL}$) is defined as the minimum of the limiting distance ($d_{\rm GW}^{\rm lim}$) or the maximum horizon distance considered for the GW detectors; that is,
\be
\label{eq:dul_gw}
d_{\rm GW}^{\rm UL} = {\rm min} \big(d_{\rm GW}^{\rm lim}, d^{\rm hor}_{\rm GW} \big)\,. 
\ee
To illustrate the dependence of the distance limits on $f_{\rm th}$ and $\delta t$, we show a density plot for the final limiting distance $d^{\rm UL}_{\rm GW}$ for CE (left), ET (middle) and the combination ET+CE (right) on the $f_{\rm th} - \delta t$ plane in Fig.~\ref{fig:et_combined_fth_deltat}. The color map is shown as a rainbow spectrum where the limiting distance increases as one goes upward from violet to red.

For CE even with the most optimistic choice of $f_{\rm th}$ and $\delta t$, we are limited to $\sim$ Gpc for the limiting distance. For ET, considering an optimistic scenario of $\delta t \sim 1$s, the distance limits can be a few Gpc assuming $f_{\rm th} \sim 0.1$\%. However for large values of $\delta t \sim 10^6$s, the limiting distance is reduced to $\sim$ a few $100$ Mpc for the same value of $f_{\rm th}$. However, for the combination of ET+CE, $\delta t \sim 1$s implies a limiting distance of $\sim 10$ Gpc for $f_{\rm th} \sim 0.01$\%. In contrast to ET, even for the not so optimistic case of $\delta t \sim 10^6$s, ET+CE can have limiting distances of up to a few Gpc. The area colored in red indicates that the limiting distance obtained from Eq.~\eqref{eq:fth} is greater than the maximum distance horizon for the GW detectors, and hence, the final limiting distance is fixed to be $d^{\rm hor}_{\rm GW}$.
\subsection{Detection prospects}
\label{subsec:det_prosp}
This section focuses on coincident detection of high-energy neutrinos from BNS mergers at the next generation neutrino detectors based on triggers from the next generation GW detectors, where the triggers are chosen based on the distance cut calculated using the fraction of sky area covered, $f_{\rm th}$.

The probability to detect more than one neutrino associated with a GW signal is given by~\cite{Matsui:2023ohr}
\be
\label{eq:qtop}
q\big( d^{\rm UL}_{\rm GW}, T_{\rm op} \big) = 1 - {\rm exp}\bigg( -T_{\rm op} I \big( d_{\rm GW}^{\rm UL} \big) \bigg)\,,
\ee
where $T_{\rm op}$ is the operation time of the GW detector. The argument of the exponential in the above expression $I$ is defined as 
\be
\label{eq:integ}
I \big( d_{\rm GW}^{\rm UL} \big) = 4\pi \int_{0}^{d_{\rm GW}^{\rm UL}} d (d_{\rm com}) \frac{T_{\rm op}}{\big( 1+z \big)} R\big( z \big) d_{\rm com}^2 P_{n \geq 1} \big( d_L \big)
\ee
The integral is performed over the comoving distance. The declination integrated (or total) probability ($P_{n \geq 1} (d_L)$) to detect at least one neutrino as a function of distance is given by
\be
P_{n \geq 1} (d_L)=\frac{1}{4 \pi} \int d\Omega\ p_{n \geq 1}(\delta,d_L)\,,
\ee
where $p_{n \geq 1}(\delta,d_L)$ is the declination ($\delta$) dependent probabililty to detect at least one neutrino is given by the Poissonian probability
\be
p_{n \geq 1}(\delta,d_L) = 1 - \exp\bigg( - N_{\nu_\mu}(\delta,d_L) \bigg)\,.
\ee
In the above expression, $N_{\nu_\mu}(\delta,d_L)$ is the number of neutrino events from a source at a given luminosity distance $d_L$. The expected number of neutrino events at a given declination is given as
\be
N_{\nu_\mu} (\delta,d_L) = \int_{E_{\nu}^{\rm LL}}^{E_{\nu}^{\rm UL}} dE_{\nu_\mu} \phi_{\nu_\mu} (E_{\nu_\mu},d_L)\mathcal{A}_{\rm eff}(E_{\nu_\mu},\delta)\,,
\ee
where $E_{\nu_\mu}$ is the neutrino energy in the observer frame, $E_{\nu}^{\rm LL}$ and $E_{\nu}^{\rm UL}$ give the lower and the upper limits of the integral and is decided based on the neutrino energy spectra from a particular source given a production channel ($pp$ or $p\gamma$ emission channels), $\phi_{\nu_\mu}$ gives the neutrino fluence from a given source at a given luminosity distance $d_L$, and $\mathcal{A}_{\rm eff}$ is the energy and declination dependent IceCube effective area.

In principle, one can consider either of the two channels for high energy neutrino production - hadronuclear ($pp$) or photomesonic processes ($p\gamma$). In general in a BNS merger the nonthermal protons produced in the collimation and internal shocks produce neutrinos through interactions with the background photons and protons. Here, we choose the more optimistic scenario where we focus on the $p\gamma$ interactions for neutrino production. The dominant channel for such process is pion production. The charged pions decay to muons and neutrinos whereas the neutral pions form gamma rays. The muons further decay to form a corresponding charged lepton and neutrinos (and antineutrinos). In this case, the upper and lower limits of the neutrino energy spectra are given by Ref.~\cite{Kimura:2018vvz}, $\varepsilon_{\nu}^{\rm min} = 10^{3}$ GeV and $\varepsilon_{\nu}^{\rm max} = 10^6$ GeV, respectively. Note that $\varepsilon_{\nu}^{\rm min}$ and $\varepsilon_{\nu}^{\rm max}$ are defined in the source frame. The neutrino energy in the observer frame with the redshift correction is then given by, $E_\nu=\varepsilon_\nu/(1+z)$. It is important to note that this serves as a comparatively optimistic scenario. Besides, the values chosen for $\varepsilon_{\nu}^{\rm min}$ and $\varepsilon_{\nu}^{\rm max}$ can differ across various models. Considering $pp$ interactions will make our results slightly more conservative. We assume a power law neutrino spectrum with spectral index equal to 2, that is, $E_{\nu}^{-2}$ spectrum for the neutrinos. Thus, the neutrino fluence is given by
\be
\phi_\nu (\mathcal{E}_\nu^{\rm HE,iso}, E_\nu ,d_L) = \frac{(1+z)}{4 \pi d_L^2} \frac{\mathcal{E}_\nu^{\rm HE,iso}}{\rm ln \big( \varepsilon_\nu^{\rm max}/\varepsilon_{\nu}^{\rm min} \big)} E_\nu^{-2}\,.
\ee
Recall from Sec.~\ref{sec:strategy} that the total isotropic-equivalent energy emitted in high-energy neutrinos is given by $\mathcal{E}_\nu^{\rm HE,iso}$. This depends on the parameters $f_{\nu}$ and $f_{\rm bm}$. To illustrate our results we choose $f_{\rm bm} = 1$\% and $f_\nu \sim \mathcal{O}(10^{-5})$. The muon neutrino fluence is given by, $\phi_{\nu_\mu} (E_\nu ,r) = (1/3)\phi_\nu(\mathcal{E}_\nu^{\rm iso-HE}, E_\nu ,r)$. Finally, for this work, we use the 10-year point source (PS) effective area from IceCube~\cite{IceCube:2021xar}, scaled with a factor of $10^{2/3}$ in accordance with the estimates of IceCube-Gen2.
\begin{figure*}
\centering
\includegraphics[width=0.98\textwidth]{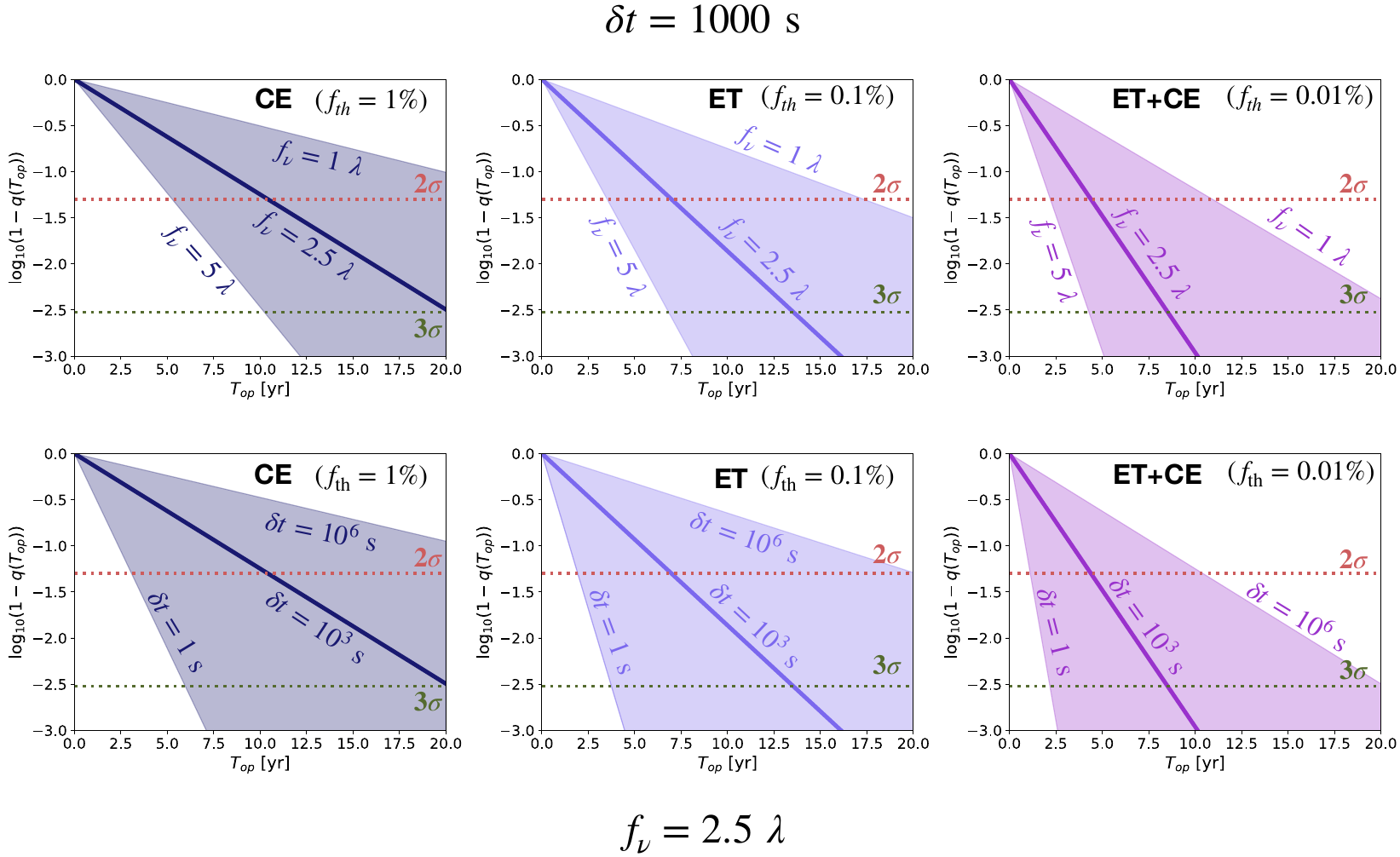}
\caption{\label{fig:ce_et_comb_fnu_delt} Probability of neutrino detection ($q$) with the operation time $T_{\rm op}$ for a range of $f_\nu$ (\emph{top row}) and $\delta t$ (\emph{bottom row}) for the different GW detectors. The fiducial case for each panel is shown as a solid line. The $2\sigma$ and $3\sigma$ C.L.s are also shown with dashed lines. For each case, ${\mathcal E}_{\rm GW} = 5 \times 10^{52}$ erg and $\lambda = 10^{-5}$. See the text for details on parameters.
}
\end{figure*}

In summary, evaluating $q\big( d^{\rm lim}_{\rm GW}, T_{\rm op} \big)$ from Eq.~\eqref{eq:qtop} enables us to calculate the probability to detect high energy neutrinos associated with GWs. However, it is important to understand the time-scales over which the next generation GW detectors need to operate to either enable coincident neutrino detections or constrain the physical parameter space at a given confidence level in case of non-detections. We address this question in Fig.~\ref{fig:ce_et_comb_fnu_delt}, where we plot the probability of neutrino detection at IceCube-Gen2 with $T_{\rm op}$ of the GW detectors, which shows one of our main results. The two primary parameters are $f_\nu$ and $\delta t$. While the former decides the total energy emitted in high energy neutrinos, the latter allows us to put a distance limit to reduce backgrounds to facilitate detections. 

In the figure, we show the plots for $1-q\big( d^{\rm lim}_{\rm GW}, T_{\rm op} \big)$ with the operation time of the GW detectors in years. The fiducial parameters for each case in the top panel is chosen as $\delta t  = 1000$ s and for the bottom panel as $f_\nu = 2.5 \times 10^{-5}$. The choice of $f_\nu$ is motivated from physical models as discussed in Sec.~\ref{sec:strategy}, where we conservatively choose the fiducial case to be $20$\% of that obtained from physical models. The total energy emitted from the system of BNS merger is fixed to be ${\mathcal E}^{\rm tot} = 5 \times 10^{54}$ erg, which is typical for two $1.4 M_\odot$ neutron star mergers~\cite{LIGOScientific:2017vwq}. The parameter $\alpha$ is fixed to be $1$\% implying $\mathcal{E}_{\rm GW} = 5 \times 10^{52}$ erg. The fiducial cases are shown by the thick solid line in each panel. The dotted horizontal lines show the confidence levels corresponding to $2 \sigma$ (red) and $3\sigma$ (green) respectively. For each of the GW detectors, we choose a different value of $f_{\rm th}$. This is a reasonable choice since each detector has a different localization capability as discussed in Sec.~\ref{subsec:frac_sky} and shown in Fig.~\ref{fig:delt_allskyerr}. For CE $f_{\rm th} = 1$\% is chosen due to its comparatively poor localization capability, ET motivates $f_{\rm th} = 0.1$\% making the situation better. Finally the combination ET+CE gives the best case of choosing the fraction of sky localization to be, $f_{\rm th} = 0.01$\%. As mentioned earlier, given a demand of signal to background ratio and an operation time for the GW detectors these choices are not arbitrary as will be discussed in Sec.~\ref{subsec:bkg}. 

In the top row of Fig.~\ref{fig:ce_et_comb_fnu_delt}, we vary $f_\nu$ between $10^{-5}$ and $5 \times 10^{-5}$, where the lower limit is an order of magnitude less than the predictions from physical models and the upper limit denotes a value that is $50$\% of the physical models to still remain in the conservative yet optimistic regime. A lower value of $f_\nu$ implies less energy in emitted in high-energy neutrinos, which implies a lower value of neutrino fluence. Hence, in all the top three panels, this gives the upper boundary of the shaded region, which is a not so optimistic scenario, requiring longer timescales of operation to reach the same confidence levels. A higher value of $f_\nu$ correspondingly gives the lower boundary of the shaded region. We see that for the fiducial case, CE requires an operation time of $\sim 20$ years to make a constraint on the parameter space at the $3 \sigma$ C.L. ET would take $\sim 14$ years to put constraints at the $3 \sigma$ C.L. but can reach the $2 \sigma$ C.L. over a timescale of $\sim 17.5$ years even for the lowest value of $f_\nu$ considered for this work. The combination ET+CE can reach the $3\sigma$ level constraints in a timescale of $\lesssim 30$ years even for the most conservative choice of $f_\nu$. 

The bottom row of Fig.~\ref{fig:ce_et_comb_fnu_delt} shows the variation in $\delta t$ between $1$ s and $10^6$ s. In this case, a longer $\delta t$ leads to less optimistic results. This is because with larger values of $\delta t$, the number of triggers within $\delta t$ increases, and hence, the distance limit for reaching a fixed value of $f_{\rm th}$ is reduced (as seen from Eq.~\eqref{eq:fth} and Fig.~\ref{fig:et_combined_fth_deltat}). This results in a smaller value of the upper limit of the integral in Eq.~\eqref{eq:integ}. Thus, a longer operation time is required to reach the same level of sensitivity. For small values of $\delta t \sim 1$s, a $3\sigma$ constraint based on non-detection is possible in $\lesssim 10$ years using any of the detectors and/or the combination. However, for large values of $\delta t$, CE may reach the $2 \sigma$ level of constraints in a time scale $> 20$ years. While the combination of ET+CE can reach a $3 \sigma$ level in around $20$ years, ET by itself would need $\sim 20$ years to reach the $2 \sigma$ level of constraints.
\subsection{Backgrounds}
\label{subsec:bkg}
\begin{figure}
\centering
\includegraphics[width=0.48\textwidth]{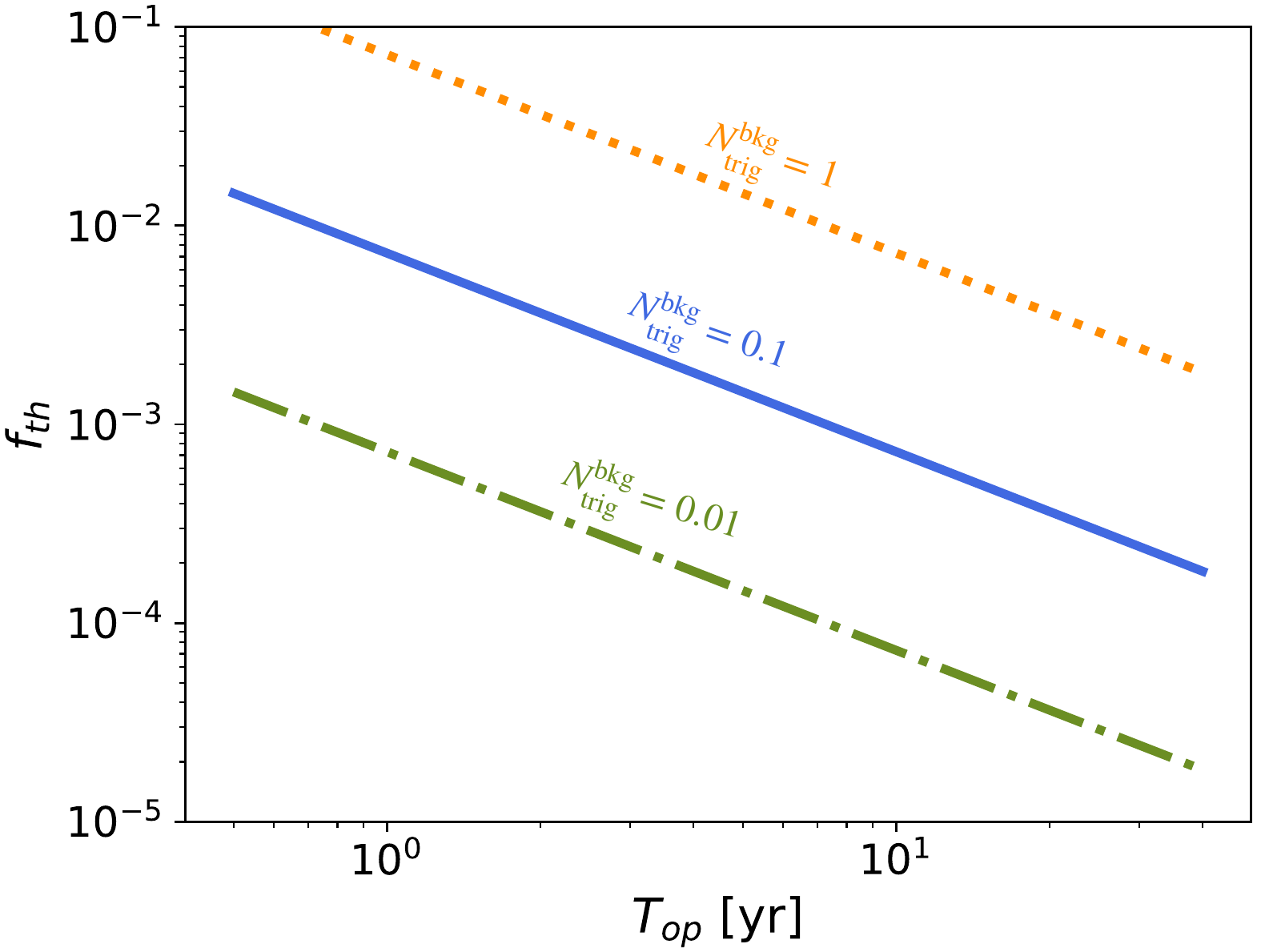}
\caption{\label{fig:fth_top} The fraction of sky area covered $f_{\rm th}$ with operation time $T_{\rm op}$ for different values of triggered background events $N_{\rm bkg}^{\rm trig}$ at IceCube-Gen2. 
}
\end{figure}
In all the discussions above we ignored the backgrounds in the context of the neutrino detectors. Constraining the distance to collect triggers from using sky-localization definitely helps in reducing the background, but it is also important to roughly estimate the backgrounds associated with such triggered searches including the distance limits. This is the focus of the current section.

The relevant backgrounds for high energy neutrino detectors, such as IceCube or KM3NeT, are the conventional and prompt atmospheric neutrinos and the diffuse astrophysical neutrinos. In this work we provide a rough estimate of resulting backgrounds from the search method presented here. For the conventional atmospheric neutrino flux, we use the model in Ref.~\cite{Honda:2006qj}. The prompt atmospheric background neutrino fluence is taken from Ref.~\cite{Enberg:2008te}. A more advanced and realistic estimate for the atmospheric backgrounds was performed in Refs.~\cite{Bhattacharya:2015jpa,Bhattacharya:2016jce}, but we use the simple model to give a rough estimate of the atmospheric backgrounds.

The diffuse astrophysical neutrino flux Ref.~\cite{IceCube:2021uhz} is assumed to be of the form $\Phi^{\rm bkg}_{\rm astro} = \Phi_{\rm astro} (E_\nu/100 \rm TeV)^{-\gamma}$, where the normalization $\Phi_{\rm astro} = 1.44 \times 10^{-18}\ \rm GeV^{-1} cm^{-2} s^{-1} sr^{-1}$ and the power law index, $\gamma = 2.37$. We follow the same prescription outlined in Sec.~\ref{subsec:det_prosp} to calculate the number of triggered background events ($N_{\rm trig}^{\rm bkg}$) using the 10-year point source effective area scaled by a factor of $10^{2/3}$ for IceCube-Gen2.

In Fig.~\ref{fig:fth_top} we show the contours for the number of triggered background events, on the the operation time ($T_{\rm op}$) and the fraction of sky area covered $f_{\rm th}$ plane. We assume that for large number of triggers the total time in triggers for the neutrino detectors will be comparable to the operation time of the GW detectors. This is a reasonable assumption owing to the next generation of GW detectors having a large distance horizon. With the inclusion of triggered background events, the figure presents an estimate for the choice of fraction of sky area covered given an operation time and a desired level of signal to background ratio. For example, fixing an operation of time of $10$ years and choosing $N_{\rm trig}^{\rm bkg} = 1, 0.1, 0.01$ dictates that $f_{\rm th}$ be chosen as $0.8$\%, $0.08$\%, $0.008$\% respectively.

Let us now discuss the implications of Fig.~\ref{fig:fth_top} on the main results presented in Fig.~\ref{fig:ce_et_comb_fnu_delt}. In Fig.~\ref{fig:ce_et_comb_fnu_delt} we assumed $f_{\rm th} = 1$\%, $0.1$\% and $0.01$\%, respectively, for CE, ET, and ET+CE. This translates to choosing $T_{\rm op} \sim 8$ years corresponding to $N_{\rm trig}^{\rm bkg} = 1, 0.1 , 0.01$ for CE, ET, and ET+CE, respectively. It is also important to highlight the fact that this implies that the fraction of sky area covered $f_{\rm th}$ in the presence of background is not an arbitrarily chosen value but is constrained by the operating time of the GW detector and the choice of signal to background ratio. For claiming a $3\sigma$ C.L. detection, we should restrict ourselves to $N_{\rm trig}^{\rm bkg} \lesssim 3 \times 10^{-3}$. However, constraints on the parameter space as a result of non-detection can be put using $N_{\rm trig}^{\rm bkg} \sim 0.1 - 0.01$.
\section{Summary}
\label{sec:conclsn}
With the advent of the next generation of GW and neutrino detectors, it would be crucial to lay down strategies to use their potential to perform multimessenger studies of astrophysical sources. In this work, we examined the prospects of performing triggered-stacking searches for coincident high-energy neutrinos from BNS mergers with the next generation GW detectors like CE, ET, and a combination of ET+CE in the context of the neutrino detector IceCube-Gen2. However, the next generation GW detectors would be sensitive to large distances, which makes it important to investigate which triggers would be optimal to perform such searches for coincident high-energy neutrinos. This is because sensitivities to large distances imply a large number of triggers in turn leading to large backgrounds in the neutrino searches which is undesired. 

In this work, we addressed this issue and showed that a limiting distance can be used for the triggers given a chosen threshold of sky localization ($f_{\rm th}$). We obtained the limiting distance based on $f_{\rm th}$ and the sky localization capability for a given GW detector. In Fig.~\ref{fig:angerr_rates}, we show the size of the localization area for a single BNS merger event with the luminosity distance for each of the detectors and infer that the combination ET+CE has the best localization capability, followed by ET and CE. The redshift dependent rate of BNS merger, shown in Fig.~\ref{fig:angerr_rates}, also helps in deciding the number of triggers. We use a fiducial rate of $300\ \rm{Mpc^{-3} yr^{-1}}$ for this work. The fraction of sky area covered obtained using Eq.~\eqref{eq:fth} is shown in Fig.~\ref{fig:delt_allskyerr}, where we see for CE choosing $f_{\rm th} \sim 1$\% gives a limiting distance of $\sim 0.72$ Gpc, for ET a choice of $f_{\rm th} \sim 0.1$\% provides a limiting distance of $\sim 1.12$ Gpc, and choosing $f_{\rm th} \sim 0.01$\% for ET+CE leads to a limiting distance of $\sim 1.85$ Gpc.

The limiting distance depends on the choice of $f_{\rm th}$ and also the time interval between the merger and the neutrino emission peak $\delta t$. In Fig.~\ref{fig:et_combined_fth_deltat}, we show the limiting distance as a density plot on the $f_{\rm th} - \delta t$ plane. We note that the optimistic scenario is given by small values of $\delta t \sim 1 - 10$ s where a limiting distance of $\sim$ a few Gpc (a few ten Gpc) is possible for ET (ET+CE), which is plausible for some physical models for high-energy neutrino emission from BNS mergers. However, for large values of $\delta t$, which can be realized in magnetar-powered scenarios, the limiting distance reduces to $\sim$ a few hundred Mpc for ET and a few Gpc for ET+CE. 

Our main result for detection prospects is shown in Fig.~\ref{fig:ce_et_comb_fnu_delt}. We considered a variation in the total energy emitted in high-energy neutrinos by varying $f_\nu$ between $10^{-5}$ and $5 \times 10^{-5}$ and choose $f_\nu = 2.5 \times 10^{-5}$ as the fiducial value. We also vary $\delta t$ between $1$ s and $10^6$ s while choosing $\delta t = 1000$ s as the fiducial case. For the fiducial parameters, we find that CE can make a $2 \sigma$ level constraint in case of non-detection in a time span of $\sim 11$ years. ET can lead to $3 \sigma$ level constraints in a time span of $14$ years, while the combination of ET+CE does the same in an operation time scale of $\sim 10$ years. For the less optimistic cases, ET+CE, owing to its excellent sky localization capabilities, can still lead to $3\sigma$ level of detection or constraints in $\lesssim 20$ years, while ET would take $\sim 20$ years to reach the $2 \sigma$ C.L..

Even in the case of non-detection of coincident high-energy neutrino events, our analysis can constrain $f_\nu$, that is, the total energy emitted in high-energy neutrinos. Such constraints can be very useful for understanding the emission mechanisms associated with high-energy neutrinos and gamma-rays. This provides further insights into understanding neutrinos from choked jets, from the interaction of the jet with the ejecta besides providing information about the neutrino emission sites and mechanisms associated with BNS mergers. For example, in the optimistic scenario of Ref.~\cite{Kimura:2018vvz} where $f_\nu \sim 10^{-4}$ and $\delta t \sim 2$ s, a $3 \sigma$ C.L. constraint is possible using CE, ET, and ET+CE in a timescale of $\sim 7,4.5,$ and $2.5$ years respectively. However, the scenario considered in Ref.~\cite{Kimura:2018vvz} is optimistic and has been disfavored in subsequent works~\cite{Gottlieb:2021pzr}. The method of gravitational wave triggered searches for high energy neutrinos from BNS mergers can thus provide observational evidence to confirm or constrain the models. Another model for extended neutrino emission from short GRBs was considered in Ref.~\cite{Kimura:2017kan}. In this case, $\delta t \sim 100$ s and $f_\nu \sim 10^{-5}$ for the optimistic scenario. This implies a $3\sigma$ level constraint can be obtained in $\sim 4,2.5,$ and $1.5$ years, using CE, ET and CE+ET respectively. Similarly, for the less optimistic case of Ref.~\cite{Fang:2017tla} where $f_\nu \sim 10^{-4}$ and $\delta \sim 10^6$ s, ET can give $2 \sigma$ level constraints on a timescale of $\sim 20$ years, while the combination of ET+CE can lead to $3 \sigma$ level constraints in an operating time of $\lesssim 30$ years.

We also provide an estimate of the backgrounds in IceCube-Gen2 associated with the search method presented in this work. Considering backgrounds constrain the choice of $f_{\rm th}$ given a desired value of signal to background ratio and the desired operation time for the GW detectors. We show the contours of the number of triggered background events in Fig.~\ref{fig:fth_top}. From the figure, it is also evident that operation timescales of $> 10$ years are not reasonable due to the extremely small value of $f_{\rm th}$ required.

\section{Discussion and Implications}
\label{sec:disc}

It is important to note that our current work is complementary to the likelihood searches performed by the high energy neutrino detectors, which also use triggered stacking searches to look for high energy neutrino events associated with merger event signals from the GW detectors. This is because our projections of a limiting distance for the next generation detectors may allow for efficient likelihood searches to then be performed on the meaningful triggers obtained from the distance cuts proposed in this work. Having an appropriate distance cut will also help reduce the computational costs without losing much of the signals and hence complement the likelihood analyses by IceCube or KM3NeT.

In fact, our method in its essence is similar to the one used by LLAMA~\cite{Countryman:2019pqq} by incorporating distance as an important astrophysical prior for the analysis. For a coincident high energy neutrino and GW event, the analysis in LLAMA considers a time-window of $\sim 1000$ s around the GW event similar to our fiducial value and a statistically significant overlap between the localization areas of the GW and high energy neutrino signal takes into account the distance as a part of the signal likelihood. However, our analysis differs from LLAMA in this way that we consider the time window post the GW trigger and not prior to that and, we only use the localization information from the GW event and do not consider the sky localization associated with the neutrino event. This implies we do not have the likelihood function involving the spatial signal and background probability distribution functions (PDFs) as is used in the analysis for LLAMA. Such a technique involving the construction of a likelihood density using both the localization information from GW and neutrino events would be important for identifying individual sources which is not the main focus of our work.

The uncertainties in this work include the fiducial rate of BNS mergers since it can vary over an order of magnitude compared to our fiducial rate. Moreover, the redshift dependent rate for BNS mergers is assumed to be the same as that of short GRBs. A higher fiducial rate would lead to even more optimistic results, i.e., associated neutrino detection or constraint in shorter operation timescales.
Another important aspect that can further improve our analysis is the quality of triggers obtained from the GW detectors. For the present analysis, we treat every trigger to be of the same quality which is far from reality. Since this search method need not be in real time, in principle, detailed analysis of the signal events would be done by CE and ET and one can then only select triggers above a threshold quality. This would also lead to more improved results in shorter operation timescales. 

We also ignore the downtime of the detectors for simplicity and assume the duty cycle or detection efficiency to be unity. Realistically, the neutrino detectors like IceCube would have a duty cycle of $\sim 100$\%~\cite{IceCube:2016zyt}, whereas the gravitational-wave detectors can generally have a duty cycle $\sim 70$\%~\cite{LIGO:2021ppb}. This would only affect our results in a minor way. Another important factor that we neglect in this work is the beaming associated with the gravitational wave emission itself. It is expected that a fraction of gravitational wave events that are associated to on-axis jets can be much larger than that for the isotropic case (see figure 4 in Ref.~\cite{Schutz:2011tw}). Such events will have a lower associated error region. Since we do not consider this effect our estimates are on the conservative side.

The potential of follow-up or simultaneous electromagnetic (EM) observation can also help reduce the localization area for distant events, which can then improve the results. A BNS merger can be followed by a short GRB, a kilonova, afterglows, and magnetar wind nebulae~\cite{Baiotti:2016qnr,Radice:2020ddv,10.3389/fspas.2020.609460}. 
A promising scenario would be an optical detection of kilonova. Optical detectors like Rubin~\cite{Blum:2022dxi} have very good angular resolution $\sim 1$ arcsecond, resulting in extremely good sky localization. In case of sGRBs, the Gamma-ray Burst Monitor (GBM) onboard the Fermi satellite~\cite{FermiGBM,Goldstein:2019pcj} may detect the gamma-rays from the prompt phase. This does not help with significant reduction in the sky localization area, because the angular resolution can be comparable to that by GW detectors. Besides, the fraction of BNS mergers estimated to have sGRB jets is low $\sim 1$\%. A possible next generation GRB detector on the satellite Galileo G2~\cite{grb_satellite} would also significantly improve the localization prospects given its capability of localization up to $\sim 1^\circ$ at $1 \sigma$ C.L.. Another next generation GRB detector, Space-based multi-band astronomical Variable Objects Monitor (SVOM)~\cite{Wei:2016eox} can help detect GRBs upto very high redshifts of $z > 5$ complementing the next-generation of gravitational wave detectors. Swift BAT~\cite{Barthelmy:2005hs} and future wide-field soft x-ray detectors, such as Einstein Probe~\cite{EinsteinProbeTeam:2015bcj} and HiZ-GUNDAM~\cite{2020SPIE11444E..2ZY} will have minutes-scale angular resolution, which should help reducing the localization.

Electromagnetic triggered searches can also be used to look for associated high-energy neutrino events from BNS mergers. The typical maximum redshift for kilonova detection for the future optical, infrared telescopes like Roman~\cite{DES:2017dgt} is $z \sim 0.2$, the Vera C. Rubin Observatory's Legacy Survey of Space and Time (LSST)~\cite{LSST:2008ijt} is $z \sim 0.1$, and the Zwicky Transient Facility (ZTF)~\cite{2019PASP..131a8002B} is $z\sim 0.02$ (see~\cite{Chase:2021ood} for details). In particular, Roman has a distance horizon for kilonova detection comparable to LIGO A+~\cite{Cahillane:2022pqm}. LSST's kilonova detection horizon is comparable to the planned upgrade of advanced LIGO (aLIGO)~\cite{LIGOScientific:2014pky}. The kilonova detection horizons using such wide-field instruments vary significantly with its associated properties like ejecta velocity and mass. However, these kilonova detection horizons from optical and infrared telescopes, are less than the projected reach for the third generation GW detectors: ET and CE.  

Although we focus on the next generation GW detectors in the current work, the upcoming improvements to the current generation of GW detectors might also lead to some constraints for the optimistic scenarios. A detailed study regarding the sky-localization capabilities of the next generation of ground based detectors -- LIGO, VIRGO, KAGRA, LIGO-India network was performed in~\cite{Gupta:2023evt}.

The second generation of the GW detectors would be able to detect mergers up to $\sim 200 - 300$ Mpc, implying the detection horizon to be $\sim 3$ times lower than what CE by itself is capable of. Furthermore, the current generation of IceCube has an effective area that is $\sim 5$ times smaller than the planned IceCube-Gen2. Plugging in the factors in Eqs.~\eqref{eq:qtop} and~\eqref{eq:integ}, we can roughly estimate the required operation timescale for the second generation of GW detectors along with current generation IceCube to be $\sim 15$ times longer than the timescales proposed using the next generation GW detectors and IceCube-Gen2 used in this work. This explains the absence of detection of high-energy neutrinos from BNS mergers using the current GW detectors~\cite{ANTARES:2017bia,IceCube:2020xks,ANTARES:2023wcj} and also illustrates the inability to place constraints in reasonable timescales. 

The distance sensitivities of the improved detectors, such as LIGO A+~\cite{KAGRA:2013rdx} and LIGO A\# (A-sharp)~\cite{T2200287}\footnote{Also see Fig. 1 in Ref.~\cite{Gupta:2023evt} for a comparison.},
are lesser than the proposed next generation detectors like ET and CE, but they can still help with the nearby detections reducing the sky-localization area and also providing some level of constraints until ET and CE are operational. For example, LIGO A\# with its improved sensitivity would still have a distance horizon upto $\sim 500$ Mpc where the rate of BNS mergers is low and hence the chances of seeing a high energy neutrino event associated with BNS mergers is low, leading to weaker constraints or longer time scales of operation to detect a significant event. On the other hand, for the neutrino detectors, a similar analysis can be performed for other next generation of planned neutrino detectors: KM3NeT~\cite{KM3Net:2016zxf}, Baikal GVD~\cite{BAIKAL:2013jko}, P-ONE~\cite{P-ONE:2020ljt}, and TRIDENT~\cite{Ye:2022vbk}. Furthermore, various proposed radio detectors like IceCube-Gen2 Radio~\cite{IceCube-Gen2:2020qha}, GRAND~\cite{GRAND:2018iaj}, RNO-G~\cite{RNO-G:2020rmc}, and optical air-shower ultra-high energy neutrino detectors like Trinity~\cite{Otte:2019aaf}, can also help with performing triggered stacking searches for high energy neutrinos from BNS mergers.

We stress that since we did not focus on collecting neutrino events from individual sources in this work, the sky localization area was a good quantity to consider for deciding a distance threshold. In most cases, the BNS merger can be identified as the high-energy neutrino source for nearby events, because two error regions associated with the GW detection have a low probability of overlapping. For the worst case, involving unlikely but large overlaps, another strategy might have to be adapted to decide a distance limit based on triggers from the GW detectors such that the total number of sources at any given instant is less than $2$. 

Ideally, a more complete strategy would be to consider any generic astrophysical source and understand whether a gravitational or neutrino trigger would be more useful in that particular context. Since the GW detectors have a much larger horizon as compared to neutrino detectors even if a source emits less energy in GWs, it might still be useful to use the GW detectors as triggers. This would also help understand what sources are ideal candidates for collecting a good sample of high energy neutrinos. Although a universally approximate neutrino spectra can be constructed for such sources, the GW templates differ between various sources depending on the physical instabilities and existing numerical models. Thus, we leave such a generic and evidently powerful analysis for future work.

With the next generation of planned detectors in neutrino, GW and electromagnetic channels the multimessenger landscape will be filled with new discovery potentials and detecting high-energy neutrino events coincident with BNS mergers will definitely be a crucial goal. Our work provides a perspective and lays down a strategy for optimum synergic operation of the GW detectors along with the neutrino detectors.

\acknowledgements
We thank Imre Bartos and Bangalore Sathyaprakash for useful comments and discussion.
M.\,M. wishes to thank the Astronomical Institute at Tohoku University, the Yukawa Institute for Theoretical Physics (YITP), Kyoto University for their hospitality where a major part of this work was completed.
M.\,M. and K.\,M. are supported by NSF Grant No. AST-2108466. M.\,M. also acknowledges support from the 
Institute for Gravitation and the Cosmos (IGC) Postdoctoral Fellowship.
S.S.K. acknowledges the support by KAKENHI No. 22K14028, No. 21H04487, No. 23H04899, and the Tohoku Initiative for Fostering Global Researchers for Interdisciplinary Sciences (TI-FRIS) of MEXT’s Strategic Professional Development Program for Young Researchers.
The work of K.M. is supported by the NSF Grant No.~AST-2108467, and KAKENHI No.~20H01901 and No.~20H05852.

\bibstyle{aps}
\bibliography{refs}

\begin{thebibliography}{86}%
\makeatletter
\providecommand \@ifxundefined [1]{%
 \@ifx{#1\undefined}
}%
\providecommand \@ifnum [1]{%
 \ifnum #1\expandafter \@firstoftwo
 \else \expandafter \@secondoftwo
 \fi
}%
\providecommand \@ifx [1]{%
 \ifx #1\expandafter \@firstoftwo
 \else \expandafter \@secondoftwo
 \fi
}%
\providecommand \natexlab [1]{#1}%
\providecommand \enquote  [1]{``#1''}%
\providecommand \bibnamefont  [1]{#1}%
\providecommand \bibfnamefont [1]{#1}%
\providecommand \citenamefont [1]{#1}%
\providecommand \href@noop [0]{\@secondoftwo}%
\providecommand \href [0]{\begingroup \@sanitize@url \@href}%
\providecommand \@href[1]{\@@startlink{#1}\@@href}%
\providecommand \@@href[1]{\endgroup#1\@@endlink}%
\providecommand \@sanitize@url [0]{\catcode `\\12\catcode `\$12\catcode
  `\&12\catcode `\#12\catcode `\^12\catcode `\_12\catcode `\%12\relax}%
\providecommand \@@startlink[1]{}%
\providecommand \@@endlink[0]{}%
\providecommand \url  [0]{\begingroup\@sanitize@url \@url }%
\providecommand \@url [1]{\endgroup\@href {#1}{\urlprefix }}%
\providecommand \urlprefix  [0]{URL }%
\providecommand \Eprint [0]{\href }%
\providecommand \doibase [0]{https://doi.org/}%
\providecommand \selectlanguage [0]{\@gobble}%
\providecommand \bibinfo  [0]{\@secondoftwo}%
\providecommand \bibfield  [0]{\@secondoftwo}%
\providecommand \translation [1]{[#1]}%
\providecommand \BibitemOpen [0]{}%
\providecommand \bibitemStop [0]{}%
\providecommand \bibitemNoStop [0]{.\EOS\space}%
\providecommand \EOS [0]{\spacefactor3000\relax}%
\providecommand \BibitemShut  [1]{\csname bibitem#1\endcsname}%
\let\auto@bib@innerbib\@empty
\bibitem [{\citenamefont {Aartsen}\ \emph {et~al.}(2021)\citenamefont {Aartsen}
  \emph {et~al.}}]{IceCube-Gen2:2020qha}%
  \BibitemOpen
  \bibfield  {author} {\bibinfo {author} {\bibfnamefont {M.~G.}\ \bibnamefont
  {Aartsen}} \emph {et~al.} (\bibinfo {collaboration} {IceCube-Gen2}),\
  }\bibfield  {title} {\bibinfo {title} {{IceCube-Gen2: the window to the
  extreme Universe}},\ }\href {https://doi.org/10.1088/1361-6471/abbd48}
  {\bibfield  {journal} {\bibinfo  {journal} {J. Phys. G}\ }\textbf {\bibinfo
  {volume} {48}},\ \bibinfo {pages} {060501} (\bibinfo {year} {2021})},\
  \Eprint {https://arxiv.org/abs/2008.04323} {arXiv:2008.04323 [astro-ph.HE]}
  \BibitemShut {NoStop}%
\bibitem [{\citenamefont {Adrian-Martinez}\ \emph {et~al.}(2016)\citenamefont
  {Adrian-Martinez} \emph {et~al.}}]{KM3Net:2016zxf}%
  \BibitemOpen
  \bibfield  {author} {\bibinfo {author} {\bibfnamefont {S.}~\bibnamefont
  {Adrian-Martinez}} \emph {et~al.} (\bibinfo {collaboration} {KM3Net}),\
  }\bibfield  {title} {\bibinfo {title} {{Letter of intent for KM3NeT 2.0}},\
  }\href {https://doi.org/10.1088/0954-3899/43/8/084001} {\bibfield  {journal}
  {\bibinfo  {journal} {J. Phys. G}\ }\textbf {\bibinfo {volume} {43}},\
  \bibinfo {pages} {084001} (\bibinfo {year} {2016})},\ \Eprint
  {https://arxiv.org/abs/1601.07459} {arXiv:1601.07459 [astro-ph.IM]}
  \BibitemShut {NoStop}%
\bibitem [{\citenamefont {Agostini}\ \emph {et~al.}(2020)\citenamefont
  {Agostini} \emph {et~al.}}]{P-ONE:2020ljt}%
  \BibitemOpen
  \bibfield  {author} {\bibinfo {author} {\bibfnamefont {M.}~\bibnamefont
  {Agostini}} \emph {et~al.} (\bibinfo {collaboration} {P-ONE}),\ }\bibfield
  {title} {\bibinfo {title} {{The Pacific Ocean Neutrino Experiment}},\ }\href
  {https://doi.org/10.1038/s41550-020-1182-4} {\bibfield  {journal} {\bibinfo
  {journal} {Nature Astron.}\ }\textbf {\bibinfo {volume} {4}},\ \bibinfo
  {pages} {913} (\bibinfo {year} {2020})},\ \Eprint
  {https://arxiv.org/abs/2005.09493} {arXiv:2005.09493 [astro-ph.HE]}
  \BibitemShut {NoStop}%
\bibitem [{\citenamefont {{Ye}}\ \emph {et~al.}(2022)\citenamefont {{Ye}},
  \citenamefont {{Hu}}, \citenamefont {{Tian}}, \citenamefont {{Chang}},
  \citenamefont {{Chang}}, \citenamefont {{Cheng}}, \citenamefont {{Gao}},
  \citenamefont {{Ge}}, \citenamefont {{Gong}}, \citenamefont {{Guo}},
  \citenamefont {{Guo}}, \citenamefont {{He}}, \citenamefont {{Huang}},
  \citenamefont {{Jiang}}, \citenamefont {{Jiang}}, \citenamefont {{Jing}},
  \citenamefont {{Li}}, \citenamefont {{Li}}, \citenamefont {{Li}},
  \citenamefont {{Li}}, \citenamefont {{Li}}, \citenamefont {{Liao}},
  \citenamefont {{Lin}}, \citenamefont {{Liu}}, \citenamefont {{Liu}},
  \citenamefont {{Liu}}, \citenamefont {{Miao}}, \citenamefont {{Mo}},
  \citenamefont {{Morton-Blake}}, \citenamefont {{Peng}}, \citenamefont
  {{Sun}}, \citenamefont {{Tang}}, \citenamefont {{Tang}}, \citenamefont
  {{Tao}}, \citenamefont {{Tian}}, \citenamefont {{Wang}}, \citenamefont
  {{Wang}}, \citenamefont {{Wang}}, \citenamefont {{Wei}}, \citenamefont
  {{Wei}}, \citenamefont {{Wu}}, \citenamefont {{Xian}}, \citenamefont
  {{Xiang}}, \citenamefont {{Xu}}, \citenamefont {{Xue}}, \citenamefont
  {{Yang}}, \citenamefont {{Yang}}, \citenamefont {{Yu}}, \citenamefont
  {{Zeng}}, \citenamefont {{Zhang}}, \citenamefont {{Zhang}}, \citenamefont
  {{Zhang}}, \citenamefont {{Zhang}}, \citenamefont {{Zhi}}, \citenamefont
  {{Zhong}}, \citenamefont {{Zhou}}, \citenamefont {{Zhu}},\ and\ \citenamefont
  {{Zhuang}}}]{Ye:2022vbk}%
  \BibitemOpen
  \bibfield  {author} {\bibinfo {author} {\bibfnamefont {Z.~P.}\ \bibnamefont
  {{Ye}}}, \bibinfo {author} {\bibfnamefont {F.}~\bibnamefont {{Hu}}}, \bibinfo
  {author} {\bibfnamefont {W.}~\bibnamefont {{Tian}}}, \bibinfo {author}
  {\bibfnamefont {Q.~C.}\ \bibnamefont {{Chang}}}, \bibinfo {author}
  {\bibfnamefont {Y.~L.}\ \bibnamefont {{Chang}}}, \bibinfo {author}
  {\bibfnamefont {Z.~S.}\ \bibnamefont {{Cheng}}}, \bibinfo {author}
  {\bibfnamefont {J.}~\bibnamefont {{Gao}}}, \bibinfo {author} {\bibfnamefont
  {T.}~\bibnamefont {{Ge}}}, \bibinfo {author} {\bibfnamefont {G.~H.}\
  \bibnamefont {{Gong}}}, \bibinfo {author} {\bibfnamefont {J.}~\bibnamefont
  {{Guo}}}, \bibinfo {author} {\bibfnamefont {X.~X.}\ \bibnamefont {{Guo}}},
  \bibinfo {author} {\bibfnamefont {X.~G.}\ \bibnamefont {{He}}}, \bibinfo
  {author} {\bibfnamefont {J.~T.}\ \bibnamefont {{Huang}}}, \bibinfo {author}
  {\bibfnamefont {K.}~\bibnamefont {{Jiang}}}, \bibinfo {author} {\bibfnamefont
  {P.~K.}\ \bibnamefont {{Jiang}}}, \bibinfo {author} {\bibfnamefont {Y.~P.}\
  \bibnamefont {{Jing}}}, \bibinfo {author} {\bibfnamefont {H.~L.}\
  \bibnamefont {{Li}}}, \bibinfo {author} {\bibfnamefont {J.~L.}\ \bibnamefont
  {{Li}}}, \bibinfo {author} {\bibfnamefont {L.}~\bibnamefont {{Li}}}, \bibinfo
  {author} {\bibfnamefont {W.~L.}\ \bibnamefont {{Li}}}, \bibinfo {author}
  {\bibfnamefont {Z.}~\bibnamefont {{Li}}}, \bibinfo {author} {\bibfnamefont
  {N.~Y.}\ \bibnamefont {{Liao}}}, \bibinfo {author} {\bibfnamefont
  {Q.}~\bibnamefont {{Lin}}}, \bibinfo {author} {\bibfnamefont
  {F.}~\bibnamefont {{Liu}}}, \bibinfo {author} {\bibfnamefont {J.~L.}\
  \bibnamefont {{Liu}}}, \bibinfo {author} {\bibfnamefont {X.~H.}\ \bibnamefont
  {{Liu}}}, \bibinfo {author} {\bibfnamefont {P.}~\bibnamefont {{Miao}}},
  \bibinfo {author} {\bibfnamefont {C.}~\bibnamefont {{Mo}}}, \bibinfo {author}
  {\bibfnamefont {I.}~\bibnamefont {{Morton-Blake}}}, \bibinfo {author}
  {\bibfnamefont {T.}~\bibnamefont {{Peng}}}, \bibinfo {author} {\bibfnamefont
  {Z.~Y.}\ \bibnamefont {{Sun}}}, \bibinfo {author} {\bibfnamefont {J.~N.}\
  \bibnamefont {{Tang}}}, \bibinfo {author} {\bibfnamefont {Z.~B.}\
  \bibnamefont {{Tang}}}, \bibinfo {author} {\bibfnamefont {C.~H.}\
  \bibnamefont {{Tao}}}, \bibinfo {author} {\bibfnamefont {X.~L.}\ \bibnamefont
  {{Tian}}}, \bibinfo {author} {\bibfnamefont {M.~X.}\ \bibnamefont {{Wang}}},
  \bibinfo {author} {\bibfnamefont {Y.}~\bibnamefont {{Wang}}}, \bibinfo
  {author} {\bibfnamefont {Y.}~\bibnamefont {{Wang}}}, \bibinfo {author}
  {\bibfnamefont {H.~D.}\ \bibnamefont {{Wei}}}, \bibinfo {author}
  {\bibfnamefont {Z.~Y.}\ \bibnamefont {{Wei}}}, \bibinfo {author}
  {\bibfnamefont {W.~H.}\ \bibnamefont {{Wu}}}, \bibinfo {author}
  {\bibfnamefont {S.~S.}\ \bibnamefont {{Xian}}}, \bibinfo {author}
  {\bibfnamefont {D.}~\bibnamefont {{Xiang}}}, \bibinfo {author} {\bibfnamefont
  {D.~L.}\ \bibnamefont {{Xu}}}, \bibinfo {author} {\bibfnamefont
  {Q.}~\bibnamefont {{Xue}}}, \bibinfo {author} {\bibfnamefont {J.~H.}\
  \bibnamefont {{Yang}}}, \bibinfo {author} {\bibfnamefont {J.~M.}\
  \bibnamefont {{Yang}}}, \bibinfo {author} {\bibfnamefont {W.~B.}\
  \bibnamefont {{Yu}}}, \bibinfo {author} {\bibfnamefont {C.}~\bibnamefont
  {{Zeng}}}, \bibinfo {author} {\bibfnamefont {F.~Y.~D.}\ \bibnamefont
  {{Zhang}}}, \bibinfo {author} {\bibfnamefont {T.}~\bibnamefont {{Zhang}}},
  \bibinfo {author} {\bibfnamefont {X.~T.}\ \bibnamefont {{Zhang}}}, \bibinfo
  {author} {\bibfnamefont {Y.~Y.}\ \bibnamefont {{Zhang}}}, \bibinfo {author}
  {\bibfnamefont {W.}~\bibnamefont {{Zhi}}}, \bibinfo {author} {\bibfnamefont
  {Y.~S.}\ \bibnamefont {{Zhong}}}, \bibinfo {author} {\bibfnamefont
  {M.}~\bibnamefont {{Zhou}}}, \bibinfo {author} {\bibfnamefont {X.~H.}\
  \bibnamefont {{Zhu}}},\ and\ \bibinfo {author} {\bibfnamefont {G.~J.}\
  \bibnamefont {{Zhuang}}},\ }\bibfield  {title} {\bibinfo {title} {{Proposal
  for a neutrino telescope in South China Sea}},\ }\href
  {https://doi.org/10.48550/arXiv.2207.04519} {\bibfield  {journal} {\bibinfo
  {journal} {arXiv e-prints}\ ,\ \bibinfo {eid} {arXiv:2207.04519}} (\bibinfo
  {year} {2022})},\ \Eprint {https://arxiv.org/abs/2207.04519}
  {arXiv:2207.04519 [astro-ph.HE]} \BibitemShut {NoStop}%
\bibitem [{\citenamefont {Avrorin}\ \emph {et~al.}(2014)\citenamefont {Avrorin}
  \emph {et~al.}}]{BAIKAL:2013jko}%
  \BibitemOpen
  \bibfield  {author} {\bibinfo {author} {\bibfnamefont {A.~D.}\ \bibnamefont
  {Avrorin}} \emph {et~al.} (\bibinfo {collaboration} {BAIKAL}),\ }\bibfield
  {title} {\bibinfo {title} {{The prototyping/early construction phase of the
  BAIKAL-GVD project}},\ }\href {https://doi.org/10.1016/j.nima.2013.10.064}
  {\bibfield  {journal} {\bibinfo  {journal} {Nucl. Instrum. Meth. A}\ }\textbf
  {\bibinfo {volume} {742}},\ \bibinfo {pages} {82} (\bibinfo {year} {2014})},\
  \Eprint {https://arxiv.org/abs/1308.1833} {arXiv:1308.1833 [astro-ph.IM]}
  \BibitemShut {NoStop}%
\bibitem [{\citenamefont {Maggiore}\ \emph {et~al.}(2020)\citenamefont
  {Maggiore} \emph {et~al.}}]{Maggiore:2019uih}%
  \BibitemOpen
  \bibfield  {author} {\bibinfo {author} {\bibfnamefont {M.}~\bibnamefont
  {Maggiore}} \emph {et~al.},\ }\bibfield  {title} {\bibinfo {title} {{Science
  Case for the Einstein Telescope}},\ }\href
  {https://doi.org/10.1088/1475-7516/2020/03/050} {\bibfield  {journal}
  {\bibinfo  {journal} {JCAP}\ }\textbf {\bibinfo {volume} {03}},\ \bibinfo
  {pages} {050}},\ \Eprint {https://arxiv.org/abs/1912.02622} {arXiv:1912.02622
  [astro-ph.CO]} \BibitemShut {NoStop}%
\bibitem [{\citenamefont {Reitze}\ \emph {et~al.}(2019)\citenamefont {Reitze}
  \emph {et~al.}}]{Reitze:2019iox}%
  \BibitemOpen
  \bibfield  {author} {\bibinfo {author} {\bibfnamefont {D.}~\bibnamefont
  {Reitze}} \emph {et~al.},\ }\bibfield  {title} {\bibinfo {title} {{Cosmic
  Explorer: The U.S. Contribution to Gravitational-Wave Astronomy beyond
  LIGO}},\ }\href {https://doi.org/10.48550/arXiv.1907.04833} {\bibfield
  {journal} {\bibinfo  {journal} {Bull. Am. Astron. Soc.}\ }\textbf {\bibinfo
  {volume} {51}},\ \bibinfo {pages} {035} (\bibinfo {year} {2019})},\ \Eprint
  {https://arxiv.org/abs/1907.04833} {arXiv:1907.04833 [astro-ph.IM]}
  \BibitemShut {NoStop}%
\bibitem [{\citenamefont {Abbott}\ \emph
  {et~al.}(2017{\natexlab{a}})\citenamefont {Abbott} \emph
  {et~al.}}]{LIGOScientific:2017vwq}%
  \BibitemOpen
  \bibfield  {author} {\bibinfo {author} {\bibfnamefont {B.~P.}\ \bibnamefont
  {Abbott}} \emph {et~al.} (\bibinfo {collaboration} {LIGO Scientific,
  Virgo}),\ }\bibfield  {title} {\bibinfo {title} {{GW170817: Observation of
  Gravitational Waves from a Binary Neutron Star Inspiral}},\ }\href
  {https://doi.org/10.1103/PhysRevLett.119.161101} {\bibfield  {journal}
  {\bibinfo  {journal} {Phys. Rev. Lett.}\ }\textbf {\bibinfo {volume} {119}},\
  \bibinfo {pages} {161101} (\bibinfo {year} {2017}{\natexlab{a}})},\ \Eprint
  {https://arxiv.org/abs/1710.05832} {arXiv:1710.05832 [gr-qc]} \BibitemShut
  {NoStop}%
\bibitem [{\citenamefont {Goldstein}\ \emph {et~al.}(2017)\citenamefont
  {Goldstein} \emph {et~al.}}]{Goldstein:2017mmi}%
  \BibitemOpen
  \bibfield  {author} {\bibinfo {author} {\bibfnamefont {A.}~\bibnamefont
  {Goldstein}} \emph {et~al.},\ }\bibfield  {title} {\bibinfo {title} {{An
  Ordinary Short Gamma-Ray Burst with Extraordinary Implications: Fermi-GBM
  Detection of GRB 170817A}},\ }\href
  {https://doi.org/10.3847/2041-8213/aa8f41} {\bibfield  {journal} {\bibinfo
  {journal} {Astrophys. J. Lett.}\ }\textbf {\bibinfo {volume} {848}},\
  \bibinfo {pages} {L14} (\bibinfo {year} {2017})},\ \Eprint
  {https://arxiv.org/abs/1710.05446} {arXiv:1710.05446 [astro-ph.HE]}
  \BibitemShut {NoStop}%
\bibitem [{\citenamefont {Abbott}\ \emph
  {et~al.}(2017{\natexlab{b}})\citenamefont {Abbott} \emph
  {et~al.}}]{LIGOScientific:2017zic}%
  \BibitemOpen
  \bibfield  {author} {\bibinfo {author} {\bibfnamefont {B.~P.}\ \bibnamefont
  {Abbott}} \emph {et~al.} (\bibinfo {collaboration} {LIGO Scientific, Virgo,
  Fermi-GBM, INTEGRAL}),\ }\bibfield  {title} {\bibinfo {title} {{Gravitational
  Waves and Gamma-rays from a Binary Neutron Star Merger: GW170817 and GRB
  170817A}},\ }\href {https://doi.org/10.3847/2041-8213/aa920c} {\bibfield
  {journal} {\bibinfo  {journal} {Astrophys. J. Lett.}\ }\textbf {\bibinfo
  {volume} {848}},\ \bibinfo {pages} {L13} (\bibinfo {year}
  {2017}{\natexlab{b}})},\ \Eprint {https://arxiv.org/abs/1710.05834}
  {arXiv:1710.05834 [astro-ph.HE]} \BibitemShut {NoStop}%
\bibitem [{\citenamefont {Abbott}\ \emph
  {et~al.}(2017{\natexlab{c}})\citenamefont {Abbott} \emph
  {et~al.}}]{LIGOScientific:2017ync}%
  \BibitemOpen
  \bibfield  {author} {\bibinfo {author} {\bibfnamefont {B.~P.}\ \bibnamefont
  {Abbott}} \emph {et~al.} (\bibinfo {collaboration} {LIGO Scientific, Virgo,
  Fermi GBM, INTEGRAL, IceCube, AstroSat Cadmium Zinc Telluride Imager Team,
  IPN, Insight-Hxmt, ANTARES, Swift, AGILE Team, 1M2H Team, Dark Energy Camera
  GW-EM, DES, DLT40, GRAWITA, Fermi-LAT, ATCA, ASKAP, Las Cumbres Observatory
  Group, OzGrav, DWF (Deeper Wider Faster Program), AST3, CAASTRO, VINROUGE,
  MASTER, J-GEM, GROWTH, JAGWAR, CaltechNRAO, TTU-NRAO, NuSTAR, Pan-STARRS,
  MAXI Team, TZAC Consortium, KU, Nordic Optical Telescope, ePESSTO, GROND,
  Texas Tech University, SALT Group, TOROS, BOOTES, MWA, CALET, IKI-GW
  Follow-up, H.E.S.S., LOFAR, LWA, HAWC, Pierre Auger, ALMA, Euro VLBI Team, Pi
  of Sky, Chandra Team at McGill University, DFN, ATLAS Telescopes, High Time
  Resolution Universe Survey, RIMAS, RATIR, SKA South Africa/MeerKAT}),\
  }\bibfield  {title} {\bibinfo {title} {{Multi-messenger Observations of a
  Binary Neutron Star Merger}},\ }\href
  {https://doi.org/10.3847/2041-8213/aa91c9} {\bibfield  {journal} {\bibinfo
  {journal} {Astrophys. J. Lett.}\ }\textbf {\bibinfo {volume} {848}},\
  \bibinfo {pages} {L12} (\bibinfo {year} {2017}{\natexlab{c}})},\ \Eprint
  {https://arxiv.org/abs/1710.05833} {arXiv:1710.05833 [astro-ph.HE]}
  \BibitemShut {NoStop}%
\bibitem [{\citenamefont {Soares-Santos}\ \emph {et~al.}(2017)\citenamefont
  {Soares-Santos} \emph {et~al.}}]{DES:2017kbs}%
  \BibitemOpen
  \bibfield  {author} {\bibinfo {author} {\bibfnamefont {M.}~\bibnamefont
  {Soares-Santos}} \emph {et~al.} (\bibinfo {collaboration} {DES, Dark Energy
  Camera GW-EM}),\ }\bibfield  {title} {\bibinfo {title} {{The Electromagnetic
  Counterpart of the Binary Neutron Star Merger LIGO/Virgo GW170817. I.
  Discovery of the Optical Counterpart Using the Dark Energy Camera}},\ }\href
  {https://doi.org/10.3847/2041-8213/aa9059} {\bibfield  {journal} {\bibinfo
  {journal} {Astrophys. J. Lett.}\ }\textbf {\bibinfo {volume} {848}},\
  \bibinfo {pages} {L16} (\bibinfo {year} {2017})},\ \Eprint
  {https://arxiv.org/abs/1710.05459} {arXiv:1710.05459 [astro-ph.HE]}
  \BibitemShut {NoStop}%
\bibitem [{\citenamefont {Coulter}\ \emph {et~al.}(2017)\citenamefont {Coulter}
  \emph {et~al.}}]{Coulter:2017wya}%
  \BibitemOpen
  \bibfield  {author} {\bibinfo {author} {\bibfnamefont {D.~A.}\ \bibnamefont
  {Coulter}} \emph {et~al.},\ }\bibfield  {title} {\bibinfo {title} {{Swope
  Supernova Survey 2017a (SSS17a), the Optical Counterpart to a Gravitational
  Wave Source}},\ }\href {https://doi.org/10.1126/science.aap9811} {\bibfield
  {journal} {\bibinfo  {journal} {Science}\ }\textbf {\bibinfo {volume}
  {358}},\ \bibinfo {pages} {1556} (\bibinfo {year} {2017})},\ \Eprint
  {https://arxiv.org/abs/1710.05452} {arXiv:1710.05452 [astro-ph.HE]}
  \BibitemShut {NoStop}%
\bibitem [{\citenamefont {Utsumi}\ \emph {et~al.}(2017)\citenamefont {Utsumi}
  \emph {et~al.}}]{J-GEM:2017tyx}%
  \BibitemOpen
  \bibfield  {author} {\bibinfo {author} {\bibfnamefont {Y.}~\bibnamefont
  {Utsumi}} \emph {et~al.} (\bibinfo {collaboration} {J-GEM}),\ }\bibfield
  {title} {\bibinfo {title} {{J-GEM observations of an electromagnetic
  counterpart to the neutron star merger GW170817}},\ }\href
  {https://doi.org/10.1093/pasj/psx118} {\bibfield  {journal} {\bibinfo
  {journal} {Publ. Astron. Soc. Jap.}\ }\textbf {\bibinfo {volume} {69}},\
  \bibinfo {pages} {101} (\bibinfo {year} {2017})},\ \Eprint
  {https://arxiv.org/abs/1710.05848} {arXiv:1710.05848 [astro-ph.HE]}
  \BibitemShut {NoStop}%
\bibitem [{\citenamefont {Valenti}\ \emph {et~al.}(2017)\citenamefont
  {Valenti}, \citenamefont {Sand}, \citenamefont {Yang}, \citenamefont
  {Cappellaro}, \citenamefont {Tartaglia}, \citenamefont {Corsi}, \citenamefont
  {Jha}, \citenamefont {Reichart}, \citenamefont {Haislip},\ and\ \citenamefont
  {Kouprianov}}]{Valenti:2017ngx}%
  \BibitemOpen
  \bibfield  {author} {\bibinfo {author} {\bibfnamefont {S.}~\bibnamefont
  {Valenti}}, \bibinfo {author} {\bibfnamefont {D.~J.}\ \bibnamefont {Sand}},
  \bibinfo {author} {\bibfnamefont {S.}~\bibnamefont {Yang}}, \bibinfo {author}
  {\bibfnamefont {E.}~\bibnamefont {Cappellaro}}, \bibinfo {author}
  {\bibfnamefont {L.}~\bibnamefont {Tartaglia}}, \bibinfo {author}
  {\bibfnamefont {A.}~\bibnamefont {Corsi}}, \bibinfo {author} {\bibfnamefont
  {S.~W.}\ \bibnamefont {Jha}}, \bibinfo {author} {\bibfnamefont {D.~E.}\
  \bibnamefont {Reichart}}, \bibinfo {author} {\bibfnamefont {J.}~\bibnamefont
  {Haislip}},\ and\ \bibinfo {author} {\bibfnamefont {V.}~\bibnamefont
  {Kouprianov}},\ }\bibfield  {title} {\bibinfo {title} {{The discovery of the
  electromagnetic counterpart of GW170817: kilonova AT 2017gfo/DLT17ck}},\
  }\href {https://doi.org/10.3847/2041-8213/aa8edf} {\bibfield  {journal}
  {\bibinfo  {journal} {Astrophys. J. Lett.}\ }\textbf {\bibinfo {volume}
  {848}},\ \bibinfo {pages} {L24} (\bibinfo {year} {2017})},\ \Eprint
  {https://arxiv.org/abs/1710.05854} {arXiv:1710.05854 [astro-ph.HE]}
  \BibitemShut {NoStop}%
\bibitem [{\citenamefont {Lipunov}\ \emph {et~al.}(2017)\citenamefont {Lipunov}
  \emph {et~al.}}]{Lipunov:2017dwd}%
  \BibitemOpen
  \bibfield  {author} {\bibinfo {author} {\bibfnamefont {V.~M.}\ \bibnamefont
  {Lipunov}} \emph {et~al.},\ }\bibfield  {title} {\bibinfo {title} {{MASTER
  Optical Detection of the First LIGO/Virgo Neutron Star Binary Merger
  GW170817}},\ }\href {https://doi.org/10.3847/2041-8213/aa92c0} {\bibfield
  {journal} {\bibinfo  {journal} {Astrophys. J. Lett.}\ }\textbf {\bibinfo
  {volume} {850}},\ \bibinfo {pages} {L1} (\bibinfo {year} {2017})},\ \Eprint
  {https://arxiv.org/abs/1710.05461} {arXiv:1710.05461 [astro-ph.HE]}
  \BibitemShut {NoStop}%
\bibitem [{\citenamefont {Albert}\ \emph {et~al.}(2017)\citenamefont {Albert}
  \emph {et~al.}}]{ANTARES:2017bia}%
  \BibitemOpen
  \bibfield  {author} {\bibinfo {author} {\bibfnamefont {A.}~\bibnamefont
  {Albert}} \emph {et~al.} (\bibinfo {collaboration} {ANTARES, IceCube, Pierre
  Auger, LIGO Scientific, Virgo}),\ }\bibfield  {title} {\bibinfo {title}
  {{Search for High-energy Neutrinos from Binary Neutron Star Merger GW170817
  with ANTARES, IceCube, and the Pierre Auger Observatory}},\ }\href
  {https://doi.org/10.3847/2041-8213/aa9aed} {\bibfield  {journal} {\bibinfo
  {journal} {Astrophys. J. Lett.}\ }\textbf {\bibinfo {volume} {850}},\
  \bibinfo {pages} {L35} (\bibinfo {year} {2017})},\ \Eprint
  {https://arxiv.org/abs/1710.05839} {arXiv:1710.05839 [astro-ph.HE]}
  \BibitemShut {NoStop}%
\bibitem [{\citenamefont {Abdalla}\ \emph {et~al.}(2017)\citenamefont {Abdalla}
  \emph {et~al.}}]{HESS:2017kmv}%
  \BibitemOpen
  \bibfield  {author} {\bibinfo {author} {\bibfnamefont {H.}~\bibnamefont
  {Abdalla}} \emph {et~al.} (\bibinfo {collaboration} {H.E.S.S.}),\ }\bibfield
  {title} {\bibinfo {title} {{TeV gamma-ray observations of the binary neutron
  star merger GW170817 with H.E.S.S}},\ }\href
  {https://doi.org/10.3847/2041-8213/aa97d2} {\bibfield  {journal} {\bibinfo
  {journal} {Astrophys. J. Lett.}\ }\textbf {\bibinfo {volume} {850}},\
  \bibinfo {pages} {L22} (\bibinfo {year} {2017})},\ \Eprint
  {https://arxiv.org/abs/1710.05862} {arXiv:1710.05862 [astro-ph.HE]}
  \BibitemShut {NoStop}%
\bibitem [{\citenamefont {Hayato}\ \emph {et~al.}(2018)\citenamefont {Hayato}
  \emph {et~al.}}]{Super-Kamiokande:2018dbf}%
  \BibitemOpen
  \bibfield  {author} {\bibinfo {author} {\bibfnamefont {Y.}~\bibnamefont
  {Hayato}} \emph {et~al.} (\bibinfo {collaboration} {Super-Kamiokande}),\
  }\bibfield  {title} {\bibinfo {title} {{Search for Neutrinos in
  Super-Kamiokande Associated with the GW170817 Neutron-star Merger}},\ }\href
  {https://doi.org/10.3847/2041-8213/aabaca} {\bibfield  {journal} {\bibinfo
  {journal} {Astrophys. J. Lett.}\ }\textbf {\bibinfo {volume} {857}},\
  \bibinfo {pages} {L4} (\bibinfo {year} {2018})},\ \Eprint
  {https://arxiv.org/abs/1802.04379} {arXiv:1802.04379 [astro-ph.HE]}
  \BibitemShut {NoStop}%
\bibitem [{\citenamefont {Kimura}\ \emph {et~al.}(2017)\citenamefont {Kimura},
  \citenamefont {Murase}, \citenamefont {M\'esz\'aros},\ and\ \citenamefont
  {Kiuchi}}]{Kimura:2017kan}%
  \BibitemOpen
  \bibfield  {author} {\bibinfo {author} {\bibfnamefont {S.~S.}\ \bibnamefont
  {Kimura}}, \bibinfo {author} {\bibfnamefont {K.}~\bibnamefont {Murase}},
  \bibinfo {author} {\bibfnamefont {P.}~\bibnamefont {M\'esz\'aros}},\ and\
  \bibinfo {author} {\bibfnamefont {K.}~\bibnamefont {Kiuchi}},\ }\bibfield
  {title} {\bibinfo {title} {{High-Energy Neutrino Emission from Short
  Gamma-Ray Bursts: Prospects for Coincident Detection with Gravitational
  Waves}},\ }\href {https://doi.org/10.3847/2041-8213/aa8d14} {\bibfield
  {journal} {\bibinfo  {journal} {Astrophys. J. Lett.}\ }\textbf {\bibinfo
  {volume} {848}},\ \bibinfo {pages} {L4} (\bibinfo {year} {2017})},\ \Eprint
  {https://arxiv.org/abs/1708.07075} {arXiv:1708.07075 [astro-ph.HE]}
  \BibitemShut {NoStop}%
\bibitem [{\citenamefont {Biehl}\ \emph {et~al.}(2018)\citenamefont {Biehl},
  \citenamefont {Heinze},\ and\ \citenamefont {Winter}}]{Biehl:2017qen}%
  \BibitemOpen
  \bibfield  {author} {\bibinfo {author} {\bibfnamefont {D.}~\bibnamefont
  {Biehl}}, \bibinfo {author} {\bibfnamefont {J.}~\bibnamefont {Heinze}},\ and\
  \bibinfo {author} {\bibfnamefont {W.}~\bibnamefont {Winter}},\ }\bibfield
  {title} {\bibinfo {title} {{Expected neutrino fluence from short Gamma-Ray
  Burst 170817A and off-axis angle constraints}},\ }\href
  {https://doi.org/10.1093/mnras/sty285} {\bibfield  {journal} {\bibinfo
  {journal} {Mon. Not. Roy. Astron. Soc.}\ }\textbf {\bibinfo {volume} {476}},\
  \bibinfo {pages} {1191} (\bibinfo {year} {2018})},\ \Eprint
  {https://arxiv.org/abs/1712.00449} {arXiv:1712.00449 [astro-ph.HE]}
  \BibitemShut {NoStop}%
\bibitem [{\citenamefont {Ahlers}\ and\ \citenamefont
  {Halser}(2019)}]{Ahlers:2019fwz}%
  \BibitemOpen
  \bibfield  {author} {\bibinfo {author} {\bibfnamefont {M.}~\bibnamefont
  {Ahlers}}\ and\ \bibinfo {author} {\bibfnamefont {L.}~\bibnamefont
  {Halser}},\ }\bibfield  {title} {\bibinfo {title} {{Neutrino Fluence from
  Gamma-Ray Bursts: Off-Axis View of Structured Jets}},\ }\href
  {https://doi.org/10.1093/mnras/stz2980} {\bibfield  {journal} {\bibinfo
  {journal} {Mon. Not. Roy. Astron. Soc.}\ }\textbf {\bibinfo {volume} {490}},\
  \bibinfo {pages} {4935} (\bibinfo {year} {2019})},\ \Eprint
  {https://arxiv.org/abs/1908.06953} {arXiv:1908.06953 [astro-ph.HE]}
  \BibitemShut {NoStop}%
\bibitem [{\citenamefont {Kimura}\ \emph {et~al.}(2018)\citenamefont {Kimura},
  \citenamefont {Murase}, \citenamefont {Bartos}, \citenamefont {Ioka},
  \citenamefont {Heng},\ and\ \citenamefont {M\'esz\'aros}}]{Kimura:2018vvz}%
  \BibitemOpen
  \bibfield  {author} {\bibinfo {author} {\bibfnamefont {S.~S.}\ \bibnamefont
  {Kimura}}, \bibinfo {author} {\bibfnamefont {K.}~\bibnamefont {Murase}},
  \bibinfo {author} {\bibfnamefont {I.}~\bibnamefont {Bartos}}, \bibinfo
  {author} {\bibfnamefont {K.}~\bibnamefont {Ioka}}, \bibinfo {author}
  {\bibfnamefont {I.~S.}\ \bibnamefont {Heng}},\ and\ \bibinfo {author}
  {\bibfnamefont {P.}~\bibnamefont {M\'esz\'aros}},\ }\bibfield  {title}
  {\bibinfo {title} {{Transejecta high-energy neutrino emission from binary
  neutron star mergers}},\ }\href {https://doi.org/10.1103/PhysRevD.98.043020}
  {\bibfield  {journal} {\bibinfo  {journal} {Phys. Rev. D}\ }\textbf {\bibinfo
  {volume} {98}},\ \bibinfo {pages} {043020} (\bibinfo {year} {2018})},\
  \Eprint {https://arxiv.org/abs/1805.11613} {arXiv:1805.11613 [astro-ph.HE]}
  \BibitemShut {NoStop}%
\bibitem [{\citenamefont {Carpio}\ and\ \citenamefont
  {Murase}(2020)}]{Carpio:2020app}%
  \BibitemOpen
  \bibfield  {author} {\bibinfo {author} {\bibfnamefont {J.}~\bibnamefont
  {Carpio}}\ and\ \bibinfo {author} {\bibfnamefont {K.}~\bibnamefont
  {Murase}},\ }\bibfield  {title} {\bibinfo {title} {{Oscillation of
  high-energy neutrinos from choked jets in stellar and merger ejecta}},\
  }\href {https://doi.org/10.1103/PhysRevD.101.123002} {\bibfield  {journal}
  {\bibinfo  {journal} {Phys. Rev. D}\ }\textbf {\bibinfo {volume} {101}},\
  \bibinfo {pages} {123002} (\bibinfo {year} {2020})},\ \Eprint
  {https://arxiv.org/abs/2002.10575} {arXiv:2002.10575 [astro-ph.HE]}
  \BibitemShut {NoStop}%
\bibitem [{\citenamefont {Fang}\ and\ \citenamefont
  {Metzger}(2017)}]{Fang:2017tla}%
  \BibitemOpen
  \bibfield  {author} {\bibinfo {author} {\bibfnamefont {K.}~\bibnamefont
  {Fang}}\ and\ \bibinfo {author} {\bibfnamefont {B.~D.}\ \bibnamefont
  {Metzger}},\ }\bibfield  {title} {\bibinfo {title} {{High-Energy Neutrinos
  from Millisecond Magnetars formed from the Merger of Binary Neutron Stars}},\
  }\href {https://doi.org/10.3847/1538-4357/aa8b6a} {\bibfield  {journal}
  {\bibinfo  {journal} {Astrophys. J.}\ }\textbf {\bibinfo {volume} {849}},\
  \bibinfo {pages} {153} (\bibinfo {year} {2017})},\ \Eprint
  {https://arxiv.org/abs/1707.04263} {arXiv:1707.04263 [astro-ph.HE]}
  \BibitemShut {NoStop}%
\bibitem [{\citenamefont {Carpio}\ \emph {et~al.}(2020)\citenamefont {Carpio},
  \citenamefont {Murase}, \citenamefont {Reno}, \citenamefont {Sarcevic},\ and\
  \citenamefont {Stasto}}]{Carpio:2020wzg}%
  \BibitemOpen
  \bibfield  {author} {\bibinfo {author} {\bibfnamefont {J.~A.}\ \bibnamefont
  {Carpio}}, \bibinfo {author} {\bibfnamefont {K.}~\bibnamefont {Murase}},
  \bibinfo {author} {\bibfnamefont {M.~H.}\ \bibnamefont {Reno}}, \bibinfo
  {author} {\bibfnamefont {I.}~\bibnamefont {Sarcevic}},\ and\ \bibinfo
  {author} {\bibfnamefont {A.}~\bibnamefont {Stasto}},\ }\bibfield  {title}
  {\bibinfo {title} {{Charm contribution to ultrahigh-energy neutrinos from
  newborn magnetars}},\ }\href {https://doi.org/10.1103/PhysRevD.102.103001}
  {\bibfield  {journal} {\bibinfo  {journal} {Phys. Rev. D}\ }\textbf {\bibinfo
  {volume} {102}},\ \bibinfo {pages} {103001} (\bibinfo {year} {2020})},\
  \Eprint {https://arxiv.org/abs/2007.07945} {arXiv:2007.07945 [astro-ph.HE]}
  \BibitemShut {NoStop}%
\bibitem [{\citenamefont {Decoene}\ \emph {et~al.}(2020)\citenamefont
  {Decoene}, \citenamefont {Gu\'epin}, \citenamefont {Fang}, \citenamefont
  {Kotera},\ and\ \citenamefont {Metzger}}]{Decoene:2019eux}%
  \BibitemOpen
  \bibfield  {author} {\bibinfo {author} {\bibfnamefont {V.}~\bibnamefont
  {Decoene}}, \bibinfo {author} {\bibfnamefont {C.}~\bibnamefont {Gu\'epin}},
  \bibinfo {author} {\bibfnamefont {K.}~\bibnamefont {Fang}}, \bibinfo {author}
  {\bibfnamefont {K.}~\bibnamefont {Kotera}},\ and\ \bibinfo {author}
  {\bibfnamefont {B.~D.}\ \bibnamefont {Metzger}},\ }\bibfield  {title}
  {\bibinfo {title} {{High-energy neutrinos from fallback accretion of binary
  neutron star merger remnants}},\ }\href
  {https://doi.org/10.1088/1475-7516/2020/04/045} {\bibfield  {journal}
  {\bibinfo  {journal} {JCAP}\ }\textbf {\bibinfo {volume} {04}},\ \bibinfo
  {pages} {045}},\ \Eprint {https://arxiv.org/abs/1910.06578} {arXiv:1910.06578
  [astro-ph.HE]} \BibitemShut {NoStop}%
\bibitem [{\citenamefont {Aartsen}\ \emph {et~al.}(2020)\citenamefont {Aartsen}
  \emph {et~al.}}]{IceCube:2020xks}%
  \BibitemOpen
  \bibfield  {author} {\bibinfo {author} {\bibfnamefont {M.~G.}\ \bibnamefont
  {Aartsen}} \emph {et~al.} (\bibinfo {collaboration} {IceCube}),\ }\bibfield
  {title} {\bibinfo {title} {{IceCube Search for Neutrinos Coincident with
  Compact Binary Mergers from LIGO-Virgo\textquoteright{}s First
  Gravitational-wave Transient Catalog}},\ }\href
  {https://doi.org/10.3847/2041-8213/ab9d24} {\bibfield  {journal} {\bibinfo
  {journal} {Astrophys. J. Lett.}\ }\textbf {\bibinfo {volume} {898}},\
  \bibinfo {pages} {L10} (\bibinfo {year} {2020})},\ \Eprint
  {https://arxiv.org/abs/2004.02910} {arXiv:2004.02910 [astro-ph.HE]}
  \BibitemShut {NoStop}%
\bibitem [{\citenamefont {Bartos}\ \emph {et~al.}(2019)\citenamefont {Bartos},
  \citenamefont {Veske}, \citenamefont {Keivani}, \citenamefont {Marka},
  \citenamefont {Countryman}, \citenamefont {Blaufuss}, \citenamefont
  {Finley},\ and\ \citenamefont {Marka}}]{Bartos:2018jco}%
  \BibitemOpen
  \bibfield  {author} {\bibinfo {author} {\bibfnamefont {I.}~\bibnamefont
  {Bartos}}, \bibinfo {author} {\bibfnamefont {D.}~\bibnamefont {Veske}},
  \bibinfo {author} {\bibfnamefont {A.}~\bibnamefont {Keivani}}, \bibinfo
  {author} {\bibfnamefont {Z.}~\bibnamefont {Marka}}, \bibinfo {author}
  {\bibfnamefont {S.}~\bibnamefont {Countryman}}, \bibinfo {author}
  {\bibfnamefont {E.}~\bibnamefont {Blaufuss}}, \bibinfo {author}
  {\bibfnamefont {C.}~\bibnamefont {Finley}},\ and\ \bibinfo {author}
  {\bibfnamefont {S.}~\bibnamefont {Marka}},\ }\bibfield  {title} {\bibinfo
  {title} {{Bayesian Multi-Messenger Search Method for Common Sources of
  Gravitational Waves and High-Energy Neutrinos}},\ }\href
  {https://doi.org/10.1103/PhysRevD.100.083017} {\bibfield  {journal} {\bibinfo
   {journal} {Phys. Rev. D}\ }\textbf {\bibinfo {volume} {100}},\ \bibinfo
  {pages} {083017} (\bibinfo {year} {2019})},\ \Eprint
  {https://arxiv.org/abs/1810.11467} {arXiv:1810.11467 [astro-ph.HE]}
  \BibitemShut {NoStop}%
\bibitem [{\citenamefont {Abbasi}\ \emph {et~al.}(2023)\citenamefont {Abbasi}
  \emph {et~al.}}]{IceCube:2022mma}%
  \BibitemOpen
  \bibfield  {author} {\bibinfo {author} {\bibfnamefont {R.}~\bibnamefont
  {Abbasi}} \emph {et~al.} (\bibinfo {collaboration} {IceCube}),\ }\bibfield
  {title} {\bibinfo {title} {{IceCube Search for Neutrinos Coincident with
  Gravitational Wave Events from LIGO/Virgo Run O3}},\ }\href
  {https://doi.org/10.3847/1538-4357/aca5fc} {\bibfield  {journal} {\bibinfo
  {journal} {Astrophys. J.}\ }\textbf {\bibinfo {volume} {944}},\ \bibinfo
  {pages} {80} (\bibinfo {year} {2023})},\ \Eprint
  {https://arxiv.org/abs/2208.09532} {arXiv:2208.09532 [astro-ph.HE]}
  \BibitemShut {NoStop}%
\bibitem [{\citenamefont {Countryman}\ \emph {et~al.}(2019)\citenamefont
  {Countryman}, \citenamefont {Keivani}, \citenamefont {Bartos}, \citenamefont
  {Marka}, \citenamefont {Kintscher}, \citenamefont {Corley}, \citenamefont
  {Blaufuss}, \citenamefont {Finley},\ and\ \citenamefont
  {Marka}}]{Countryman:2019pqq}%
  \BibitemOpen
  \bibfield  {author} {\bibinfo {author} {\bibfnamefont {S.}~\bibnamefont
  {Countryman}}, \bibinfo {author} {\bibfnamefont {A.}~\bibnamefont {Keivani}},
  \bibinfo {author} {\bibfnamefont {I.}~\bibnamefont {Bartos}}, \bibinfo
  {author} {\bibfnamefont {Z.}~\bibnamefont {Marka}}, \bibinfo {author}
  {\bibfnamefont {T.}~\bibnamefont {Kintscher}}, \bibinfo {author}
  {\bibfnamefont {R.}~\bibnamefont {Corley}}, \bibinfo {author} {\bibfnamefont
  {E.}~\bibnamefont {Blaufuss}}, \bibinfo {author} {\bibfnamefont
  {C.}~\bibnamefont {Finley}},\ and\ \bibinfo {author} {\bibfnamefont
  {S.}~\bibnamefont {Marka}},\ }\bibfield  {title} {\bibinfo {title}
  {{Low-Latency Algorithm for Multi-messenger Astrophysics (LLAMA) with
  Gravitational-Wave and High-Energy Neutrino Candidates}},\ }\href@noop {} {\
  (\bibinfo {year} {2019})},\ \Eprint {https://arxiv.org/abs/1901.05486}
  {arXiv:1901.05486 [astro-ph.HE]} \BibitemShut {NoStop}%
\bibitem [{\citenamefont {Mukhopadhyay}\ \emph {et~al.}(2022)\citenamefont
  {Mukhopadhyay}, \citenamefont {Lin},\ and\ \citenamefont
  {Lunardini}}]{Mukhopadhyay:2021gox}%
  \BibitemOpen
  \bibfield  {author} {\bibinfo {author} {\bibfnamefont {M.}~\bibnamefont
  {Mukhopadhyay}}, \bibinfo {author} {\bibfnamefont {Z.}~\bibnamefont {Lin}},\
  and\ \bibinfo {author} {\bibfnamefont {C.}~\bibnamefont {Lunardini}},\
  }\bibfield  {title} {\bibinfo {title} {{Memory-triggered supernova neutrino
  detection}},\ }\href {https://doi.org/10.1103/PhysRevD.106.043020} {\bibfield
   {journal} {\bibinfo  {journal} {Phys. Rev. D}\ }\textbf {\bibinfo {volume}
  {106}},\ \bibinfo {pages} {043020} (\bibinfo {year} {2022})},\ \Eprint
  {https://arxiv.org/abs/2110.14657} {arXiv:2110.14657 [astro-ph.HE]}
  \BibitemShut {NoStop}%
\bibitem [{\citenamefont {Mukhopadhyay}(2022)}]{Mukhopadhyay:2022qmo}%
  \BibitemOpen
  \bibfield  {author} {\bibinfo {author} {\bibfnamefont {M.}~\bibnamefont
  {Mukhopadhyay}},\ }\bibfield  {title} {\bibinfo {title} {{Searching for
  supernova neutrinos with GW memory triggers}},\ }in\ \href@noop {} {\emph
  {\bibinfo {booktitle} {{56th Rencontres de Moriond on Electroweak
  Interactions and Unified Theories}}}}\ (\bibinfo {year} {2022})\ \Eprint
  {https://arxiv.org/abs/2205.07966} {arXiv:2205.07966 [astro-ph.HE]}
  \BibitemShut {NoStop}%
\bibitem [{\citenamefont {Kyutoku}\ and\ \citenamefont
  {Kashiyama}(2018)}]{Kyutoku:2017wnb}%
  \BibitemOpen
  \bibfield  {author} {\bibinfo {author} {\bibfnamefont {K.}~\bibnamefont
  {Kyutoku}}\ and\ \bibinfo {author} {\bibfnamefont {K.}~\bibnamefont
  {Kashiyama}},\ }\bibfield  {title} {\bibinfo {title} {{Detectability of
  thermal neutrinos from binary-neutron-star mergers and implication to
  neutrino physics}},\ }\href {https://doi.org/10.1103/PhysRevD.97.103001}
  {\bibfield  {journal} {\bibinfo  {journal} {Phys. Rev. D}\ }\textbf {\bibinfo
  {volume} {97}},\ \bibinfo {pages} {103001} (\bibinfo {year} {2018})},\
  \Eprint {https://arxiv.org/abs/1710.05922} {arXiv:1710.05922 [astro-ph.HE]}
  \BibitemShut {NoStop}%
\bibitem [{\citenamefont {Lin}\ and\ \citenamefont
  {Lunardini}(2020)}]{Lin:2019piz}%
  \BibitemOpen
  \bibfield  {author} {\bibinfo {author} {\bibfnamefont {Z.}~\bibnamefont
  {Lin}}\ and\ \bibinfo {author} {\bibfnamefont {C.}~\bibnamefont
  {Lunardini}},\ }\bibfield  {title} {\bibinfo {title} {{Observing cosmological
  binary mergers with next generation neutrino and gravitational wave
  detectors}},\ }\href {https://doi.org/10.1103/PhysRevD.101.023016} {\bibfield
   {journal} {\bibinfo  {journal} {Phys. Rev. D}\ }\textbf {\bibinfo {volume}
  {101}},\ \bibinfo {pages} {023016} (\bibinfo {year} {2020})},\ \Eprint
  {https://arxiv.org/abs/1907.00034} {arXiv:1907.00034 [astro-ph.HE]}
  \BibitemShut {NoStop}%
\bibitem [{\citenamefont {Murase}\ \emph {et~al.}(2013)\citenamefont {Murase},
  \citenamefont {Ahlers},\ and\ \citenamefont {Lacki}}]{Murase:2013rfa}%
  \BibitemOpen
  \bibfield  {author} {\bibinfo {author} {\bibfnamefont {K.}~\bibnamefont
  {Murase}}, \bibinfo {author} {\bibfnamefont {M.}~\bibnamefont {Ahlers}},\
  and\ \bibinfo {author} {\bibfnamefont {B.~C.}\ \bibnamefont {Lacki}},\
  }\bibfield  {title} {\bibinfo {title} {{Testing the Hadronuclear Origin of
  PeV Neutrinos Observed with IceCube}},\ }\href
  {https://doi.org/10.1103/PhysRevD.88.121301} {\bibfield  {journal} {\bibinfo
  {journal} {Phys. Rev. D}\ }\textbf {\bibinfo {volume} {88}},\ \bibinfo
  {pages} {121301} (\bibinfo {year} {2013})},\ \Eprint
  {https://arxiv.org/abs/1306.3417} {arXiv:1306.3417 [astro-ph.HE]}
  \BibitemShut {NoStop}%
\bibitem [{\citenamefont {Shibata}\ and\ \citenamefont
  {Uryu}(2002)}]{Shibata:2002jb}%
  \BibitemOpen
  \bibfield  {author} {\bibinfo {author} {\bibfnamefont {M.}~\bibnamefont
  {Shibata}}\ and\ \bibinfo {author} {\bibfnamefont {K.}~\bibnamefont {Uryu}},\
  }\bibfield  {title} {\bibinfo {title} {{Gravitational waves from the merger
  of binary neutron stars in a fully general relativistic simulation}},\ }\href
  {https://doi.org/10.1143/PTP.107.265} {\bibfield  {journal} {\bibinfo
  {journal} {Prog. Theor. Phys.}\ }\textbf {\bibinfo {volume} {107}},\ \bibinfo
  {pages} {265} (\bibinfo {year} {2002})},\ \Eprint
  {https://arxiv.org/abs/gr-qc/0203037} {arXiv:gr-qc/0203037} \BibitemShut
  {NoStop}%
\bibitem [{\citenamefont {Radice}\ \emph {et~al.}(2017)\citenamefont {Radice},
  \citenamefont {Bernuzzi}, \citenamefont {Del~Pozzo}, \citenamefont
  {Roberts},\ and\ \citenamefont {Ott}}]{Radice:2016rys}%
  \BibitemOpen
  \bibfield  {author} {\bibinfo {author} {\bibfnamefont {D.}~\bibnamefont
  {Radice}}, \bibinfo {author} {\bibfnamefont {S.}~\bibnamefont {Bernuzzi}},
  \bibinfo {author} {\bibfnamefont {W.}~\bibnamefont {Del~Pozzo}}, \bibinfo
  {author} {\bibfnamefont {L.~F.}\ \bibnamefont {Roberts}},\ and\ \bibinfo
  {author} {\bibfnamefont {C.~D.}\ \bibnamefont {Ott}},\ }\bibfield  {title}
  {\bibinfo {title} {{Probing Extreme-Density Matter with Gravitational Wave
  Observations of Binary Neutron Star Merger Remnants}},\ }\href
  {https://doi.org/10.3847/2041-8213/aa775f} {\bibfield  {journal} {\bibinfo
  {journal} {Astrophys. J. Lett.}\ }\textbf {\bibinfo {volume} {842}},\
  \bibinfo {pages} {L10} (\bibinfo {year} {2017})},\ \Eprint
  {https://arxiv.org/abs/1612.06429} {arXiv:1612.06429 [astro-ph.HE]}
  \BibitemShut {NoStop}%
\bibitem [{\citenamefont {Shibata}\ and\ \citenamefont
  {Kiuchi}(2017)}]{Shibata:2017xht}%
  \BibitemOpen
  \bibfield  {author} {\bibinfo {author} {\bibfnamefont {M.}~\bibnamefont
  {Shibata}}\ and\ \bibinfo {author} {\bibfnamefont {K.}~\bibnamefont
  {Kiuchi}},\ }\bibfield  {title} {\bibinfo {title} {{Gravitational waves from
  remnant massive neutron stars of binary neutron star merger: Viscous
  hydrodynamics effects}},\ }\href {https://doi.org/10.1103/PhysRevD.95.123003}
  {\bibfield  {journal} {\bibinfo  {journal} {Phys. Rev. D}\ }\textbf {\bibinfo
  {volume} {95}},\ \bibinfo {pages} {123003} (\bibinfo {year} {2017})},\
  \Eprint {https://arxiv.org/abs/1705.06142} {arXiv:1705.06142 [astro-ph.HE]}
  \BibitemShut {NoStop}%
\bibitem [{\citenamefont {Zappa}\ \emph {et~al.}(2018)\citenamefont {Zappa},
  \citenamefont {Bernuzzi}, \citenamefont {Radice}, \citenamefont {Perego},\
  and\ \citenamefont {Dietrich}}]{PhysRevLett.120.111101}%
  \BibitemOpen
  \bibfield  {author} {\bibinfo {author} {\bibfnamefont {F.}~\bibnamefont
  {Zappa}}, \bibinfo {author} {\bibfnamefont {S.}~\bibnamefont {Bernuzzi}},
  \bibinfo {author} {\bibfnamefont {D.}~\bibnamefont {Radice}}, \bibinfo
  {author} {\bibfnamefont {A.}~\bibnamefont {Perego}},\ and\ \bibinfo {author}
  {\bibfnamefont {T.}~\bibnamefont {Dietrich}},\ }\bibfield  {title} {\bibinfo
  {title} {Gravitational-wave luminosity of binary neutron stars mergers},\
  }\href {https://doi.org/10.1103/PhysRevLett.120.111101} {\bibfield  {journal}
  {\bibinfo  {journal} {Phys. Rev. Lett.}\ }\textbf {\bibinfo {volume} {120}},\
  \bibinfo {pages} {111101} (\bibinfo {year} {2018})}\BibitemShut {NoStop}%
\bibitem [{\citenamefont {Shibata}\ and\ \citenamefont
  {Hotokezaka}(2019)}]{Shibata:2019wef}%
  \BibitemOpen
  \bibfield  {author} {\bibinfo {author} {\bibfnamefont {M.}~\bibnamefont
  {Shibata}}\ and\ \bibinfo {author} {\bibfnamefont {K.}~\bibnamefont
  {Hotokezaka}},\ }\bibfield  {title} {\bibinfo {title} {{Merger and Mass
  Ejection of Neutron-Star Binaries}},\ }\href
  {https://doi.org/10.1146/annurev-nucl-101918-023625} {\bibfield  {journal}
  {\bibinfo  {journal} {Ann. Rev. Nucl. Part. Sci.}\ }\textbf {\bibinfo
  {volume} {69}},\ \bibinfo {pages} {41} (\bibinfo {year} {2019})},\ \Eprint
  {https://arxiv.org/abs/1908.02350} {arXiv:1908.02350 [astro-ph.HE]}
  \BibitemShut {NoStop}%
\bibitem [{\citenamefont {Radice}\ \emph {et~al.}(2020)\citenamefont {Radice},
  \citenamefont {Bernuzzi},\ and\ \citenamefont {Perego}}]{Radice:2020ddv}%
  \BibitemOpen
  \bibfield  {author} {\bibinfo {author} {\bibfnamefont {D.}~\bibnamefont
  {Radice}}, \bibinfo {author} {\bibfnamefont {S.}~\bibnamefont {Bernuzzi}},\
  and\ \bibinfo {author} {\bibfnamefont {A.}~\bibnamefont {Perego}},\
  }\bibfield  {title} {\bibinfo {title} {{The Dynamics of Binary Neutron Star
  Mergers and GW170817}},\ }\href
  {https://doi.org/10.1146/annurev-nucl-013120-114541} {\bibfield  {journal}
  {\bibinfo  {journal} {Ann. Rev. Nucl. Part. Sci.}\ }\textbf {\bibinfo
  {volume} {70}},\ \bibinfo {pages} {95} (\bibinfo {year} {2020})},\ \Eprint
  {https://arxiv.org/abs/2002.03863} {arXiv:2002.03863 [astro-ph.HE]}
  \BibitemShut {NoStop}%
\bibitem [{\citenamefont {Fong}\ \emph {et~al.}(2015)\citenamefont {Fong},
  \citenamefont {Berger}, \citenamefont {Margutti},\ and\ \citenamefont
  {Zauderer}}]{Fong:2015oha}%
  \BibitemOpen
  \bibfield  {author} {\bibinfo {author} {\bibfnamefont {W.-f.}\ \bibnamefont
  {Fong}}, \bibinfo {author} {\bibfnamefont {E.}~\bibnamefont {Berger}},
  \bibinfo {author} {\bibfnamefont {R.}~\bibnamefont {Margutti}},\ and\
  \bibinfo {author} {\bibfnamefont {B.~A.}\ \bibnamefont {Zauderer}},\
  }\bibfield  {title} {\bibinfo {title} {{A Decade of Short-duration Gamma-ray
  Burst Broadband Afterglows: Energetics, Circumburst Densities, and jet
  Opening Angles}},\ }\href {https://doi.org/10.1088/0004-637X/815/2/102}
  {\bibfield  {journal} {\bibinfo  {journal} {Astrophys. J.}\ }\textbf
  {\bibinfo {volume} {815}},\ \bibinfo {pages} {102} (\bibinfo {year}
  {2015})},\ \Eprint {https://arxiv.org/abs/1509.02922} {arXiv:1509.02922
  [astro-ph.HE]} \BibitemShut {NoStop}%
\bibitem [{\citenamefont {Chan}\ \emph {et~al.}(2018)\citenamefont {Chan},
  \citenamefont {Messenger}, \citenamefont {Heng},\ and\ \citenamefont
  {Hendry}}]{Chan:2018csa}%
  \BibitemOpen
  \bibfield  {author} {\bibinfo {author} {\bibfnamefont {M.~L.}\ \bibnamefont
  {Chan}}, \bibinfo {author} {\bibfnamefont {C.}~\bibnamefont {Messenger}},
  \bibinfo {author} {\bibfnamefont {I.~S.}\ \bibnamefont {Heng}},\ and\
  \bibinfo {author} {\bibfnamefont {M.}~\bibnamefont {Hendry}},\ }\bibfield
  {title} {\bibinfo {title} {{Binary Neutron Star Mergers and Third Generation
  Detectors: Localization and Early Warning}},\ }\href
  {https://doi.org/10.1103/PhysRevD.97.123014} {\bibfield  {journal} {\bibinfo
  {journal} {Phys. Rev. D}\ }\textbf {\bibinfo {volume} {97}},\ \bibinfo
  {pages} {123014} (\bibinfo {year} {2018})},\ \Eprint
  {https://arxiv.org/abs/1803.09680} {arXiv:1803.09680 [astro-ph.HE]}
  \BibitemShut {NoStop}%
\bibitem [{\citenamefont {Regimbau}\ \emph {et~al.}(2012)\citenamefont
  {Regimbau} \emph {et~al.}}]{Regimbau:2012ir}%
  \BibitemOpen
  \bibfield  {author} {\bibinfo {author} {\bibfnamefont {T.}~\bibnamefont
  {Regimbau}} \emph {et~al.},\ }\bibfield  {title} {\bibinfo {title} {{A Mock
  Data Challenge for the Einstein Gravitational-Wave Telescope}},\ }\href
  {https://doi.org/10.1103/PhysRevD.86.122001} {\bibfield  {journal} {\bibinfo
  {journal} {Phys. Rev. D}\ }\textbf {\bibinfo {volume} {86}},\ \bibinfo
  {pages} {122001} (\bibinfo {year} {2012})},\ \Eprint
  {https://arxiv.org/abs/1201.3563} {arXiv:1201.3563 [gr-qc]} \BibitemShut
  {NoStop}%
\bibitem [{\citenamefont {Evans}\ \emph {et~al.}(2021)\citenamefont {Evans}
  \emph {et~al.}}]{Evans:2021gyd}%
  \BibitemOpen
  \bibfield  {author} {\bibinfo {author} {\bibfnamefont {M.}~\bibnamefont
  {Evans}} \emph {et~al.},\ }\bibfield  {title} {\bibinfo {title} {{A Horizon
  Study for Cosmic Explorer: Science, Observatories, and Community}},\
  }\href@noop {} {\  (\bibinfo {year} {2021})},\ \Eprint
  {https://arxiv.org/abs/2109.09882} {arXiv:2109.09882 [astro-ph.IM]}
  \BibitemShut {NoStop}%
\bibitem [{\citenamefont {Baral}\ \emph {et~al.}(2023)\citenamefont {Baral},
  \citenamefont {Morisaki}, \citenamefont {Maga\~na Hernandez},\ and\
  \citenamefont {Creighton}}]{Baral:2023xst}%
  \BibitemOpen
  \bibfield  {author} {\bibinfo {author} {\bibfnamefont {P.}~\bibnamefont
  {Baral}}, \bibinfo {author} {\bibfnamefont {S.}~\bibnamefont {Morisaki}},
  \bibinfo {author} {\bibfnamefont {I.}~\bibnamefont {Maga\~na Hernandez}},\
  and\ \bibinfo {author} {\bibfnamefont {J.}~\bibnamefont {Creighton}},\
  }\bibfield  {title} {\bibinfo {title} {{Localization of binary neutron star
  mergers with a single cosmic explorer}},\ }\href
  {https://doi.org/10.1103/PhysRevD.108.043010} {\bibfield  {journal} {\bibinfo
   {journal} {Phys. Rev. D}\ }\textbf {\bibinfo {volume} {108}},\ \bibinfo
  {pages} {043010} (\bibinfo {year} {2023})},\ \Eprint
  {https://arxiv.org/abs/2304.09889} {arXiv:2304.09889 [astro-ph.HE]}
  \BibitemShut {NoStop}%
\bibitem [{\citenamefont {Abbasi}\ \emph {et~al.}(2021)\citenamefont {Abbasi}
  \emph {et~al.}}]{IceCube:2021xar}%
  \BibitemOpen
  \bibfield  {author} {\bibinfo {author} {\bibfnamefont {R.}~\bibnamefont
  {Abbasi}} \emph {et~al.} (\bibinfo {collaboration} {IceCube}),\ }\bibfield
  {title} {\bibinfo {title} {{IceCube Data for Neutrino Point-Source Searches
  Years 2008-2018}}\ }\href {https://doi.org/10.21234/CPKQ-K003}
  {10.21234/CPKQ-K003} (\bibinfo {year} {2021}),\ \Eprint
  {https://arxiv.org/abs/2101.09836} {arXiv:2101.09836 [astro-ph.HE]}
  \BibitemShut {NoStop}%
\bibitem [{\citenamefont {Wanderman}\ and\ \citenamefont
  {Piran}(2015)}]{Wanderman:2014eza}%
  \BibitemOpen
  \bibfield  {author} {\bibinfo {author} {\bibfnamefont {D.}~\bibnamefont
  {Wanderman}}\ and\ \bibinfo {author} {\bibfnamefont {T.}~\bibnamefont
  {Piran}},\ }\bibfield  {title} {\bibinfo {title} {{The rate, luminosity
  function and time delay of non-Collapsar short GRBs}},\ }\href
  {https://doi.org/10.1093/mnras/stv123} {\bibfield  {journal} {\bibinfo
  {journal} {Mon. Not. Roy. Astron. Soc.}\ }\textbf {\bibinfo {volume} {448}},\
  \bibinfo {pages} {3026} (\bibinfo {year} {2015})},\ \Eprint
  {https://arxiv.org/abs/1405.5878} {arXiv:1405.5878 [astro-ph.HE]}
  \BibitemShut {NoStop}%
\bibitem [{\citenamefont {Abbott}\ \emph {et~al.}(2021)\citenamefont {Abbott}
  \emph {et~al.}}]{LIGOScientific:2021djp}%
  \BibitemOpen
  \bibfield  {author} {\bibinfo {author} {\bibfnamefont {R.}~\bibnamefont
  {Abbott}} \emph {et~al.} (\bibinfo {collaboration} {LIGO Scientific, VIRGO,
  KAGRA}),\ }\bibfield  {title} {\bibinfo {title} {{GWTC-3: Compact Binary
  Coalescences Observed by LIGO and Virgo During the Second Part of the Third
  Observing Run}},\ }\href@noop {} {\  (\bibinfo {year} {2021})},\ \Eprint
  {https://arxiv.org/abs/2111.03606} {arXiv:2111.03606 [gr-qc]} \BibitemShut
  {NoStop}%
\bibitem [{\citenamefont {Abbott}\ \emph {et~al.}(2023)\citenamefont {Abbott}
  \emph {et~al.}}]{KAGRA:2021duu}%
  \BibitemOpen
  \bibfield  {author} {\bibinfo {author} {\bibfnamefont {R.}~\bibnamefont
  {Abbott}} \emph {et~al.} (\bibinfo {collaboration} {KAGRA, VIRGO, LIGO
  Scientific}),\ }\bibfield  {title} {\bibinfo {title} {{Population of Merging
  Compact Binaries Inferred Using Gravitational Waves through GWTC-3}},\ }\href
  {https://doi.org/10.1103/PhysRevX.13.011048} {\bibfield  {journal} {\bibinfo
  {journal} {Phys. Rev. X}\ }\textbf {\bibinfo {volume} {13}},\ \bibinfo
  {pages} {011048} (\bibinfo {year} {2023})},\ \Eprint
  {https://arxiv.org/abs/2111.03634} {arXiv:2111.03634 [astro-ph.HE]}
  \BibitemShut {NoStop}%
\bibitem [{\citenamefont {Hall.}(2022)}]{galaxies10040090}%
  \BibitemOpen
  \bibfield  {author} {\bibinfo {author} {\bibfnamefont {E.~D.}\ \bibnamefont
  {Hall.}},\ }\bibfield  {title} {\bibinfo {title} {Cosmic explorer: A
  next-generation ground-based gravitational-wave observatory},\ }\bibfield
  {journal} {\bibinfo  {journal} {Galaxies}\ }\textbf {\bibinfo {volume}
  {10}},\ \href {https://doi.org/10.3390/galaxies10040090}
  {10.3390/galaxies10040090} (\bibinfo {year} {2022})\BibitemShut {NoStop}%
\bibitem [{\citenamefont {Matsui}\ \emph {et~al.}(2023)\citenamefont {Matsui},
  \citenamefont {Kimura}, \citenamefont {Toma},\ and\ \citenamefont
  {Murase}}]{Matsui:2023ohr}%
  \BibitemOpen
  \bibfield  {author} {\bibinfo {author} {\bibfnamefont {R.}~\bibnamefont
  {Matsui}}, \bibinfo {author} {\bibfnamefont {S.~S.}\ \bibnamefont {Kimura}},
  \bibinfo {author} {\bibfnamefont {K.}~\bibnamefont {Toma}},\ and\ \bibinfo
  {author} {\bibfnamefont {K.}~\bibnamefont {Murase}},\ }\bibfield  {title}
  {\bibinfo {title} {{High-energy Neutrino Emission Associated with
  Gravitational-wave Signals: Effects of Cocoon Photons and Constraints on
  Late-time Emission}},\ }\href {https://doi.org/10.3847/1538-4357/acd004}
  {\bibfield  {journal} {\bibinfo  {journal} {Astrophys. J.}\ }\textbf
  {\bibinfo {volume} {950}},\ \bibinfo {pages} {190} (\bibinfo {year}
  {2023})},\ \Eprint {https://arxiv.org/abs/2302.04130} {arXiv:2302.04130
  [astro-ph.HE]} \BibitemShut {NoStop}%
\bibitem [{\citenamefont {Baret}\ \emph {et~al.}(2011)\citenamefont {Baret}
  \emph {et~al.}}]{Baret:2011tk}%
  \BibitemOpen
  \bibfield  {author} {\bibinfo {author} {\bibfnamefont {B.}~\bibnamefont
  {Baret}} \emph {et~al.},\ }\bibfield  {title} {\bibinfo {title} {{Bounding
  the Time Delay between High-energy Neutrinos and Gravitational-wave
  Transients from Gamma-ray Bursts}},\ }\href
  {https://doi.org/10.1016/j.astropartphys.2011.04.001} {\bibfield  {journal}
  {\bibinfo  {journal} {Astropart. Phys.}\ }\textbf {\bibinfo {volume} {35}},\
  \bibinfo {pages} {1} (\bibinfo {year} {2011})},\ \Eprint
  {https://arxiv.org/abs/1101.4669} {arXiv:1101.4669 [astro-ph.HE]}
  \BibitemShut {NoStop}%
\bibitem [{\citenamefont {Albert}\ \emph {et~al.}(2023)\citenamefont {Albert}
  \emph {et~al.}}]{ANTARES:2023wcj}%
  \BibitemOpen
  \bibfield  {author} {\bibinfo {author} {\bibfnamefont {A.}~\bibnamefont
  {Albert}} \emph {et~al.} (\bibinfo {collaboration} {ANTARES}),\ }\bibfield
  {title} {\bibinfo {title} {{Search for neutrino counterparts to the
  gravitational wave sources from LIGO/Virgo O3 run with the ANTARES
  detector}},\ }\href {https://doi.org/10.1088/1475-7516/2023/04/004}
  {\bibfield  {journal} {\bibinfo  {journal} {JCAP}\ }\textbf {\bibinfo
  {volume} {04}},\ \bibinfo {pages} {004}},\ \Eprint
  {https://arxiv.org/abs/2302.07723} {arXiv:2302.07723 [astro-ph.HE]}
  \BibitemShut {NoStop}%
\bibitem [{\citenamefont {Honda}\ \emph {et~al.}(2007)\citenamefont {Honda},
  \citenamefont {Kajita}, \citenamefont {Kasahara}, \citenamefont
  {Midorikawa},\ and\ \citenamefont {Sanuki}}]{Honda:2006qj}%
  \BibitemOpen
  \bibfield  {author} {\bibinfo {author} {\bibfnamefont {M.}~\bibnamefont
  {Honda}}, \bibinfo {author} {\bibfnamefont {T.}~\bibnamefont {Kajita}},
  \bibinfo {author} {\bibfnamefont {K.}~\bibnamefont {Kasahara}}, \bibinfo
  {author} {\bibfnamefont {S.}~\bibnamefont {Midorikawa}},\ and\ \bibinfo
  {author} {\bibfnamefont {T.}~\bibnamefont {Sanuki}},\ }\bibfield  {title}
  {\bibinfo {title} {{Calculation of atmospheric neutrino flux using the
  interaction model calibrated with atmospheric muon data}},\ }\href
  {https://doi.org/10.1103/PhysRevD.75.043006} {\bibfield  {journal} {\bibinfo
  {journal} {Phys. Rev. D}\ }\textbf {\bibinfo {volume} {75}},\ \bibinfo
  {pages} {043006} (\bibinfo {year} {2007})},\ \Eprint
  {https://arxiv.org/abs/astro-ph/0611418} {arXiv:astro-ph/0611418}
  \BibitemShut {NoStop}%
\bibitem [{\citenamefont {Enberg}\ \emph {et~al.}(2008)\citenamefont {Enberg},
  \citenamefont {Reno},\ and\ \citenamefont {Sarcevic}}]{Enberg:2008te}%
  \BibitemOpen
  \bibfield  {author} {\bibinfo {author} {\bibfnamefont {R.}~\bibnamefont
  {Enberg}}, \bibinfo {author} {\bibfnamefont {M.~H.}\ \bibnamefont {Reno}},\
  and\ \bibinfo {author} {\bibfnamefont {I.}~\bibnamefont {Sarcevic}},\
  }\bibfield  {title} {\bibinfo {title} {{Prompt neutrino fluxes from
  atmospheric charm}},\ }\href {https://doi.org/10.1103/PhysRevD.78.043005}
  {\bibfield  {journal} {\bibinfo  {journal} {Phys. Rev. D}\ }\textbf {\bibinfo
  {volume} {78}},\ \bibinfo {pages} {043005} (\bibinfo {year} {2008})},\
  \Eprint {https://arxiv.org/abs/0806.0418} {arXiv:0806.0418 [hep-ph]}
  \BibitemShut {NoStop}%
\bibitem [{\citenamefont {Bhattacharya}\ \emph {et~al.}(2015)\citenamefont
  {Bhattacharya}, \citenamefont {Enberg}, \citenamefont {Reno}, \citenamefont
  {Sarcevic},\ and\ \citenamefont {Stasto}}]{Bhattacharya:2015jpa}%
  \BibitemOpen
  \bibfield  {author} {\bibinfo {author} {\bibfnamefont {A.}~\bibnamefont
  {Bhattacharya}}, \bibinfo {author} {\bibfnamefont {R.}~\bibnamefont
  {Enberg}}, \bibinfo {author} {\bibfnamefont {M.~H.}\ \bibnamefont {Reno}},
  \bibinfo {author} {\bibfnamefont {I.}~\bibnamefont {Sarcevic}},\ and\
  \bibinfo {author} {\bibfnamefont {A.}~\bibnamefont {Stasto}},\ }\bibfield
  {title} {\bibinfo {title} {{Perturbative charm production and the prompt
  atmospheric neutrino flux in light of RHIC and LHC}},\ }\href
  {https://doi.org/10.1007/JHEP06(2015)110} {\bibfield  {journal} {\bibinfo
  {journal} {JHEP}\ }\textbf {\bibinfo {volume} {06}},\ \bibinfo {pages}
  {110}},\ \Eprint {https://arxiv.org/abs/1502.01076} {arXiv:1502.01076
  [hep-ph]} \BibitemShut {NoStop}%
\bibitem [{\citenamefont {Bhattacharya}\ \emph {et~al.}(2016)\citenamefont
  {Bhattacharya}, \citenamefont {Enberg}, \citenamefont {Jeong}, \citenamefont
  {Kim}, \citenamefont {Reno}, \citenamefont {Sarcevic},\ and\ \citenamefont
  {Stasto}}]{Bhattacharya:2016jce}%
  \BibitemOpen
  \bibfield  {author} {\bibinfo {author} {\bibfnamefont {A.}~\bibnamefont
  {Bhattacharya}}, \bibinfo {author} {\bibfnamefont {R.}~\bibnamefont
  {Enberg}}, \bibinfo {author} {\bibfnamefont {Y.~S.}\ \bibnamefont {Jeong}},
  \bibinfo {author} {\bibfnamefont {C.~S.}\ \bibnamefont {Kim}}, \bibinfo
  {author} {\bibfnamefont {M.~H.}\ \bibnamefont {Reno}}, \bibinfo {author}
  {\bibfnamefont {I.}~\bibnamefont {Sarcevic}},\ and\ \bibinfo {author}
  {\bibfnamefont {A.}~\bibnamefont {Stasto}},\ }\bibfield  {title} {\bibinfo
  {title} {{Prompt atmospheric neutrino fluxes: perturbative QCD models and
  nuclear effects}},\ }\href {https://doi.org/10.1007/JHEP11(2016)167}
  {\bibfield  {journal} {\bibinfo  {journal} {JHEP}\ }\textbf {\bibinfo
  {volume} {11}},\ \bibinfo {pages} {167}},\ \Eprint
  {https://arxiv.org/abs/1607.00193} {arXiv:1607.00193 [hep-ph]} \BibitemShut
  {NoStop}%
\bibitem [{\citenamefont {Abbasi}\ \emph {et~al.}(2022)\citenamefont {Abbasi}
  \emph {et~al.}}]{IceCube:2021uhz}%
  \BibitemOpen
  \bibfield  {author} {\bibinfo {author} {\bibfnamefont {R.}~\bibnamefont
  {Abbasi}} \emph {et~al.} (\bibinfo {collaboration} {IceCube}),\ }\bibfield
  {title} {\bibinfo {title} {{Improved Characterization of the Astrophysical
  Muon\textendash{}neutrino Flux with 9.5 Years of IceCube Data}},\ }\href
  {https://doi.org/10.3847/1538-4357/ac4d29} {\bibfield  {journal} {\bibinfo
  {journal} {Astrophys. J.}\ }\textbf {\bibinfo {volume} {928}},\ \bibinfo
  {pages} {50} (\bibinfo {year} {2022})},\ \Eprint
  {https://arxiv.org/abs/2111.10299} {arXiv:2111.10299 [astro-ph.HE]}
  \BibitemShut {NoStop}%
\bibitem [{\citenamefont {Gottlieb}\ and\ \citenamefont
  {Globus}(2021)}]{Gottlieb:2021pzr}%
  \BibitemOpen
  \bibfield  {author} {\bibinfo {author} {\bibfnamefont {O.}~\bibnamefont
  {Gottlieb}}\ and\ \bibinfo {author} {\bibfnamefont {N.}~\bibnamefont
  {Globus}},\ }\bibfield  {title} {\bibinfo {title} {{The Role of
  Jet\textendash{}Cocoon Mixing, Magnetization, and Shock Breakout in Neutrino
  and Cosmic-Ray Emission from Short Gamma-Ray Bursts}},\ }\href
  {https://doi.org/10.3847/2041-8213/ac05c5} {\bibfield  {journal} {\bibinfo
  {journal} {Astrophys. J. Lett.}\ }\textbf {\bibinfo {volume} {915}},\
  \bibinfo {pages} {L4} (\bibinfo {year} {2021})},\ \Eprint
  {https://arxiv.org/abs/2105.01076} {arXiv:2105.01076 [astro-ph.HE]}
  \BibitemShut {NoStop}%
\bibitem [{\citenamefont {Aartsen}\ \emph {et~al.}(2017)\citenamefont {Aartsen}
  \emph {et~al.}}]{IceCube:2016zyt}%
  \BibitemOpen
  \bibfield  {author} {\bibinfo {author} {\bibfnamefont {M.~G.}\ \bibnamefont
  {Aartsen}} \emph {et~al.} (\bibinfo {collaboration} {IceCube}),\ }\bibfield
  {title} {\bibinfo {title} {{The IceCube Neutrino Observatory: Instrumentation
  and Online Systems}},\ }\href
  {https://doi.org/10.1088/1748-0221/12/03/P03012} {\bibfield  {journal}
  {\bibinfo  {journal} {JINST}\ }\textbf {\bibinfo {volume} {12}}\bibfield
  {number} {\bibinfo  {number} { (03)},\ \bibinfo {pages} {P03012}},\ }\Eprint
  {https://arxiv.org/abs/1612.05093} {arXiv:1612.05093 [astro-ph.IM]}
  \BibitemShut {NoStop}%
\bibitem [{\citenamefont {Davis}\ \emph {et~al.}(2021)\citenamefont {Davis}
  \emph {et~al.}}]{LIGO:2021ppb}%
  \BibitemOpen
  \bibfield  {author} {\bibinfo {author} {\bibfnamefont {D.}~\bibnamefont
  {Davis}} \emph {et~al.} (\bibinfo {collaboration} {LIGO}),\ }\bibfield
  {title} {\bibinfo {title} {{LIGO detector characterization in the second and
  third observing runs}},\ }\href {https://doi.org/10.1088/1361-6382/abfd85}
  {\bibfield  {journal} {\bibinfo  {journal} {Class. Quant. Grav.}\ }\textbf
  {\bibinfo {volume} {38}},\ \bibinfo {pages} {135014} (\bibinfo {year}
  {2021})},\ \Eprint {https://arxiv.org/abs/2101.11673} {arXiv:2101.11673
  [astro-ph.IM]} \BibitemShut {NoStop}%
\bibitem [{\citenamefont {Schutz}(2011)}]{Schutz:2011tw}%
  \BibitemOpen
  \bibfield  {author} {\bibinfo {author} {\bibfnamefont {B.~F.}\ \bibnamefont
  {Schutz}},\ }\bibfield  {title} {\bibinfo {title} {{Networks of gravitational
  wave detectors and three figures of merit}},\ }\href
  {https://doi.org/10.1088/0264-9381/28/12/125023} {\bibfield  {journal}
  {\bibinfo  {journal} {Class. Quant. Grav.}\ }\textbf {\bibinfo {volume}
  {28}},\ \bibinfo {pages} {125023} (\bibinfo {year} {2011})},\ \Eprint
  {https://arxiv.org/abs/1102.5421} {arXiv:1102.5421 [astro-ph.IM]}
  \BibitemShut {NoStop}%
\bibitem [{\citenamefont {Baiotti}\ and\ \citenamefont
  {Rezzolla}(2017)}]{Baiotti:2016qnr}%
  \BibitemOpen
  \bibfield  {author} {\bibinfo {author} {\bibfnamefont {L.}~\bibnamefont
  {Baiotti}}\ and\ \bibinfo {author} {\bibfnamefont {L.}~\bibnamefont
  {Rezzolla}},\ }\bibfield  {title} {\bibinfo {title} {{Binary neutron star
  mergers: a review of Einstein\textquoteright{}s richest laboratory}},\ }\href
  {https://doi.org/10.1088/1361-6633/aa67bb} {\bibfield  {journal} {\bibinfo
  {journal} {Rept. Prog. Phys.}\ }\textbf {\bibinfo {volume} {80}},\ \bibinfo
  {pages} {096901} (\bibinfo {year} {2017})},\ \Eprint
  {https://arxiv.org/abs/1607.03540} {arXiv:1607.03540 [gr-qc]} \BibitemShut
  {NoStop}%
\bibitem [{\citenamefont {Pian}(2021)}]{10.3389/fspas.2020.609460}%
  \BibitemOpen
  \bibfield  {author} {\bibinfo {author} {\bibfnamefont {E.}~\bibnamefont
  {Pian}},\ }\bibfield  {title} {\bibinfo {title} {Mergers of binary neutron
  star systems: A multimessenger revolution},\ }\href
  {https://doi.org/10.3389/fspas.2020.609460} {\bibfield  {journal} {\bibinfo
  {journal} {Frontiers in Astronomy and Space Sciences}\ }\textbf {\bibinfo
  {volume} {7}},\ \bibinfo {pages} {108} (\bibinfo {year} {2021})}\BibitemShut
  {NoStop}%
\bibitem [{\citenamefont {Blum}\ \emph {et~al.}(2022)\citenamefont {Blum} \emph
  {et~al.}}]{Blum:2022dxi}%
  \BibitemOpen
  \bibfield  {author} {\bibinfo {author} {\bibfnamefont {B.}~\bibnamefont
  {Blum}} \emph {et~al.},\ }\bibfield  {title} {\bibinfo {title} {{Snowmass2021
  Cosmic Frontier White Paper: Rubin Observatory after LSST}},\ }in\ \href@noop
  {} {\emph {\bibinfo {booktitle} {{Snowmass 2021}}}}\ (\bibinfo {year}
  {2022})\ \Eprint {https://arxiv.org/abs/2203.07220} {arXiv:2203.07220
  [astro-ph.CO]} \BibitemShut {NoStop}%
\bibitem [{\citenamefont {{Meegan}}\ \emph {et~al.}(2009)\citenamefont
  {{Meegan}}, \citenamefont {{Lichti}}, \citenamefont {{Bhat}}, \citenamefont
  {{Bissaldi}}, \citenamefont {{Briggs}}, \citenamefont {{Connaughton}},
  \citenamefont {{Diehl}}, \citenamefont {{Fishman}}, \citenamefont
  {{Greiner}}, \citenamefont {{Hoover}}, \citenamefont {{van der Horst}},
  \citenamefont {{von Kienlin}}, \citenamefont {{Kippen}}, \citenamefont
  {{Kouveliotou}}, \citenamefont {{McBreen}}, \citenamefont {{Paciesas}},
  \citenamefont {{Preece}}, \citenamefont {{Steinle}}, \citenamefont
  {{Wallace}}, \citenamefont {{Wilson}},\ and\ \citenamefont
  {{Wilson-Hodge}}}]{FermiGBM}%
  \BibitemOpen
  \bibfield  {author} {\bibinfo {author} {\bibfnamefont {C.}~\bibnamefont
  {{Meegan}}}, \bibinfo {author} {\bibfnamefont {G.}~\bibnamefont {{Lichti}}},
  \bibinfo {author} {\bibfnamefont {P.~N.}\ \bibnamefont {{Bhat}}}, \bibinfo
  {author} {\bibfnamefont {E.}~\bibnamefont {{Bissaldi}}}, \bibinfo {author}
  {\bibfnamefont {M.~S.}\ \bibnamefont {{Briggs}}}, \bibinfo {author}
  {\bibfnamefont {V.}~\bibnamefont {{Connaughton}}}, \bibinfo {author}
  {\bibfnamefont {R.}~\bibnamefont {{Diehl}}}, \bibinfo {author} {\bibfnamefont
  {G.}~\bibnamefont {{Fishman}}}, \bibinfo {author} {\bibfnamefont
  {J.}~\bibnamefont {{Greiner}}}, \bibinfo {author} {\bibfnamefont {A.~S.}\
  \bibnamefont {{Hoover}}}, \bibinfo {author} {\bibfnamefont {A.~J.}\
  \bibnamefont {{van der Horst}}}, \bibinfo {author} {\bibfnamefont
  {A.}~\bibnamefont {{von Kienlin}}}, \bibinfo {author} {\bibfnamefont {R.~M.}\
  \bibnamefont {{Kippen}}}, \bibinfo {author} {\bibfnamefont {C.}~\bibnamefont
  {{Kouveliotou}}}, \bibinfo {author} {\bibfnamefont {S.}~\bibnamefont
  {{McBreen}}}, \bibinfo {author} {\bibfnamefont {W.~S.}\ \bibnamefont
  {{Paciesas}}}, \bibinfo {author} {\bibfnamefont {R.}~\bibnamefont
  {{Preece}}}, \bibinfo {author} {\bibfnamefont {H.}~\bibnamefont {{Steinle}}},
  \bibinfo {author} {\bibfnamefont {M.~S.}\ \bibnamefont {{Wallace}}}, \bibinfo
  {author} {\bibfnamefont {R.~B.}\ \bibnamefont {{Wilson}}},\ and\ \bibinfo
  {author} {\bibfnamefont {C.}~\bibnamefont {{Wilson-Hodge}}},\ }\bibfield
  {title} {\bibinfo {title} {{The Fermi Gamma-ray Burst Monitor}},\ }\href
  {https://doi.org/10.1088/0004-637X/702/1/791} {\bibfield  {journal} {\bibinfo
   {journal} {\apj}\ }\textbf {\bibinfo {volume} {702}},\ \bibinfo {pages}
  {791} (\bibinfo {year} {2009})},\ \Eprint {https://arxiv.org/abs/0908.0450}
  {arXiv:0908.0450 [astro-ph.IM]} \BibitemShut {NoStop}%
\bibitem [{\citenamefont {Goldstein}\ \emph {et~al.}(2020)\citenamefont
  {Goldstein} \emph {et~al.}}]{Goldstein:2019pcj}%
  \BibitemOpen
  \bibfield  {author} {\bibinfo {author} {\bibfnamefont {A.}~\bibnamefont
  {Goldstein}} \emph {et~al.},\ }\bibfield  {title} {\bibinfo {title}
  {{Evaluation of Automated Fermi GBM Localizations of Gamma-Ray Bursts}},\
  }\href {https://doi.org/10.3847/1538-4357/ab8bdb} {\bibfield  {journal}
  {\bibinfo  {journal} {Astrophys. J.}\ }\textbf {\bibinfo {volume} {895}},\
  \bibinfo {pages} {40} (\bibinfo {year} {2020})},\ \Eprint
  {https://arxiv.org/abs/1909.03006} {arXiv:1909.03006 [astro-ph.IM]}
  \BibitemShut {NoStop}%
\bibitem [{\citenamefont {Greiner}\ \emph {et~al.}(2022)\citenamefont
  {Greiner}, \citenamefont {Hugentobler}, \citenamefont {Burgess},
  \citenamefont {Berlato}, \citenamefont {Rott},\ and\ \citenamefont
  {Tsvetkova}}]{grb_satellite}%
  \BibitemOpen
  \bibfield  {author} {\bibinfo {author} {\bibfnamefont {J.}~\bibnamefont
  {Greiner}}, \bibinfo {author} {\bibfnamefont {U.}~\bibnamefont
  {Hugentobler}}, \bibinfo {author} {\bibfnamefont {J.~M.}\ \bibnamefont
  {Burgess}}, \bibinfo {author} {\bibfnamefont {F.}~\bibnamefont {Berlato}},
  \bibinfo {author} {\bibfnamefont {M.}~\bibnamefont {Rott}},\ and\ \bibinfo
  {author} {\bibfnamefont {A.}~\bibnamefont {Tsvetkova}},\ }\bibfield  {title}
  {\bibinfo {title} {{A proposed network of gamma-ray burst detectors on the
  global navigation satellite system Galileo G2}},\ }\href
  {https://doi.org/10.1051/0004-6361/202142835} {\bibfield  {journal} {\bibinfo
   {journal} {Astron. Astrophys.}\ }\textbf {\bibinfo {volume} {664}},\
  \bibinfo {pages} {A131} (\bibinfo {year} {2022})},\ \Eprint
  {https://arxiv.org/abs/2205.08637} {arXiv:2205.08637 [astro-ph.IM]}
  \BibitemShut {NoStop}%
\bibitem [{\citenamefont {Wei}\ and\ \citenamefont
  {Cordier}(2016)}]{Wei:2016eox}%
  \BibitemOpen
  \bibfield  {author} {\bibinfo {author} {\bibfnamefont {J.}~\bibnamefont
  {Wei}}\ and\ \bibinfo {author} {\bibfnamefont {B.}~\bibnamefont {Cordier}},\
  }\bibfield  {title} {\bibinfo {title} {{The Deep and Transient Universe in
  the SVOM Era: New Challenges and Opportunities - Scientific prospects of the
  SVOM mission}}\ }(\bibinfo {year} {2016})\ \Eprint
  {https://arxiv.org/abs/1610.06892} {arXiv:1610.06892 [astro-ph.IM]}
  \BibitemShut {NoStop}%
\bibitem [{\citenamefont {Barthelmy}\ \emph {et~al.}(2005)\citenamefont
  {Barthelmy} \emph {et~al.}}]{Barthelmy:2005hs}%
  \BibitemOpen
  \bibfield  {author} {\bibinfo {author} {\bibfnamefont {S.~D.}\ \bibnamefont
  {Barthelmy}} \emph {et~al.},\ }\bibfield  {title} {\bibinfo {title} {{The
  Burst Alert Telescope (BAT) on the Swift MIDEX mission}},\ }\href
  {https://doi.org/10.1007/s11214-005-5096-3} {\bibfield  {journal} {\bibinfo
  {journal} {Space Sci. Rev.}\ }\textbf {\bibinfo {volume} {120}},\ \bibinfo
  {pages} {143} (\bibinfo {year} {2005})},\ \Eprint
  {https://arxiv.org/abs/astro-ph/0507410} {arXiv:astro-ph/0507410}
  \BibitemShut {NoStop}%
\bibitem [{\citenamefont {Yuan}\ \emph {et~al.}(2015)\citenamefont {Yuan} \emph
  {et~al.}}]{EinsteinProbeTeam:2015bcj}%
  \BibitemOpen
  \bibfield  {author} {\bibinfo {author} {\bibfnamefont {W.}~\bibnamefont
  {Yuan}} \emph {et~al.} (\bibinfo {collaboration} {Einstein Probe Team}),\
  }\bibfield  {title} {\bibinfo {title} {{Einstein Probe - a small mission to
  monitor and explore the dynamic X-ray Universe}},\ }\href
  {https://doi.org/10.22323/1.233.0006} {\bibfield  {journal} {\bibinfo
  {journal} {PoS}\ }\textbf {\bibinfo {volume} {SWIFT10}},\ \bibinfo {pages}
  {006} (\bibinfo {year} {2015})},\ \Eprint {https://arxiv.org/abs/1506.07735}
  {arXiv:1506.07735 [astro-ph.HE]} \BibitemShut {NoStop}%
\bibitem [{\citenamefont {{Yonetoku}}\ \emph {et~al.}(2020)\citenamefont
  {{Yonetoku}}, \citenamefont {{Mihara}}, \citenamefont {{Doi}}, \citenamefont
  {{Sakamoto}}, \citenamefont {{Tsumura}}, \citenamefont {{Ioka}},
  \citenamefont {{Amaya}}, \citenamefont {{Arimoto}}, \citenamefont {{Enoto}},
  \citenamefont {{Fujii}}, \citenamefont {{Goto}}, \citenamefont {{Gunji}},
  \citenamefont {{Hiraga}}, \citenamefont {{Ikeda}}, \citenamefont {{Kawai}},
  \citenamefont {{Kurosawa}}, \citenamefont {{Li}}, \citenamefont {{Maeda}},
  \citenamefont {{Mitsuishi}}, \citenamefont {{Murakami}}, \citenamefont
  {{Nakagawa}}, \citenamefont {{Ogino}}, \citenamefont {{Ohno}}, \citenamefont
  {{Sawano}}, \citenamefont {{Sei}}, \citenamefont {{Serino}}, \citenamefont
  {{Sugita}}, \citenamefont {{Tamagawa}}, \citenamefont {{Tamura}},
  \citenamefont {{Tanaka}}, \citenamefont {{Tanimori}}, \citenamefont
  {{Tashiro}}, \citenamefont {{Tomida}}, \citenamefont {{Wang}}, \citenamefont
  {{Yamaguchi}}, \citenamefont {{Yamamoto}}, \citenamefont {{Yamaoka}},
  \citenamefont {{Yamauchi}}, \citenamefont {{Yatsu}}, \citenamefont
  {{Yoshida}}, \citenamefont {{Yuhi}}, \citenamefont {{Akitaya}}, \citenamefont
  {{Fukui}}, \citenamefont {{Ita}}, \citenamefont {{Kaneda}}, \citenamefont
  {{Kawabata}}, \citenamefont {{Kawata}}, \citenamefont {{Kurimata}},
  \citenamefont {{Matsumoto}}, \citenamefont {{Matsuura}}, \citenamefont
  {{Miyasaka}}, \citenamefont {{Motohara}}, \citenamefont {{Narita}},
  \citenamefont {{Noda}}, \citenamefont {{Ohashi}}, \citenamefont {{Okita}},
  \citenamefont {{Sano}}, \citenamefont {{Tanaka}}, \citenamefont {{Urata}},
  \citenamefont {{Wada}}, \citenamefont {{Yamaguchi}}, \citenamefont
  {{Yanagisawa}}, \citenamefont {{Yoshida}}, \citenamefont {{Asano}},
  \citenamefont {{Inayoshi}}, \citenamefont {{Inoue}}, \citenamefont {{Ito}},
  \citenamefont {{Izumiura}}, \citenamefont {{Kawanaka}}, \citenamefont
  {{Kinugawa}}, \citenamefont {{Kisaka}}, \citenamefont {{Kiuchi}},
  \citenamefont {{Matsumoto}}, \citenamefont {{Mizuta}}, \citenamefont
  {{Murase}}, \citenamefont {{Nagakura}}, \citenamefont {{Nagataki}},
  \citenamefont {{Nakada}}, \citenamefont {{Nakamura}}, \citenamefont
  {{Niino}}, \citenamefont {{Suwa}}, \citenamefont {{Takahashi}}, \citenamefont
  {{Tanaka}}, \citenamefont {{Toma}}, \citenamefont {{Totani}}, \citenamefont
  {{Yamazaki}},\ and\ \citenamefont {{Yokoyama}}}]{2020SPIE11444E..2ZY}%
  \BibitemOpen
  \bibfield  {author} {\bibinfo {author} {\bibfnamefont {D.}~\bibnamefont
  {{Yonetoku}}}, \bibinfo {author} {\bibfnamefont {T.}~\bibnamefont
  {{Mihara}}}, \bibinfo {author} {\bibfnamefont {A.}~\bibnamefont {{Doi}}},
  \bibinfo {author} {\bibfnamefont {T.}~\bibnamefont {{Sakamoto}}}, \bibinfo
  {author} {\bibfnamefont {K.}~\bibnamefont {{Tsumura}}}, \bibinfo {author}
  {\bibfnamefont {K.}~\bibnamefont {{Ioka}}}, \bibinfo {author} {\bibfnamefont
  {Y.}~\bibnamefont {{Amaya}}}, \bibinfo {author} {\bibfnamefont
  {M.}~\bibnamefont {{Arimoto}}}, \bibinfo {author} {\bibfnamefont
  {T.}~\bibnamefont {{Enoto}}}, \bibinfo {author} {\bibfnamefont
  {T.}~\bibnamefont {{Fujii}}}, \bibinfo {author} {\bibfnamefont
  {H.}~\bibnamefont {{Goto}}}, \bibinfo {author} {\bibfnamefont
  {S.}~\bibnamefont {{Gunji}}}, \bibinfo {author} {\bibfnamefont
  {J.}~\bibnamefont {{Hiraga}}}, \bibinfo {author} {\bibfnamefont
  {H.}~\bibnamefont {{Ikeda}}}, \bibinfo {author} {\bibfnamefont
  {N.}~\bibnamefont {{Kawai}}}, \bibinfo {author} {\bibfnamefont
  {S.}~\bibnamefont {{Kurosawa}}}, \bibinfo {author} {\bibfnamefont
  {J.}~\bibnamefont {{Li}}}, \bibinfo {author} {\bibfnamefont {Y.}~\bibnamefont
  {{Maeda}}}, \bibinfo {author} {\bibfnamefont {I.}~\bibnamefont
  {{Mitsuishi}}}, \bibinfo {author} {\bibfnamefont {T.}~\bibnamefont
  {{Murakami}}}, \bibinfo {author} {\bibfnamefont {Y.}~\bibnamefont
  {{Nakagawa}}}, \bibinfo {author} {\bibfnamefont {N.}~\bibnamefont {{Ogino}}},
  \bibinfo {author} {\bibfnamefont {M.}~\bibnamefont {{Ohno}}}, \bibinfo
  {author} {\bibfnamefont {T.}~\bibnamefont {{Sawano}}}, \bibinfo {author}
  {\bibfnamefont {K.}~\bibnamefont {{Sei}}}, \bibinfo {author} {\bibfnamefont
  {M.}~\bibnamefont {{Serino}}}, \bibinfo {author} {\bibfnamefont
  {S.}~\bibnamefont {{Sugita}}}, \bibinfo {author} {\bibfnamefont
  {T.}~\bibnamefont {{Tamagawa}}}, \bibinfo {author} {\bibfnamefont
  {K.}~\bibnamefont {{Tamura}}}, \bibinfo {author} {\bibfnamefont
  {T.}~\bibnamefont {{Tanaka}}}, \bibinfo {author} {\bibfnamefont
  {T.}~\bibnamefont {{Tanimori}}}, \bibinfo {author} {\bibfnamefont {M.~S.}\
  \bibnamefont {{Tashiro}}}, \bibinfo {author} {\bibfnamefont {H.}~\bibnamefont
  {{Tomida}}}, \bibinfo {author} {\bibfnamefont {H.}~\bibnamefont {{Wang}}},
  \bibinfo {author} {\bibfnamefont {T.}~\bibnamefont {{Yamaguchi}}}, \bibinfo
  {author} {\bibfnamefont {A.}~\bibnamefont {{Yamamoto}}}, \bibinfo {author}
  {\bibfnamefont {K.}~\bibnamefont {{Yamaoka}}}, \bibinfo {author}
  {\bibfnamefont {M.}~\bibnamefont {{Yamauchi}}}, \bibinfo {author}
  {\bibfnamefont {Y.}~\bibnamefont {{Yatsu}}}, \bibinfo {author} {\bibfnamefont
  {A.}~\bibnamefont {{Yoshida}}}, \bibinfo {author} {\bibfnamefont
  {D.}~\bibnamefont {{Yuhi}}}, \bibinfo {author} {\bibfnamefont
  {H.}~\bibnamefont {{Akitaya}}}, \bibinfo {author} {\bibfnamefont
  {A.}~\bibnamefont {{Fukui}}}, \bibinfo {author} {\bibfnamefont
  {Y.}~\bibnamefont {{Ita}}}, \bibinfo {author} {\bibfnamefont
  {H.}~\bibnamefont {{Kaneda}}}, \bibinfo {author} {\bibfnamefont
  {K.}~\bibnamefont {{Kawabata}}}, \bibinfo {author} {\bibfnamefont
  {Y.}~\bibnamefont {{Kawata}}}, \bibinfo {author} {\bibfnamefont
  {M.}~\bibnamefont {{Kurimata}}}, \bibinfo {author} {\bibfnamefont
  {T.}~\bibnamefont {{Matsumoto}}}, \bibinfo {author} {\bibfnamefont
  {S.}~\bibnamefont {{Matsuura}}}, \bibinfo {author} {\bibfnamefont
  {A.}~\bibnamefont {{Miyasaka}}}, \bibinfo {author} {\bibfnamefont
  {K.}~\bibnamefont {{Motohara}}}, \bibinfo {author} {\bibfnamefont
  {N.}~\bibnamefont {{Narita}}}, \bibinfo {author} {\bibfnamefont
  {H.}~\bibnamefont {{Noda}}}, \bibinfo {author} {\bibfnamefont
  {A.}~\bibnamefont {{Ohashi}}}, \bibinfo {author} {\bibfnamefont
  {H.}~\bibnamefont {{Okita}}}, \bibinfo {author} {\bibfnamefont
  {K.}~\bibnamefont {{Sano}}}, \bibinfo {author} {\bibfnamefont
  {M.}~\bibnamefont {{Tanaka}}}, \bibinfo {author} {\bibfnamefont
  {Y.}~\bibnamefont {{Urata}}}, \bibinfo {author} {\bibfnamefont
  {T.}~\bibnamefont {{Wada}}}, \bibinfo {author} {\bibfnamefont
  {H.}~\bibnamefont {{Yamaguchi}}}, \bibinfo {author} {\bibfnamefont
  {K.}~\bibnamefont {{Yanagisawa}}}, \bibinfo {author} {\bibfnamefont
  {M.}~\bibnamefont {{Yoshida}}}, \bibinfo {author} {\bibfnamefont
  {K.}~\bibnamefont {{Asano}}}, \bibinfo {author} {\bibfnamefont
  {K.}~\bibnamefont {{Inayoshi}}}, \bibinfo {author} {\bibfnamefont
  {S.}~\bibnamefont {{Inoue}}}, \bibinfo {author} {\bibfnamefont
  {H.}~\bibnamefont {{Ito}}}, \bibinfo {author} {\bibfnamefont
  {H.}~\bibnamefont {{Izumiura}}}, \bibinfo {author} {\bibfnamefont
  {N.}~\bibnamefont {{Kawanaka}}}, \bibinfo {author} {\bibfnamefont
  {T.}~\bibnamefont {{Kinugawa}}}, \bibinfo {author} {\bibfnamefont
  {S.}~\bibnamefont {{Kisaka}}}, \bibinfo {author} {\bibfnamefont
  {K.}~\bibnamefont {{Kiuchi}}}, \bibinfo {author} {\bibfnamefont
  {J.}~\bibnamefont {{Matsumoto}}}, \bibinfo {author} {\bibfnamefont
  {A.}~\bibnamefont {{Mizuta}}}, \bibinfo {author} {\bibfnamefont
  {K.}~\bibnamefont {{Murase}}}, \bibinfo {author} {\bibfnamefont
  {H.}~\bibnamefont {{Nagakura}}}, \bibinfo {author} {\bibfnamefont
  {S.}~\bibnamefont {{Nagataki}}}, \bibinfo {author} {\bibfnamefont
  {Y.}~\bibnamefont {{Nakada}}}, \bibinfo {author} {\bibfnamefont
  {T.}~\bibnamefont {{Nakamura}}}, \bibinfo {author} {\bibfnamefont
  {Y.}~\bibnamefont {{Niino}}}, \bibinfo {author} {\bibfnamefont
  {Y.}~\bibnamefont {{Suwa}}}, \bibinfo {author} {\bibfnamefont
  {K.}~\bibnamefont {{Takahashi}}}, \bibinfo {author} {\bibfnamefont
  {T.}~\bibnamefont {{Tanaka}}}, \bibinfo {author} {\bibfnamefont
  {K.}~\bibnamefont {{Toma}}}, \bibinfo {author} {\bibfnamefont
  {T.}~\bibnamefont {{Totani}}}, \bibinfo {author} {\bibfnamefont
  {R.}~\bibnamefont {{Yamazaki}}},\ and\ \bibinfo {author} {\bibfnamefont
  {J.}~\bibnamefont {{Yokoyama}}},\ }\bibfield  {title} {\bibinfo {title}
  {{High-redshift gamma-ray burst for unraveling the Dark Ages Mission:
  HiZ-GUNDAM}},\ }in\ \href {https://doi.org/10.1117/12.2560603} {\emph
  {\bibinfo {booktitle} {Space Telescopes and Instrumentation 2020: Ultraviolet
  to Gamma Ray}}},\ \bibinfo {series} {Society of Photo-Optical Instrumentation
  Engineers (SPIE) Conference Series}, Vol.\ \bibinfo {volume} {11444},\
  \bibinfo {editor} {edited by\ \bibinfo {editor} {\bibfnamefont {J.-W.~A.}\
  \bibnamefont {{den Herder}}}, \bibinfo {editor} {\bibfnamefont
  {S.}~\bibnamefont {{Nikzad}}},\ and\ \bibinfo {editor} {\bibfnamefont
  {K.}~\bibnamefont {{Nakazawa}}}}\ (\bibinfo {year} {2020})\ p.\ \bibinfo
  {pages} {114442Z}\BibitemShut {NoStop}%
\bibitem [{\citenamefont {Scolnic}\ \emph {et~al.}(2018)\citenamefont {Scolnic}
  \emph {et~al.}}]{DES:2017dgt}%
  \BibitemOpen
  \bibfield  {author} {\bibinfo {author} {\bibfnamefont {D.}~\bibnamefont
  {Scolnic}} \emph {et~al.} (\bibinfo {collaboration} {DES}),\ }\bibfield
  {title} {\bibinfo {title} {{How Many Kilonovae Can Be Found in Past, Present,
  and Future Survey Data Sets?}},\ }\href
  {https://doi.org/10.3847/2041-8213/aa9d82} {\bibfield  {journal} {\bibinfo
  {journal} {Astrophys. J. Lett.}\ }\textbf {\bibinfo {volume} {852}},\
  \bibinfo {pages} {L3} (\bibinfo {year} {2018})},\ \Eprint
  {https://arxiv.org/abs/1710.05845} {arXiv:1710.05845 [astro-ph.IM]}
  \BibitemShut {NoStop}%
\bibitem [{\citenamefont {Ivezi\'c}\ \emph {et~al.}(2019)\citenamefont
  {Ivezi\'c} \emph {et~al.}}]{LSST:2008ijt}%
  \BibitemOpen
  \bibfield  {author} {\bibinfo {author} {\bibfnamefont {v.}~\bibnamefont
  {Ivezi\'c}} \emph {et~al.} (\bibinfo {collaboration} {LSST}),\ }\bibfield
  {title} {\bibinfo {title} {{LSST: from Science Drivers to Reference Design
  and Anticipated Data Products}},\ }\href
  {https://doi.org/10.3847/1538-4357/ab042c} {\bibfield  {journal} {\bibinfo
  {journal} {Astrophys. J.}\ }\textbf {\bibinfo {volume} {873}},\ \bibinfo
  {pages} {111} (\bibinfo {year} {2019})},\ \Eprint
  {https://arxiv.org/abs/0805.2366} {arXiv:0805.2366 [astro-ph]} \BibitemShut
  {NoStop}%
\bibitem [{\citenamefont {{Bellm}}\ \emph {et~al.}(2019)\citenamefont
  {{Bellm}}, \citenamefont {{Kulkarni}}, \citenamefont {{Graham}},
  \citenamefont {{Dekany}}, \citenamefont {{Smith}}, \citenamefont {{Riddle}},
  \citenamefont {{Masci}}, \citenamefont {{Helou}}, \citenamefont {{Prince}},
  \citenamefont {{Adams}}, \citenamefont {{Barbarino}}, \citenamefont
  {{Barlow}}, \citenamefont {{Bauer}}, \citenamefont {{Beck}}, \citenamefont
  {{Belicki}}, \citenamefont {{Biswas}}, \citenamefont {{Blagorodnova}},
  \citenamefont {{Bodewits}}, \citenamefont {{Bolin}}, \citenamefont
  {{Brinnel}}, \citenamefont {{Brooke}}, \citenamefont {{Bue}}, \citenamefont
  {{Bulla}}, \citenamefont {{Burruss}}, \citenamefont {{Cenko}}, \citenamefont
  {{Chang}}, \citenamefont {{Connolly}}, \citenamefont {{Coughlin}},
  \citenamefont {{Cromer}}, \citenamefont {{Cunningham}}, \citenamefont {{De}},
  \citenamefont {{Delacroix}}, \citenamefont {{Desai}}, \citenamefont {{Duev}},
  \citenamefont {{Eadie}}, \citenamefont {{Farnham}}, \citenamefont {{Feeney}},
  \citenamefont {{Feindt}}, \citenamefont {{Flynn}}, \citenamefont
  {{Franckowiak}}, \citenamefont {{Frederick}}, \citenamefont {{Fremling}},
  \citenamefont {{Gal-Yam}}, \citenamefont {{Gezari}}, \citenamefont {{Giomi}},
  \citenamefont {{Goldstein}}, \citenamefont {{Golkhou}}, \citenamefont
  {{Goobar}}, \citenamefont {{Groom}}, \citenamefont {{Hacopians}},
  \citenamefont {{Hale}}, \citenamefont {{Henning}}, \citenamefont {{Ho}},
  \citenamefont {{Hover}}, \citenamefont {{Howell}}, \citenamefont {{Hung}},
  \citenamefont {{Huppenkothen}}, \citenamefont {{Imel}}, \citenamefont {{Ip}},
  \citenamefont {{Ivezi{\'c}}}, \citenamefont {{Jackson}}, \citenamefont
  {{Jones}}, \citenamefont {{Juric}}, \citenamefont {{Kasliwal}}, \citenamefont
  {{Kaspi}}, \citenamefont {{Kaye}}, \citenamefont {{Kelley}}, \citenamefont
  {{Kowalski}}, \citenamefont {{Kramer}}, \citenamefont {{Kupfer}},
  \citenamefont {{Landry}}, \citenamefont {{Laher}}, \citenamefont {{Lee}},
  \citenamefont {{Lin}}, \citenamefont {{Lin}}, \citenamefont {{Lunnan}},
  \citenamefont {{Giomi}}, \citenamefont {{Mahabal}}, \citenamefont {{Mao}},
  \citenamefont {{Miller}}, \citenamefont {{Monkewitz}}, \citenamefont
  {{Murphy}}, \citenamefont {{Ngeow}}, \citenamefont {{Nordin}}, \citenamefont
  {{Nugent}}, \citenamefont {{Ofek}}, \citenamefont {{Patterson}},
  \citenamefont {{Penprase}}, \citenamefont {{Porter}}, \citenamefont
  {{Rauch}}, \citenamefont {{Rebbapragada}}, \citenamefont {{Reiley}},
  \citenamefont {{Rigault}}, \citenamefont {{Rodriguez}}, \citenamefont {{van
  Roestel}}, \citenamefont {{Rusholme}}, \citenamefont {{van Santen}},
  \citenamefont {{Schulze}}, \citenamefont {{Shupe}}, \citenamefont {{Singer}},
  \citenamefont {{Soumagnac}}, \citenamefont {{Stein}}, \citenamefont
  {{Surace}}, \citenamefont {{Sollerman}}, \citenamefont {{Szkody}},
  \citenamefont {{Taddia}}, \citenamefont {{Terek}}, \citenamefont {{Van
  Sistine}}, \citenamefont {{van Velzen}}, \citenamefont {{Vestrand}},
  \citenamefont {{Walters}}, \citenamefont {{Ward}}, \citenamefont {{Ye}},
  \citenamefont {{Yu}}, \citenamefont {{Yan}},\ and\ \citenamefont
  {{Zolkower}}}]{2019PASP..131a8002B}%
  \BibitemOpen
  \bibfield  {author} {\bibinfo {author} {\bibfnamefont {E.~C.}\ \bibnamefont
  {{Bellm}}}, \bibinfo {author} {\bibfnamefont {S.~R.}\ \bibnamefont
  {{Kulkarni}}}, \bibinfo {author} {\bibfnamefont {M.~J.}\ \bibnamefont
  {{Graham}}}, \bibinfo {author} {\bibfnamefont {R.}~\bibnamefont {{Dekany}}},
  \bibinfo {author} {\bibfnamefont {R.~M.}\ \bibnamefont {{Smith}}}, \bibinfo
  {author} {\bibfnamefont {R.}~\bibnamefont {{Riddle}}}, \bibinfo {author}
  {\bibfnamefont {F.~J.}\ \bibnamefont {{Masci}}}, \bibinfo {author}
  {\bibfnamefont {G.}~\bibnamefont {{Helou}}}, \bibinfo {author} {\bibfnamefont
  {T.~A.}\ \bibnamefont {{Prince}}}, \bibinfo {author} {\bibfnamefont {S.~M.}\
  \bibnamefont {{Adams}}}, \bibinfo {author} {\bibfnamefont {C.}~\bibnamefont
  {{Barbarino}}}, \bibinfo {author} {\bibfnamefont {T.}~\bibnamefont
  {{Barlow}}}, \bibinfo {author} {\bibfnamefont {J.}~\bibnamefont {{Bauer}}},
  \bibinfo {author} {\bibfnamefont {R.}~\bibnamefont {{Beck}}}, \bibinfo
  {author} {\bibfnamefont {J.}~\bibnamefont {{Belicki}}}, \bibinfo {author}
  {\bibfnamefont {R.}~\bibnamefont {{Biswas}}}, \bibinfo {author}
  {\bibfnamefont {N.}~\bibnamefont {{Blagorodnova}}}, \bibinfo {author}
  {\bibfnamefont {D.}~\bibnamefont {{Bodewits}}}, \bibinfo {author}
  {\bibfnamefont {B.}~\bibnamefont {{Bolin}}}, \bibinfo {author} {\bibfnamefont
  {V.}~\bibnamefont {{Brinnel}}}, \bibinfo {author} {\bibfnamefont
  {T.}~\bibnamefont {{Brooke}}}, \bibinfo {author} {\bibfnamefont
  {B.}~\bibnamefont {{Bue}}}, \bibinfo {author} {\bibfnamefont
  {M.}~\bibnamefont {{Bulla}}}, \bibinfo {author} {\bibfnamefont
  {R.}~\bibnamefont {{Burruss}}}, \bibinfo {author} {\bibfnamefont {S.~B.}\
  \bibnamefont {{Cenko}}}, \bibinfo {author} {\bibfnamefont {C.-K.}\
  \bibnamefont {{Chang}}}, \bibinfo {author} {\bibfnamefont {A.}~\bibnamefont
  {{Connolly}}}, \bibinfo {author} {\bibfnamefont {M.}~\bibnamefont
  {{Coughlin}}}, \bibinfo {author} {\bibfnamefont {J.}~\bibnamefont
  {{Cromer}}}, \bibinfo {author} {\bibfnamefont {V.}~\bibnamefont
  {{Cunningham}}}, \bibinfo {author} {\bibfnamefont {K.}~\bibnamefont {{De}}},
  \bibinfo {author} {\bibfnamefont {A.}~\bibnamefont {{Delacroix}}}, \bibinfo
  {author} {\bibfnamefont {V.}~\bibnamefont {{Desai}}}, \bibinfo {author}
  {\bibfnamefont {D.~A.}\ \bibnamefont {{Duev}}}, \bibinfo {author}
  {\bibfnamefont {G.}~\bibnamefont {{Eadie}}}, \bibinfo {author} {\bibfnamefont
  {T.~L.}\ \bibnamefont {{Farnham}}}, \bibinfo {author} {\bibfnamefont
  {M.}~\bibnamefont {{Feeney}}}, \bibinfo {author} {\bibfnamefont
  {U.}~\bibnamefont {{Feindt}}}, \bibinfo {author} {\bibfnamefont
  {D.}~\bibnamefont {{Flynn}}}, \bibinfo {author} {\bibfnamefont
  {A.}~\bibnamefont {{Franckowiak}}}, \bibinfo {author} {\bibfnamefont
  {S.}~\bibnamefont {{Frederick}}}, \bibinfo {author} {\bibfnamefont
  {C.}~\bibnamefont {{Fremling}}}, \bibinfo {author} {\bibfnamefont
  {A.}~\bibnamefont {{Gal-Yam}}}, \bibinfo {author} {\bibfnamefont
  {S.}~\bibnamefont {{Gezari}}}, \bibinfo {author} {\bibfnamefont
  {M.}~\bibnamefont {{Giomi}}}, \bibinfo {author} {\bibfnamefont {D.~A.}\
  \bibnamefont {{Goldstein}}}, \bibinfo {author} {\bibfnamefont {V.~Z.}\
  \bibnamefont {{Golkhou}}}, \bibinfo {author} {\bibfnamefont {A.}~\bibnamefont
  {{Goobar}}}, \bibinfo {author} {\bibfnamefont {S.}~\bibnamefont {{Groom}}},
  \bibinfo {author} {\bibfnamefont {E.}~\bibnamefont {{Hacopians}}}, \bibinfo
  {author} {\bibfnamefont {D.}~\bibnamefont {{Hale}}}, \bibinfo {author}
  {\bibfnamefont {J.}~\bibnamefont {{Henning}}}, \bibinfo {author}
  {\bibfnamefont {A.~Y.~Q.}\ \bibnamefont {{Ho}}}, \bibinfo {author}
  {\bibfnamefont {D.}~\bibnamefont {{Hover}}}, \bibinfo {author} {\bibfnamefont
  {J.}~\bibnamefont {{Howell}}}, \bibinfo {author} {\bibfnamefont
  {T.}~\bibnamefont {{Hung}}}, \bibinfo {author} {\bibfnamefont
  {D.}~\bibnamefont {{Huppenkothen}}}, \bibinfo {author} {\bibfnamefont
  {D.}~\bibnamefont {{Imel}}}, \bibinfo {author} {\bibfnamefont {W.-H.}\
  \bibnamefont {{Ip}}}, \bibinfo {author} {\bibfnamefont
  {{\v{Z}}.}~\bibnamefont {{Ivezi{\'c}}}}, \bibinfo {author} {\bibfnamefont
  {E.}~\bibnamefont {{Jackson}}}, \bibinfo {author} {\bibfnamefont
  {L.}~\bibnamefont {{Jones}}}, \bibinfo {author} {\bibfnamefont
  {M.}~\bibnamefont {{Juric}}}, \bibinfo {author} {\bibfnamefont {M.~M.}\
  \bibnamefont {{Kasliwal}}}, \bibinfo {author} {\bibfnamefont
  {S.}~\bibnamefont {{Kaspi}}}, \bibinfo {author} {\bibfnamefont
  {S.}~\bibnamefont {{Kaye}}}, \bibinfo {author} {\bibfnamefont {M.~S.~P.}\
  \bibnamefont {{Kelley}}}, \bibinfo {author} {\bibfnamefont {M.}~\bibnamefont
  {{Kowalski}}}, \bibinfo {author} {\bibfnamefont {E.}~\bibnamefont
  {{Kramer}}}, \bibinfo {author} {\bibfnamefont {T.}~\bibnamefont {{Kupfer}}},
  \bibinfo {author} {\bibfnamefont {W.}~\bibnamefont {{Landry}}}, \bibinfo
  {author} {\bibfnamefont {R.~R.}\ \bibnamefont {{Laher}}}, \bibinfo {author}
  {\bibfnamefont {C.-D.}\ \bibnamefont {{Lee}}}, \bibinfo {author}
  {\bibfnamefont {H.~W.}\ \bibnamefont {{Lin}}}, \bibinfo {author}
  {\bibfnamefont {Z.-Y.}\ \bibnamefont {{Lin}}}, \bibinfo {author}
  {\bibfnamefont {R.}~\bibnamefont {{Lunnan}}}, \bibinfo {author}
  {\bibfnamefont {M.}~\bibnamefont {{Giomi}}}, \bibinfo {author} {\bibfnamefont
  {A.}~\bibnamefont {{Mahabal}}}, \bibinfo {author} {\bibfnamefont
  {P.}~\bibnamefont {{Mao}}}, \bibinfo {author} {\bibfnamefont {A.~A.}\
  \bibnamefont {{Miller}}}, \bibinfo {author} {\bibfnamefont {S.}~\bibnamefont
  {{Monkewitz}}}, \bibinfo {author} {\bibfnamefont {P.}~\bibnamefont
  {{Murphy}}}, \bibinfo {author} {\bibfnamefont {C.-C.}\ \bibnamefont
  {{Ngeow}}}, \bibinfo {author} {\bibfnamefont {J.}~\bibnamefont {{Nordin}}},
  \bibinfo {author} {\bibfnamefont {P.}~\bibnamefont {{Nugent}}}, \bibinfo
  {author} {\bibfnamefont {E.}~\bibnamefont {{Ofek}}}, \bibinfo {author}
  {\bibfnamefont {M.~T.}\ \bibnamefont {{Patterson}}}, \bibinfo {author}
  {\bibfnamefont {B.}~\bibnamefont {{Penprase}}}, \bibinfo {author}
  {\bibfnamefont {M.}~\bibnamefont {{Porter}}}, \bibinfo {author}
  {\bibfnamefont {L.}~\bibnamefont {{Rauch}}}, \bibinfo {author} {\bibfnamefont
  {U.}~\bibnamefont {{Rebbapragada}}}, \bibinfo {author} {\bibfnamefont
  {D.}~\bibnamefont {{Reiley}}}, \bibinfo {author} {\bibfnamefont
  {M.}~\bibnamefont {{Rigault}}}, \bibinfo {author} {\bibfnamefont
  {H.}~\bibnamefont {{Rodriguez}}}, \bibinfo {author} {\bibfnamefont
  {J.}~\bibnamefont {{van Roestel}}}, \bibinfo {author} {\bibfnamefont
  {B.}~\bibnamefont {{Rusholme}}}, \bibinfo {author} {\bibfnamefont
  {J.}~\bibnamefont {{van Santen}}}, \bibinfo {author} {\bibfnamefont
  {S.}~\bibnamefont {{Schulze}}}, \bibinfo {author} {\bibfnamefont {D.~L.}\
  \bibnamefont {{Shupe}}}, \bibinfo {author} {\bibfnamefont {L.~P.}\
  \bibnamefont {{Singer}}}, \bibinfo {author} {\bibfnamefont {M.~T.}\
  \bibnamefont {{Soumagnac}}}, \bibinfo {author} {\bibfnamefont
  {R.}~\bibnamefont {{Stein}}}, \bibinfo {author} {\bibfnamefont
  {J.}~\bibnamefont {{Surace}}}, \bibinfo {author} {\bibfnamefont
  {J.}~\bibnamefont {{Sollerman}}}, \bibinfo {author} {\bibfnamefont
  {P.}~\bibnamefont {{Szkody}}}, \bibinfo {author} {\bibfnamefont
  {F.}~\bibnamefont {{Taddia}}}, \bibinfo {author} {\bibfnamefont
  {S.}~\bibnamefont {{Terek}}}, \bibinfo {author} {\bibfnamefont
  {A.}~\bibnamefont {{Van Sistine}}}, \bibinfo {author} {\bibfnamefont
  {S.}~\bibnamefont {{van Velzen}}}, \bibinfo {author} {\bibfnamefont {W.~T.}\
  \bibnamefont {{Vestrand}}}, \bibinfo {author} {\bibfnamefont
  {R.}~\bibnamefont {{Walters}}}, \bibinfo {author} {\bibfnamefont
  {C.}~\bibnamefont {{Ward}}}, \bibinfo {author} {\bibfnamefont {Q.-Z.}\
  \bibnamefont {{Ye}}}, \bibinfo {author} {\bibfnamefont {P.-C.}\ \bibnamefont
  {{Yu}}}, \bibinfo {author} {\bibfnamefont {L.}~\bibnamefont {{Yan}}},\ and\
  \bibinfo {author} {\bibfnamefont {J.}~\bibnamefont {{Zolkower}}},\ }\bibfield
   {title} {\bibinfo {title} {{The Zwicky Transient Facility: System Overview,
  Performance, and First Results}},\ }\href
  {https://doi.org/10.1088/1538-3873/aaecbe} {\ \textbf {\bibinfo {volume}
  {131}},\ \bibinfo {pages} {018002} (\bibinfo {year} {2019})},\ \Eprint
  {https://arxiv.org/abs/1902.01932} {arXiv:1902.01932 [astro-ph.IM]}
  \BibitemShut {NoStop}%
\bibitem [{\citenamefont {Chase}\ \emph {et~al.}(2022)\citenamefont {Chase},
  \citenamefont {O'Connor}, \citenamefont {Fryer}, \citenamefont {Troja},
  \citenamefont {Korobkin}, \citenamefont {Wollaeger}, \citenamefont {Ristic},
  \citenamefont {Fontes}, \citenamefont {Hungerford},\ and\ \citenamefont
  {Herring}}]{Chase:2021ood}%
  \BibitemOpen
  \bibfield  {author} {\bibinfo {author} {\bibfnamefont {E.~A.}\ \bibnamefont
  {Chase}}, \bibinfo {author} {\bibfnamefont {B.}~\bibnamefont {O'Connor}},
  \bibinfo {author} {\bibfnamefont {C.~L.}\ \bibnamefont {Fryer}}, \bibinfo
  {author} {\bibfnamefont {E.}~\bibnamefont {Troja}}, \bibinfo {author}
  {\bibfnamefont {O.}~\bibnamefont {Korobkin}}, \bibinfo {author}
  {\bibfnamefont {R.~T.}\ \bibnamefont {Wollaeger}}, \bibinfo {author}
  {\bibfnamefont {M.}~\bibnamefont {Ristic}}, \bibinfo {author} {\bibfnamefont
  {C.~J.}\ \bibnamefont {Fontes}}, \bibinfo {author} {\bibfnamefont {A.~L.}\
  \bibnamefont {Hungerford}},\ and\ \bibinfo {author} {\bibfnamefont {A.~M.}\
  \bibnamefont {Herring}},\ }\bibfield  {title} {\bibinfo {title} {{Kilonova
  Detectability with Wide-field Instruments}},\ }\href
  {https://doi.org/10.3847/1538-4357/ac3d25} {\bibfield  {journal} {\bibinfo
  {journal} {Astrophys. J.}\ }\textbf {\bibinfo {volume} {927}},\ \bibinfo
  {pages} {163} (\bibinfo {year} {2022})},\ \Eprint
  {https://arxiv.org/abs/2105.12268} {arXiv:2105.12268 [astro-ph.HE]}
  \BibitemShut {NoStop}%
\bibitem [{\citenamefont {Cahillane}\ and\ \citenamefont
  {Mansell}(2022)}]{Cahillane:2022pqm}%
  \BibitemOpen
  \bibfield  {author} {\bibinfo {author} {\bibfnamefont {C.}~\bibnamefont
  {Cahillane}}\ and\ \bibinfo {author} {\bibfnamefont {G.}~\bibnamefont
  {Mansell}},\ }\bibfield  {title} {\bibinfo {title} {{Review of the Advanced
  LIGO Gravitational Wave Observatories Leading to Observing Run Four}},\
  }\href {https://doi.org/10.3390/galaxies10010036} {\bibfield  {journal}
  {\bibinfo  {journal} {Galaxies}\ }\textbf {\bibinfo {volume} {10}},\ \bibinfo
  {pages} {36} (\bibinfo {year} {2022})},\ \Eprint
  {https://arxiv.org/abs/2202.00847} {arXiv:2202.00847 [gr-qc]} \BibitemShut
  {NoStop}%
\bibitem [{\citenamefont {Aasi}\ \emph {et~al.}(2015)\citenamefont {Aasi} \emph
  {et~al.}}]{LIGOScientific:2014pky}%
  \BibitemOpen
  \bibfield  {author} {\bibinfo {author} {\bibfnamefont {J.}~\bibnamefont
  {Aasi}} \emph {et~al.} (\bibinfo {collaboration} {LIGO Scientific}),\
  }\bibfield  {title} {\bibinfo {title} {{Advanced LIGO}},\ }\href
  {https://doi.org/10.1088/0264-9381/32/7/074001} {\bibfield  {journal}
  {\bibinfo  {journal} {Class. Quant. Grav.}\ }\textbf {\bibinfo {volume}
  {32}},\ \bibinfo {pages} {074001} (\bibinfo {year} {2015})},\ \Eprint
  {https://arxiv.org/abs/1411.4547} {arXiv:1411.4547 [gr-qc]} \BibitemShut
  {NoStop}%
\bibitem [{\citenamefont {Gupta}\ \emph {et~al.}(2023)\citenamefont {Gupta},
  \citenamefont {Borhanian}, \citenamefont {Dhani}, \citenamefont
  {Chattopadhyay}, \citenamefont {Kashyap}, \citenamefont {Villar},\ and\
  \citenamefont {Sathyaprakash}}]{Gupta:2023evt}%
  \BibitemOpen
  \bibfield  {author} {\bibinfo {author} {\bibfnamefont {I.}~\bibnamefont
  {Gupta}}, \bibinfo {author} {\bibfnamefont {S.}~\bibnamefont {Borhanian}},
  \bibinfo {author} {\bibfnamefont {A.}~\bibnamefont {Dhani}}, \bibinfo
  {author} {\bibfnamefont {D.}~\bibnamefont {Chattopadhyay}}, \bibinfo {author}
  {\bibfnamefont {R.}~\bibnamefont {Kashyap}}, \bibinfo {author} {\bibfnamefont
  {V.~A.}\ \bibnamefont {Villar}},\ and\ \bibinfo {author} {\bibfnamefont
  {B.~S.}\ \bibnamefont {Sathyaprakash}},\ }\bibfield  {title} {\bibinfo
  {title} {{Neutron star-black hole mergers in next generation
  gravitational-wave observatories}},\ }\href
  {https://doi.org/10.1103/PhysRevD.107.124007} {\bibfield  {journal} {\bibinfo
   {journal} {Phys. Rev. D}\ }\textbf {\bibinfo {volume} {107}},\ \bibinfo
  {pages} {124007} (\bibinfo {year} {2023})},\ \Eprint
  {https://arxiv.org/abs/2301.08763} {arXiv:2301.08763 [gr-qc]} \BibitemShut
  {NoStop}%
\bibitem [{\citenamefont {Abbott}\ \emph {et~al.}(2018)\citenamefont {Abbott}
  \emph {et~al.}}]{KAGRA:2013rdx}%
  \BibitemOpen
  \bibfield  {author} {\bibinfo {author} {\bibfnamefont {B.~P.}\ \bibnamefont
  {Abbott}} \emph {et~al.} (\bibinfo {collaboration} {KAGRA, LIGO Scientific,
  Virgo, VIRGO}),\ }\bibfield  {title} {\bibinfo {title} {{Prospects for
  observing and localizing gravitational-wave transients with Advanced LIGO,
  Advanced Virgo and KAGRA}},\ }\href
  {https://doi.org/10.1007/s41114-020-00026-9} {\bibfield  {journal} {\bibinfo
  {journal} {Living Rev. Rel.}\ }\textbf {\bibinfo {volume} {21}},\ \bibinfo
  {pages} {3} (\bibinfo {year} {2018})},\ \Eprint
  {https://arxiv.org/abs/1304.0670} {arXiv:1304.0670 [gr-qc]} \BibitemShut
  {NoStop}%
\bibitem [{\citenamefont {Fritschel}\ \emph {et~al.}(2022)\citenamefont
  {Fritschel}, \citenamefont {Kuns}, \citenamefont {Driggers}, \citenamefont
  {Effler}, \citenamefont {Lantz}, \citenamefont {Ottaway}, \citenamefont
  {Ballmer}, \citenamefont {Dooley}, \citenamefont {Adhikari}, \citenamefont
  {Evans}, \citenamefont {Farr}, \citenamefont {Gonzalez}, \citenamefont
  {Schmidt},\ and\ \citenamefont {Raja}}]{T2200287}%
  \BibitemOpen
  \bibfield  {author} {\bibinfo {author} {\bibfnamefont {P.}~\bibnamefont
  {Fritschel}}, \bibinfo {author} {\bibfnamefont {K.}~\bibnamefont {Kuns}},
  \bibinfo {author} {\bibfnamefont {J.}~\bibnamefont {Driggers}}, \bibinfo
  {author} {\bibfnamefont {A.}~\bibnamefont {Effler}}, \bibinfo {author}
  {\bibfnamefont {B.}~\bibnamefont {Lantz}}, \bibinfo {author} {\bibfnamefont
  {D.}~\bibnamefont {Ottaway}}, \bibinfo {author} {\bibfnamefont
  {S.}~\bibnamefont {Ballmer}}, \bibinfo {author} {\bibfnamefont
  {K.}~\bibnamefont {Dooley}}, \bibinfo {author} {\bibfnamefont
  {R.}~\bibnamefont {Adhikari}}, \bibinfo {author} {\bibfnamefont
  {M.}~\bibnamefont {Evans}}, \bibinfo {author} {\bibfnamefont
  {B.}~\bibnamefont {Farr}}, \bibinfo {author} {\bibfnamefont {G.}~\bibnamefont
  {Gonzalez}}, \bibinfo {author} {\bibfnamefont {P.}~\bibnamefont {Schmidt}},\
  and\ \bibinfo {author} {\bibfnamefont {S.}~\bibnamefont {Raja}},\ }\href
  {https://dcc.ligo.org/LIGO-T2200287/public} {\emph {\bibinfo {title} {Report
  from the LSC Post-O5 Study Group}}},\ \bibinfo {type} {Tech. Rep.}\ \bibinfo
  {number} {T2200287}\ (\bibinfo  {institution} {LIGO},\ \bibinfo {year}
  {2022})\BibitemShut {NoStop}%
\bibitem [{\citenamefont {\'Alvarez-Mu\~niz}\ \emph {et~al.}(2020)\citenamefont
  {\'Alvarez-Mu\~niz} \emph {et~al.}}]{GRAND:2018iaj}%
  \BibitemOpen
  \bibfield  {author} {\bibinfo {author} {\bibfnamefont {J.}~\bibnamefont
  {\'Alvarez-Mu\~niz}} \emph {et~al.} (\bibinfo {collaboration} {GRAND}),\
  }\bibfield  {title} {\bibinfo {title} {{The Giant Radio Array for Neutrino
  Detection (GRAND): Science and Design}},\ }\href
  {https://doi.org/10.1007/s11433-018-9385-7} {\bibfield  {journal} {\bibinfo
  {journal} {Sci. China Phys. Mech. Astron.}\ }\textbf {\bibinfo {volume}
  {63}},\ \bibinfo {pages} {219501} (\bibinfo {year} {2020})},\ \Eprint
  {https://arxiv.org/abs/1810.09994} {arXiv:1810.09994 [astro-ph.HE]}
  \BibitemShut {NoStop}%
\bibitem [{\citenamefont {Aguilar}\ \emph {et~al.}(2021)\citenamefont {Aguilar}
  \emph {et~al.}}]{RNO-G:2020rmc}%
  \BibitemOpen
  \bibfield  {author} {\bibinfo {author} {\bibfnamefont {J.~A.}\ \bibnamefont
  {Aguilar}} \emph {et~al.} (\bibinfo {collaboration} {RNO-G}),\ }\bibfield
  {title} {\bibinfo {title} {{Design and Sensitivity of the Radio Neutrino
  Observatory in Greenland (RNO-G)}},\ }\href
  {https://doi.org/10.1088/1748-0221/16/03/P03025} {\bibfield  {journal}
  {\bibinfo  {journal} {JINST}\ }\textbf {\bibinfo {volume} {16}}\bibfield
  {number} {\bibinfo  {number} { (03)},\ \bibinfo {pages} {P03025}},\ }\bibinfo
  {note} {[Erratum: JINST 18, E03001 (2023)]},\ \Eprint
  {https://arxiv.org/abs/2010.12279} {arXiv:2010.12279 [astro-ph.IM]}
  \BibitemShut {NoStop}%
\bibitem [{\citenamefont {{Otte}}\ \emph {et~al.}(2019)\citenamefont {{Otte}},
  \citenamefont {{Brown}}, \citenamefont {{Doro}}, \citenamefont {{Falcone}},
  \citenamefont {{Holder}}, \citenamefont {{Judd}}, \citenamefont {{Kaaret}},
  \citenamefont {{Mariotti}}, \citenamefont {{Murase}},\ and\ \citenamefont
  {{Taboada}}}]{Otte:2019aaf}%
  \BibitemOpen
  \bibfield  {author} {\bibinfo {author} {\bibfnamefont {N.}~\bibnamefont
  {{Otte}}}, \bibinfo {author} {\bibfnamefont {A.~M.}\ \bibnamefont {{Brown}}},
  \bibinfo {author} {\bibfnamefont {M.}~\bibnamefont {{Doro}}}, \bibinfo
  {author} {\bibfnamefont {A.}~\bibnamefont {{Falcone}}}, \bibinfo {author}
  {\bibfnamefont {J.}~\bibnamefont {{Holder}}}, \bibinfo {author}
  {\bibfnamefont {E.}~\bibnamefont {{Judd}}}, \bibinfo {author} {\bibfnamefont
  {P.}~\bibnamefont {{Kaaret}}}, \bibinfo {author} {\bibfnamefont
  {M.}~\bibnamefont {{Mariotti}}}, \bibinfo {author} {\bibfnamefont
  {K.}~\bibnamefont {{Murase}}},\ and\ \bibinfo {author} {\bibfnamefont
  {I.}~\bibnamefont {{Taboada}}},\ }\bibfield  {title} {\bibinfo {title}
  {{Trinity: An Air-Shower Imaging Instrument to detect Ultrahigh Energy
  Neutrinos}},\ }in\ \href {https://doi.org/10.48550/arXiv.1907.08727} {\emph
  {\bibinfo {booktitle} {Bulletin of the American Astronomical Society}}},\
  Vol.~\bibinfo {volume} {51}\ (\bibinfo {year} {2019})\ p.~\bibinfo {pages}
  {67},\ \Eprint {https://arxiv.org/abs/1907.08727} {arXiv:1907.08727
  [astro-ph.IM]} \BibitemShut {NoStop}%
\end{thebibliography}%

\end{document}